\documentclass[a4paper]{article}
\usepackage[T1]{fontenc}
\usepackage[english]{babel}
\usepackage[numbers]{natbib}
\usepackage[margin=1in]{geometry}
\usepackage{xcolor}
\usepackage{amsmath}
\usepackage{yhmath}
\usepackage{amsthm}
\usepackage{amssymb}
\usepackage{dsfont}
\usepackage{enumitem}
\usepackage{float}
\usepackage[noblocks]{authblk}
\usepackage[hidelinks]{hyperref} 
\usepackage{graphicx}
\graphicspath{{imgs/}}
\usepackage{caption}
\usepackage[ruled,linesnumbered,noresetcount]{algorithm2e}
\usepackage{color}
\usepackage{soul}
\usepackage{mathrsfs}
\usepackage{mathtools}

\usepackage{lineno}
\usepackage{xcolor}

\usepackage[normalem]{ulem}

\usepackage{chngcntr}
\usepackage{apptools}
\AtAppendix{\counterwithin{lemma}{section}}

\usepackage{setspace}
\SetKwRepeat{Do}{do}{while}%

\theoremstyle{plain}
\newtheorem{theorem}{Theorem}[section]
\newtheorem{lemma}{Lemma}
\newtheorem{assumption}{Assumption}
\newtheorem{prop}{Proposition}[section]
\newtheorem{corol}[prop]{Corollary}%
\theoremstyle{definition}
\newtheorem{definition}{Definition}[section]

\theoremstyle{remark}
\newtheorem{remark}{Remark}[section]

\theoremstyle{definition}
\author{ 
Jean-François Chassagneux\textsuperscript{1}, Stéphane Crépey\textsuperscript{2}, Grégoire Loeper\textsuperscript{3}, Justin Ruelland\textsuperscript{2,3}
}

\date{ 
\textsuperscript{1}ENSAE-CREST, Institut Polytechnique de Paris, Palaiseau, France \\
\textsuperscript{2}Laboratoire de Probabilités, Statistique et Modélisation (UMR 8001), Paris, France \\
\textsuperscript{3}BNP Paribas Global Markets, Quantitative Research, Paris, France
}

\title{Well-Posedness of Semimartingale Optimal Transport with Partial Marginal Constraints for Model Calibration\thanks{This research has benefited from the support of \textit{Chair Futures of Quantitative Finance}, a partnership between LPSM at Université Paris Cité, CERMICS at École nationale des ponts et chaussées, and BNP Paribas Global Markets.}}

\DeclareMathOperator*{\argmax}{arg\,max}
\DeclareMathOperator{\Hessian}{Hess}

\newcommand{\wtilde}{\widetilde}

\DeclareMathOperator*{\dom}{dom\,}

\DeclareMathOperator{\BUC}{BUC}

\newcommand{\demi}{\frac{1}{2}}

\newcommand{\R}{\mathbb R}

\newcommand{\E}{\mathbb E}
\newcommand{\bP}{\mathbb P}

\newcommand{\cL}{\mathcal L}
\newcommand{\cP}{\mathcal P}
\newcommand{\cM}{\mathcal M}
\newcommand{\cV}{\mathcal V}

\newcommand{\dt}{\partial_t}

\newcommand{\Dx}{\nabla_x}

\newcommand{\Dxx}{\nabla^2_x}

\begin{document}
\maketitle
\begin{abstract}
We study the well-posedness and stability of a semimartingale optimal
transport problem with partial marginal constraints at several maturities,
motivated by the calibration of financial models to finitely many
option prices. We connect its probabilistic and Eulerian formulations
within the Benamou--Brenier framework. Under convex-duality assumptions,
we establish strong duality for every price vector in the extended-value
sense and existence of a primal optimizer whenever the value is finite.
Under additional PDE assumptions, we establish dual attainment on the
relative interior of the admissible price set, together with uniqueness
of the primal optimizer. Under further non-degeneracy assumptions,
we show that this price set has nonempty interior, on which the dual
optimizer is unique and depends locally Lipschitz continuously on
the prices. The analysis combines convex methods with regularity and
local strong convexity results for Hamilton--Jacobi--Bellman solution
maps established in a companion paper. For a fixed deterministic
initial state, we also prove continuity of the optimal law on path
space in the adapted Wasserstein distance of order two as prices
vary within this interior. This stability concerns the optimizer
over the full admissible class of semimartingale laws, allowing
characteristics that depend on the past.
The framework accommodates bounded uniformly continuous payoffs,
including European puts in log-price coordinates, without payoff
regularization or spatial truncation.
\end{abstract}

\section*{Introduction}
\addcontentsline{toc}{section}{Introduction}
Calibrating a financial model to a finite set of option prices at
several maturities imposes constraints on the marginal distributions
of the underlying process. These partial marginal constraints do not,
in general, determine the marginal distributions in full, nor the law
of the process on path space.
Approaches that prescribe full marginal distributions require these
distributions to be specified from the available quotes, typically
through additional interpolation or extrapolation choices.
Here, the marginal distributions are themselves part of the
optimization: they are selected jointly with the process dynamics,
subject to the observed price constraints.
Semimartingale optimal transport provides
a variational approach to selecting a calibrated model by minimizing
a cost depending on its drift and covariance characteristics among
the laws that reproduce the prescribed prices. This leads to the
questions studied in this paper: under what conditions does an optimal
law exist, when is it unique, and how does it depend on the calibration
prices?

More precisely, fix a time horizon $T>0$, an initial law $\mu_0$,
maturity times $\tau=(\tau_1,\ldots,\tau_P)$, and a finite family
of payoffs $G$. For a vector of observed prices $\pi$, let
$\mathcal P(\mu_0,\tau,\pi)$ denote the admissible laws of continuous
semimartingales with initial law $\mu_0$ for which the expected
payoffs at the corresponding maturities equal the prescribed prices.
Given a running cost $F$, the calibration problem is
\begin{equation*}
    \mathcal V(\pi)
    :=\inf_{\mathbb P\in\mathcal P(\mu_0,\tau,\pi)}
    \mathbb E^{\mathbb P}\!\left[
        \int_0^T F(t,X_t,\alpha_t^{\mathbb P},\beta_t^{\mathbb P})\,dt
    \right].
\end{equation*}
Here $X$ is the canonical process, and $\alpha^{\mathbb P}$ and
$\beta^{\mathbb P}$ are its drift and instantaneous covariance
characteristics under $\mathbb P$.
These characteristics may depend on the entire past trajectory;
no Markovian restriction is imposed on the initial admissible class.
The running cost $F$ may take the value $+\infty$ to enforce
structural constraints on the admissible dynamics, including
drift restrictions associated with risk-neutral modelling.
These restrictions complement the option-price constraints
encoded in $\mathcal P(\mu_0,\tau,\pi)$.
The precise admissibility and integrability conditions are given
in Section~\ref{sec:framework}.

Existence and uniqueness alone do not ensure that a calibration
procedure is robust to errors in its input data. Even when the
optimal model is uniquely determined, small perturbations of the
prescribed prices could otherwise lead to substantially different
calibrated dynamics. Stability completes the well-posedness of the
problem by requiring continuous dependence of the solution on the
data. It is particularly relevant when prices are known only up to
observation or rounding errors: sufficiently small admissible
perturbations should then produce nearby optimal laws in a suitable
topology. This regularity of the calibration map also provides an
analytical foundation for numerical approximation, since computations
use data represented with finite precision and must account for the
sensitivity of the solution to these inputs.
We therefore study the dependence of the optimal law on the
price vector $\pi$, keeping the running cost $F$, the payoff family
$G$, the maturity times $\tau$, and the initial law $\mu_0$ fixed.
Our analysis also concerns the optimal Lagrange multipliers associated
with the price constraints in the dual problem. Their dependence on
the prices provides the link between the dual formulation and the
stability of the calibrated law on path space.

\medskip

Let $\mathfrak P:=\{\pi:\mathcal V(\pi)<+\infty\}$ denote the set
of admissible prices, namely those that can be calibrated with finite
cost for the chosen running cost and admissible model class.
We first connect the probabilistic and Eulerian formulations of the
calibration problem and establish strong duality under convex-duality
assumptions for every price vector, allowing infinite values.
For every $\pi\in\mathfrak P$,
the primal problem admits an optimizer
(Theorem~\ref{thm:regular-strong-duality} and
Corollary~\ref{coro:primal-minimum}). Under additional PDE assumptions,
we identify the dual problem through a cascade of
Hamilton--Jacobi--Bellman equations and establish dual attainment
for prices in the relative interior of the admissible price set
(Proposition~\ref{prop:dual-attainement}).
The resulting HJB dual problem is an optimization over finitely
many dual variables, one for each price constraint, with an
objective evaluated through the HJB cascade.
Our analysis draws on the
companion article~\cite{HJB-property}, which studies regularity,
stability, and convexity properties of solution maps for fully
nonlinear parabolic HJB equations.
Under its differentiability assumptions, the gradient of the
dual objective can be computed from the solutions of the
linearized HJB cascade. This finite-dimensional formulation
and access to its gradient offer a route to efficient numerical
calibration using gradient-based optimization methods.
We also derive optimality conditions linking the drift and covariance
characteristics of any optimal calibrated model to a solution of
the dual HJB equations.

For prices with finite value and an attained dual optimum,
we then establish uniqueness of the primal optimizer under the
corresponding PDE hypotheses, by combining
these optimality conditions with uniqueness in law for the associated
stochastic differential equation
(Corollary~\ref{coro:primal-uniqueness-pde}). Under additional
non-degeneracy assumptions, the convexity results of this companion
article yield local strong convexity of the initial HJB value
as a function of the dual variables.
We show that $\mathfrak P$ has nonempty interior in the ambient
Euclidean price space
(Corollary~\ref{coro:dual-conjugate-regularity}) and that, on this
interior, the optimal multiplier is unique and depends locally
Lipschitz continuously on the prices
(Proposition~\ref{prop:main-stability-result}).
The existence of this nonempty open region is itself part of the
conclusion: it provides a domain for local stability in which
sufficiently small perturbations of each price vector in any
direction remain admissible.

Finally, for a fixed deterministic initial state, we prove continuity
of the optimal law in the adapted Wasserstein distance of order two.
Writing $\mathbb P^{\star,\pi}$
for the unique optimal law at price $\pi$,
Theorem~\ref{thm:stability-primal} gives
\begin{equation*}
    \pi_n\longrightarrow\pi
    \quad\text{in }\operatorname{int}(\mathfrak P)
    \quad\Longrightarrow\quad
    \mathcal{AW}_2\!\left(
        \mathbb P^{\star,\pi_n},\mathbb P^{\star,\pi}
    \right)\longrightarrow0.
\end{equation*}
This distance compares the martingale and finite-variation parts
of the processes through couplings that respect the flow of
information in both directions; its definition is recalled in
Section~\ref{subsec:primal-stability}.
The result concerns the optimizer over the full probabilistic
admissible class and controls its law on path space, beyond the
convergence of its time marginals.
The conclusion for the optimal laws is continuity, without an
asserted quantitative rate.

\medskip

Our duality framework uses a convex running cost whose effective
domain imposes uniform bounds on the drift and covariance
characteristics, together with continuity assumptions on the
associated Hamiltonian. The stability results require additional
regularity of the Hamiltonian, uniform ellipticity of the associated
diffusions, and a non-degeneracy condition on the HJB solution map.
These additional conditions allow us to apply the results of the
companion article; they do not enter the strong duality theorem.
The precise assumptions are stated alongside the corresponding
results in Sections~\ref{sec:duality} and~\ref{sec:stability}.

The payoff functions are assumed to be bounded and uniformly
continuous on $\R^d$, but need not be differentiable.
This includes European put payoffs expressed in the log-price
variable: for a strike $K>0$, the function
$x\mapsto(K-e^x)^+$ is bounded and uniformly continuous on $\R$,
despite its kink at $x=\log K$.
When the forward prices and discount factors are fixed, put--call
parity allows us to incorporate call quotes through the corresponding
put-price constraints, while retaining bounded payoff functions.
The analysis is carried out on the full state space $\R^d$,
without smoothing the payoff functions or truncating the spatial
domain.

\medskip

The connection between optimal transport and stochastic control
was developed by Mikami and Thieullen~\cite{mikami-thieulen2006},
who established a duality theorem for a stochastic control problem
with convex cost, and by Tan and Touzi~\cite{tan2013optimal}, who
established a Kantorovich-type duality for transport along continuous
semimartingales with prescribed initial and terminal
distributions, allowing control of both the drift and
covariance characteristics.
Benamou and Brenier~\cite{BB99} introduced a dynamic formulation
of quadratic optimal transport, in which a time-dependent
density and velocity field connect the prescribed initial and
terminal distributions through the continuity equation,
while minimizing kinetic energy.
In the semimartingale setting, the Lagrangian description replaces
the ordinary differential equation associated with the velocity
field by a stochastic differential equation with drift and
diffusion coefficients. In the Eulerian description, which tracks
the marginal distributions, the continuity equation is correspondingly
replaced by the Fokker--Planck equation.
For continuous-time martingale optimal transport, Huesmann and
Trevisan~\cite{huesmann-trevisan} established the equivalence
between the probabilistic and Eulerian formulations and used
the latter to prove existence and strong duality, with a dual
problem expressed in terms of HJB supersolutions.

For model calibration, Guo, Loeper, and
Wang~\cite{guo2022lsv-calibration} formulated the calibration of
local-stochastic volatility models as a semimartingale optimal
transport problem constrained by finitely many European option
prices, with an Eulerian formulation and an HJB dual problem.
Applications of this approach include the joint calibration of
SPX and VIX by Guo, Loeper, Ob\l\'oj, and
Wang~\cite{guo2022jointVIX}, and the joint calibration of local
volatility and stochastic interest-rate models by Joseph, Loeper,
and Ob\l\'oj~\cite{joseph-obloj-loeper}.
These works provide the variational calibration framework in
which we study the well-posedness and stability of the optimal
model under partial marginal constraints.
For an overview of optimal transport methods for model calibration
under finitely many option-price constraints, we refer to the survey
by Guo, Loeper, Ob\l\'oj, and Wang~\cite{SurveyOT}.

Beyond existence and duality, the dependence of a calibrated model
on the observed prices is a central question in calibration.
For local volatility calibration from either a continuum or
a finite set of option prices, Cr\'epey~\cite{crepey2003calibration}
established stability of Tikhonov-regularized solutions and,
under suitable parameter choices, subsequential convergence
towards a solution of minimum distance to a prescribed reference
coefficient, assuming the data are calibrateable. Avellaneda, Friedman, Holmes, and Samperi~\cite{avellaneda1997calibration}
formulated calibration to finitely many option prices as a
constrained stochastic control problem based on relative-entropy
minimization, combining a Bellman equation with an optimization
over the calibration multipliers.
Samperi~\cite{samperi} subsequently studied solvability and
stability of a regularized version of this approach, retaining
exact pricing constraints for payout profiles smoothed in both
space and time, with the process stopped at spatial barriers.
His analysis uses the regularity and convexity of the dual value
function to relate
the dependence of the optimal multipliers on the calibration
prices to that of the optimal volatility and the induced pricing
operator. Our work follows a related dual approach, formulating
calibration as a semimartingale optimal transport problem.
Through its Eulerian formulation, we establish strong duality
for every price vector in the extended-value sense, under
assumptions separate from those needed for stability. Samperi
obtains equality of the primal and dual values on the gradient
image of a region where the dual value function has the required
regularity and convexity.
We treat bounded uniformly continuous payoffs at prescribed maturities on
the full state space, whereas Samperi's PDE analysis is carried out on
a bounded spatial domain. We also establish stability directly for
the optimal law on path space, in adapted Wasserstein distance.

Our choice of adapted Wasserstein distance is motivated by the role
of information in financial models.
Backhoff-Veraguas, Bartl, Beiglb\"ock, and
Eder~\cite{backhoff-adaptedwas} showed
how transport distances respecting the temporal information
structure yield stability results for hedging in semimartingale
models.

\medskip

The paper is organised as follows.
Section~\ref{sec:framework} introduces the calibration problem
and its probabilistic and Eulerian formulations.
Section~\ref{sec:duality} establishes strong duality, existence
of primal and dual optimizers under the respective hypotheses,
and optimality conditions.
Section~\ref{sec:stability} establishes uniqueness of primal
and dual optimizers, local Lipschitz stability of the optimal
dual variables, and continuity of the optimal laws in adapted
Wasserstein distance.
Appendices~\ref{app:convex-analysis}
and~\ref{app:feedback-coefficient-estimates} provide the auxiliary
convex-analytic result and the estimates for the HJB feedback
coefficients used in the stability analysis.

\section{The calibration problem and its formulations}\label{sec:framework}
\subsection{Notation and probabilistic framework}

Fix a time horizon $T>0$ and a state-space dimension $d\geq1$.
For vectors $x,y\in\R^n$, in any finite dimension $n$, we write
\begin{equation*}
    x\cdot y:=x^\top y,
    \qquad |x|:=(x\cdot x)^{1/2}.
\end{equation*}
For real matrices $A$ and $B$ of the same dimensions, we use
the Frobenius inner product and norm
\begin{equation*}
    A:B:=\operatorname{Tr}(A^\top B),
    \qquad \|A\|:=(A:A)^{1/2}.
\end{equation*}
Unless otherwise specified, single bars denote the absolute value
of a scalar or the Euclidean norm of a vector, and double bars
around a matrix denote its Frobenius norm.

We denote by $\mathcal S^d$ the space of real symmetric
$d\times d$ matrices and by $\mathcal S^d_+$ its cone of positive
semidefinite matrices. For $A\in\mathcal S^d_+$, the notation $A^{1/2}$ refers
to its unique symmetric positive semidefinite square root.

On $\R^d\times\mathcal S^d$, we use the product inner product
and its associated norm,
\begin{equation*}
    \langle(a,b),(a',b')\rangle:=a\cdot a'+b:b',
    \qquad |(a,b)|:=\bigl(|a|^2+\|b\|^2\bigr)^{1/2}.
\end{equation*}
Gradients with respect to a symmetric matrix variable are taken
relative to the Frobenius inner product. Thus, for a differentiable
function $f:\R^d\times\mathcal S^d\to\R$,
\begin{equation*}
    Df(a,b)[h,K]
    =\nabla_a f(a,b)\cdot h+\nabla_b f(a,b):K,
    \qquad (h,K)\in\R^d\times\mathcal S^d.
\end{equation*}
The norm of a linear operator is denoted by $\|\cdot\|_{\mathrm{op}}$
and is induced by the specified norms on its domain and codomain.
For a bilinear form, the same notation denotes the supremum of
its absolute value over pairs of unit vectors.
Other product norms used for function spaces will be specified
where they are introduced.

For a metric space $E$ and a finite-dimensional normed space $V$,
we denote by $C(E;V)$ the space of continuous functions from $E$
to $V$, by $C_b(E;V)$ its subspace of bounded functions, and by
$\BUC(E;V)$ the space of bounded uniformly continuous functions.
Uniform continuity is understood with respect to the specified
metric on $E$; subsets of Euclidean spaces carry the induced metric.
For $f:E\to V$ and a nonempty subset $D\subseteq E$, we write
\begin{equation*}
    \|f\|_{\infty;D}:=\sup_{x\in D}\|f(x)\|_V,
    \qquad \|f\|_\infty:=\|f\|_{\infty;E}.
\end{equation*}
The target norm $\|\cdot\|_V$ follows the preceding conventions
for scalars, vectors, matrices, and their products.

For an open set $D\subseteq\R^n$ and an integer $k\geq1$,
$C^k(D;V)$ denotes the space of functions with continuous
derivatives up to order $k$. For a scalar function $u(t,x)$,
we write $\partial_tu$, $\Dx u$, and $\Dxx u$ for its time
derivative, spatial gradient, and spatial Hessian, respectively.
On an open set $Q\subseteq\R\times\R^d$, the space $C^{1,2}(Q)$
consists of functions $u$ for which $u$, $\partial_tu$, $\Dx u$,
and $\Dxx u$ are continuous.
The subscript $b$ in $C_b^k$ and $C_b^{1,2}$ requires the function
and these derivatives to be bounded; the subscript $c$ indicates
compact support in the specified domain.
On a closed time interval, derivatives are understood in the
interior and continuously extended to the temporal endpoints,
with one-sided time derivatives there. Compactly supported
functions may be nonzero at these endpoints. Parabolic H\"older
norms are specified or referenced where they are used.

For a positive Borel measure $\mu$ on $E$, the space $L^1(\mu;V)$
consists of measurable functions $f:E\to V$ satisfying
\begin{equation*}
    \|f\|_{L^1(\mu;V)}:=\int_E\|f(x)\|_V\,\mu(dx)<\infty,
\end{equation*}
identified up to $\mu$-almost everywhere equality.
For real-valued functions, we omit the target space and write
$C(E)$, $C_b(E)$, $\BUC(E)$, and $L^1(\mu)$.

For a Polish space $E$ equipped with its Borel $\sigma$-algebra,
we denote by $\mathcal M(E)$ the space of finite signed Borel
measures, by $\mathcal M_+(E)$ its cone of nonnegative measures,
and by $\mathcal P(E)$ the set of Borel probability measures.
The notation $\mathcal M(E;V)$ denotes finite $V$-valued Borel
measures. For $\mu\in\mathcal M(E)$, we write $|\mu|$ for its
total variation measure and
\begin{equation*}
    \mu(f):=\int_E f\,d\mu,
    \qquad f\in L^1(|\mu|).
\end{equation*}
For a Borel measurable map $f:E\to E'$ between Polish spaces,
the pushforward measure $f_\#\mu$, also written $f\#\mu$, is
defined by $(f_\#\mu)(B)=\mu(f^{-1}(B))$ for every Borel set
$B\subseteq E'$.

We equip $\mathcal P(E)$ with the topology of narrow convergence:
$\mu_n\to\mu$ narrowly if
\begin{equation*}
    \mu_n(f)\longrightarrow\mu(f)
    \qquad\text{for every }f\in C_b(E).
\end{equation*}
Whenever a topology on signed measures is needed, its test
function space will be specified. If $E$ is also locally compact,
$C_0(E)$ denotes the continuous functions vanishing at infinity,
and $C_0(E)^*$ is identified with $\mathcal M(E)$ by integration.
We distinguish this identification from the dual space $C_b(E)^*$
of bounded linear functionals on $C_b(E)$, whose elements need
not be represented by finite Borel measures. Its weak-$*$
topology is $\sigma(C_b(E)^*,C_b(E))$.
We write $\langle\ell,f\rangle=\ell(f)$ for the pairing between
a continuous linear functional and an element of its underlying space.

For a path $a:[s,t]\to\R^d$, with $s<t$, its total variation is
\begin{equation}\label{eq:notation-path-total-variation}
    \operatorname{Var}_{[s,t]}(a)
    :=\sup_{s=t_0<\cdots<t_N=t}
        \sum_{j=1}^N |a(t_j)-a(t_{j-1})|,
\end{equation}
where the supremum is taken over all finite partitions of $[s,t]$.
The path has finite variation if this quantity is finite.
For an absolutely continuous path, it equals
$\int_s^t|\dot a(r)|\,dr$.
For a stochastic process $A$, this definition is applied to each
sample path $r\mapsto A_r(\omega)$.

Unless explicitly fixed, constants denoted by $C$ may change
from one estimate to the next. Their relevant dependencies
are indicated locally.

\medskip

We work on the canonical path space
$\Omega:=C([0,T];\R^d)$, equipped with the uniform norm and
its Borel $\sigma$-algebra. The canonical process is
$X_t(\omega):=\omega(t)$, and its raw natural filtration is
$\mathbb F=(\mathcal F_t)_{0\leq t\leq T}$, where
\begin{equation*}
    \mathcal F_t:=\sigma(X_s:0\leq s\leq t).
\end{equation*}
In particular, $\mathcal F_T$ is the Borel $\sigma$-algebra of
$\Omega$, and probability laws are defined on $(\Omega,\mathcal F_T)$.
For a fixed law $\mathbb P$, stochastic calculus is understood
with respect to the usual $\mathbb P$-augmentation of $\mathbb F$,
obtained by completion and right-continuous modification.

We consider laws $\mathbb P$ under which $X$ admits the
semimartingale decomposition
\begin{equation*}
    X_t=X_0+A_t^{\mathbb P}+M_t^{\mathbb P},
    \qquad t\in[0,T],\quad\mathbb P\text{-a.s.},
\end{equation*}
where $A^{\mathbb P}$ is a continuous adapted process of finite
variation and $M^{\mathbb P}$ is a continuous martingale, both
starting at zero. Since $M^{\mathbb P}$ is continuous, its matrix
quadratic variation $[M^{\mathbb P}]$ is predictable and coincides
with its predictable bracket $\langle M^{\mathbb P}\rangle$.
We use both notations interchangeably.
We require $A^{\mathbb P}$ and $\langle M^{\mathbb P}\rangle$
to be absolutely continuous with respect to Lebesgue measure
in time, $\mathbb P$-almost surely.
Their densities admit progressively measurable versions
$(\alpha^{\mathbb P},\beta^{\mathbb P})$ with values in
$\R^d\times\mathcal S^d_+$, so that
\begin{equation*}
    A_t^{\mathbb P}=\int_0^t\alpha_s^{\mathbb P}\,ds,
    \qquad
    \langle M^{\mathbb P}\rangle_t
    =\int_0^t\beta_s^{\mathbb P}\,ds.
\end{equation*}
We call $\alpha^{\mathbb P}$ and $\beta^{\mathbb P}$ the drift
and covariance characteristics of $\mathbb P$, respectively.
They are identified up to $dt\otimes\mathbb P$-almost everywhere
equality and may depend on the past of the process.
The associated symmetric diffusion coefficient is
$\sigma^{\mathbb P}:=(\beta^{\mathbb P})^{1/2}$, satisfying
$\sigma^{\mathbb P}(\sigma^{\mathbb P})^\top=\beta^{\mathbb P}$.

We denote by $\mathcal P\subseteq\mathcal P(\Omega)$ the class
of laws satisfying the preceding conditions, and by
$\mathcal P^1$ its subclass satisfying
\begin{equation}\label{eq:-char-integrability}
    \mathbb E^{\mathbb P}\!\left[
        \int_0^T\bigl(|\alpha_t^{\mathbb P}|
        +\|\beta_t^{\mathbb P}\|\bigr)\,dt
    \right]<\infty.
\end{equation}
We further define $\mathcal{SM}_2$ as the set of laws
$\mathbb P\in\mathcal P^1$ such that
\begin{equation*}
    \mathbb E^{\mathbb P}\!\left[
        \int_0^T\operatorname{Tr}(\beta_t^{\mathbb P})\,dt
        +\left(\int_0^T|\alpha_t^{\mathbb P}|\,dt\right)^2
    \right]<\infty.
\end{equation*}
Equivalently, this condition reads
\begin{equation*}
    \mathbb E^{\mathbb P}\!\left[
        \operatorname{Tr}\langle M^{\mathbb P}\rangle_T
        +\operatorname{Var}_{[0,T]}(A^{\mathbb P})^2
    \right]<\infty,
\end{equation*}
where total variation is defined
by~\eqref{eq:notation-path-total-variation}.
These are the two integrability terms entering the adapted
Wasserstein distance of order two in
Subsection~\ref{subsec:primal-stability}.

Finally, $\mathcal P^1_{\mathcal M}$ denotes the set of laws
$\mathbb P\in\mathcal P^1$ for which there exist Borel measurable
maps $\alpha:[0,T]\times\R^d\to\R^d$ and
$\beta:[0,T]\times\R^d\to\mathcal S^d_+$ such that
\begin{equation*}
    \alpha_t^{\mathbb P}=\alpha(t,X_t),
    \qquad \beta_t^{\mathbb P}=\beta(t,X_t),
    \qquad dt\otimes\mathbb P\text{-a.e.}
\end{equation*}
Thus, this subclass requires the characteristics to depend only
on time and the current state.

For every $\mathbb P\in\mathcal P$, the absolute continuity of
$\langle M^{\mathbb P}\rangle$ yields a Brownian representation
on an extension of the filtered probability space; see
\cite[Chapter~3, Theorem~4.2]{karatzas-shreve1991}.
The Brownian representation can be chosen with diffusion
coefficient $(\beta^{\mathbb P})^{1/2}$. Thus there exists a $d$-dimensional
Brownian motion $W^{\mathbb P}$ on this extension such that
\begin{equation}\label{eq:sde-doob-decomp}
    X_t=X_0+\int_0^t\alpha_s^{\mathbb P}\,ds
        +\int_0^t(\beta_s^{\mathbb P})^{1/2}\,dW_s^{\mathbb P},
    \qquad t\in[0,T],
\end{equation}
almost surely under the extended probability measure.
Here we retain the same notation for the lifted processes;
the law of $X$ on $\Omega$ remains $\mathbb P$.

\subsection{Calibration problem}

The calibration problem is formulated as the minimization of a cost functional. This functional can be interpreted as a “distance” to a prescribed reference model while enforcing exact calibration. Writing $\Lambda=[0,T]\times\R^d$, we introduce a running cost $F:\Lambda \times \mathbb{R}^d \times \mathcal{S}^d \to [0,+\infty]$, which is proper, convex, and lower semicontinuous with respect to the variables $(\alpha,\beta)$ for every $(t,x)\in\Lambda$. Additional assumptions on $F$ will be stated where needed; in particular, Assumption~\ref{ass:convex-duality} specifies the effective-domain bounds and Hamiltonian continuity used for strong duality. We impose $F=+\infty$ whenever $\beta \notin \mathcal{S}^d_+$ so as to guarantee that $\beta$ represents a valid covariance matrix. This convention will be used implicitly throughout the paper, for instance when restricting to $\beta\in\mathcal{S}^d_+$.

The following additional assumption is imposed only when explicitly invoked.
\begin{assumption}[Strict convexity of the running cost]
\label{ass:strict-convexity-cost}
For every $(t,x)\in\Lambda$, the function
$(\alpha,\beta)\mapsto F(t,x,\alpha,\beta)$ is strictly convex
on its effective domain.
\end{assumption}

The associated Hamiltonian is the Legendre--Fenchel transform of $F$
with respect to the drift and covariance variables
(see \cite{rockafellar-1970a}): \begin{equation*}
H(t,x,a,b):=F^*(t,x,a,b):=\sup_{\alpha\in\R^d,\beta\in\mathcal{S}_+^d}\{\alpha\cdot a + \beta : b-F(t,x,\alpha,\beta)\}.
\end{equation*} Strictly speaking, the supremum is taken over $(\alpha,\beta)\in\mathbb{R}^d\times\mathcal{S}^d$, but thanks to the convention adopted above it can be restricted to $\beta\in\mathcal{S}^d_+$. To simplify notation, when no ambiguity arises we write $F(\alpha,\beta):= F(t,x,\alpha,\beta)$ and $F^*(a,b):= F^*(t,x,a,b)$.

We now formulate the calibration objective as an optimal transport problem. 
Following \cite{guo2022lsv-calibration, joseph-obloj-loeper}, since the market 
constraints depend only on marginal distributions and the cost functional is convex, 
a combination of the Markovian projection and Jensen's inequality allow us to restrict attention to a Markovian formulation, where the characteristic coefficients are measurable functions of $(t, X_t)$. The superposition principle then yields a PDE formulation of the calibration problem, placing it within the Benamou--Brenier framework.

We aim to calibrate our model to market prices of European options with different maturities. Let $P \geq 1$ be the number of maturities, denoted by $(\tau_p)_{p=1}^P$ with $0 = \tau_0 < \tau_1 < \dots < \tau_{P} = T$. For each maturity $\tau_p$, we consider $M \geq 1$ options. We introduce the notation $i(p,m)=(p-1)\times M + m$. The $m^{\mathrm{th}}$ option at maturity $\tau_p$ is associated with a payoff $G_{i(p,m)} \in \BUC(\mathbb{R}^d)$ and a market price $\pi_{i(p,m)} \in \mathbb{R}$. The convention $G_{i(p,m)}\equiv0$ and $\pi_{i(p,m)}=0$ may be used only as a notational padding device for maturities with fewer quoted options. Such dummy coordinates are ignored in the non-degeneracy assumptions and in the uniqueness statements for the dual multipliers. Although European call payoffs are not bounded, they are equivalent calibration instruments to put options via the usual call--put parity. Put option payoffs are bounded and uniformly continuous on $\R_+$, showing that the assumption $G_{i(p,m)} \in \BUC(\mathbb{R}^d)$ is relevant in practical settings. This remains true when working with the log-spot variable, since $x\mapsto (K-\exp(x))^+$ is bounded and uniformly continuous on $\R$. In the case of put options, $G_{i(p,m)}$ is equivalently characterized by the pair $(\tau_p, K_{i(p,m)})$, where $K_{i(p,m)}$ denotes the strike. We denote $\tau=(\tau_1,\dots,\tau_P) \in \R^P$, $G(x)=(G_{1}(x),\dots,G_{M \times P}(x))\in\R^{MP}$ similarly for $\pi \in\R^{MP} $, and $K\in\R^{MP}$ when considering strike-based options.

In the following, we consider the following regularity and non-degeneracy assumption on the payoff family.

\begin{assumption}\label{hyp:G}
For every $p\in\{1,\dots,P\}$, the payoff vector satisfies
\begin{equation*}
    (G_{i(p,1)},\dots,G_{i(p,M)})
    \in\BUC(\mathbb R^d;\mathbb R^M).
\end{equation*}
Moreover, its components are linearly independent modulo constants:
if $\sum_{m=1}^M \mu_m G_{i(p,m)}(\cdot)$ is constant
on $\mathbb R^d$, then $\mu=0$.
\end{assumption}

\begin{remark}\label{rk:strikes-assumptions}
For standard strike-based put payoffs, Assumption~\ref{hyp:G} is satisfied, for
each maturity, when the strikes $(K_{i(p,m)})_{m=1}^M$ are strictly positive and pairwise distinct provided the state variable is chosen so that the payoffs are bounded on the state space, for instance for log-put payoffs $x\mapsto (K-e^x)^+$. The linear independence assumption is natural under hard no-arbitrage price
constraints, since any non-trivial affine relation among the payoffs would have
to be reflected by the corresponding market prices. Removing such redundant
coordinates leaves the calibration problem unchanged while restoring
identifiability of the dual multipliers.
\end{remark}

Given an initial distribution $\mu_0$, maturity times $\tau$, market prices $\pi$ and the associated payoffs $G$, we introduce the set of market calibrated measures\begin{equation}
    \cP(\mu_0,\tau,\pi) := \left\{ \mathbb{P}\in \cP^1 : X_0 \# \bP = \mu_0, \E^\bP\left[ G_{i(p,m)}(X_{\tau_p}) \right] = \pi_{i(p,m)} \text{ for } 1\leq m\leq M,\ 1\leq p\leq P\right\}.
\end{equation} In the following, we will assume $\mu_0 = \delta_{x_0}$ with $x_0 \in \mathbb{R}^d$. All the results remain valid for another $\mu_0$, provided appropriate integrability conditions hold.

\begin{remark}
When interest rates are deterministic, discounting can be absorbed
into the calibration prices by dividing each quoted price by the
corresponding discount factor.
Stochastic interest rates can instead be incorporated into the
probabilistic formulation by augmenting the state with the short
rate and its time integral. Discounted option payoffs then become
functions of the augmented state at the prescribed maturities.
This leads to an augmented Markovian formulation of the type
discussed below. Provided that the augmented process belongs
to $\mathcal P^1$ and the discounted payoffs are integrable,
the projection preserves the calibration constraints.
The subsequent results do not apply automatically, however:
the augmented payoffs need not be bounded, and the integrated-rate
coordinate introduces a degenerate diffusion direction.
For a formulation tailored to stochastic interest rates, based
on discounted densities, we refer to
Joseph, Loeper, and Ob{\l}\'oj~\cite{joseph-obloj-loeper}.
\end{remark}

Within this framework, we introduce the calibration problem, referred to as the 
probabilistic primal problem. It consists in selecting a calibrated model in 
$\mathcal{P}(\mu_0,\tau,\pi)$ that minimizes the cost. \begin{equation}
    \cV(\pi) := \inf_{\bP \in \cP(\mu_0,\tau,\pi)} \E^\bP \left[ \int_0^T F(t,X_t, \alpha^\bP_t,\beta^\bP_t)\,dt \right].
\end{equation}
The notation $\mathcal V(\pi)$ emphasizes dependence on the market
prices, with $F$, $\mu_0$, $\tau$, and $G$ held fixed.
We also write $\mathcal V$ when this dependence is not at issue.

\subsection{Markovian projection} 

We introduce the set of market-calibrated Markovian measures $\cP_{\mathcal{M}}(\mu_0,\tau,\pi):=\cP(\mu_0,\tau,\pi)\cap \cP^1_{\mathcal{M}}.$
By the Markovian projection (see the mimicking results of \cite{brunick-shreve}, 
which extend the work of \cite{gyongy} following a remark of N.\,V.~Krylov), 
the probabilistic primal problem can be reduced to this class of measures. Under such measures, the characteristics of the semimartingale $X$ are measurable functions of $(t,X_t)$. We formalise this result in the following theorem.

The conditional expectations in the following theorem are taken
under the fixed law $\mathbb P$. The existence of representatives
jointly measurable in $(t,x)$ is part of its conclusion.

\begin{theorem}\label{thm:markovian-proj}
    Let $\bP \in \cP^1$ and consider $(\alpha^\bP_t,\beta^\bP_t)$ the associated characteristics. Then there exists $\wtilde{\bP} \in \cP^1_{\mathcal{M}}$ such that $\forall t \in[0,T], \, X_t \# \bP = X_t \# \wtilde{\bP}$.\\
    Moreover, there exist measurable functions $\alpha:[0,T]\times \R^d \to \R^d$ and $\beta:[0,T]\times \R^d \to \mathcal{S}^d_+$ such that $\alpha(t,X_t) = \E^\bP \left[ \alpha^\bP_t | X_t \right]$ and $\beta(t,X_t)= \E^\bP \left[ \beta^\bP_t | X_t \right]$  $dt\otimes d\bP$-a.e.. The measure $\wtilde{\bP}$ has characteristics $(\alpha^{\wtilde{\bP}},\beta^{\wtilde{\bP}})$, such that $\alpha^{\wtilde{\bP}}_t = \alpha(t,X_t)$ and $\beta^{\wtilde{\bP}}_t=\beta(t,X_t)$ $dt\otimes d\wtilde{\bP}$-a.e..\\
    Furthermore, possibly on an extension $(\widehat{\mathit{\Omega}},\widehat{\mathbb{F}},\widehat{P})$, of the canonical filtered probability space $(\mathit{\Omega},\mathbb{F},\wtilde{\bP})$, there exists a $d$-dimensional $\widehat{\mathbb{F}}$ adapted Brownian motion $W^{\widehat{\mathbb{P}}}$ such that the lifted canonical process $X$ is satisfying: \begin{equation}\label{eq:thm-markovian-proj}
        X_t = X_0 + \int_0^t \alpha(s,X_s)ds + \int_0^t \left(\beta(s,X_s)\right)^{1/2} dW^{\widehat{\bP}}_s, \quad t\leq T.
    \end{equation}
\end{theorem}

\noindent Theorem~\ref{thm:markovian-proj} is a direct consequence of Corolarry~3.7 in \cite{brunick-shreve}, applied to the canonical process.\\

Since the calibration constraints depend only on the marginal
distributions of the state process at the prescribed maturities,
they are preserved by the preceding projection. We therefore
consider the following Markovian formulation:
\begin{equation}
    \cV_{\mathcal{M}} := \inf_{\bP \in \cP_{\mathcal{M}}(\mu_0,\tau,\pi)} \E^\bP \left[ \int_0^T F(t,X_t,\alpha^\bP(t,X_t),\beta^\bP(t,X_t))\,dt \right].
\end{equation}
The inclusion of the admissible classes gives
$\cV \leq \cV_{\mathcal{M}}$.
Conversely, convexity of $F$ in the drift and covariance variables
ensures, by conditional Jensen's inequality, that the projection
does not increase the cost. This yields the reverse inequality,
as stated in the following proposition.

\begin{prop}\label{prop:markovian-proj}
For $(\mu_0,\tau,\pi)$ fixed, we have $\cV = \cV_{\mathcal{M}}$.
\end{prop}

\begin{proof}
Suppose $\cP(\mu_0,\tau,\pi)\neq \emptyset$, otherwise both sides are equal to $+\infty$. Let $\bP \in \cP(\mu_0,\tau,\pi)$, by Theorem~\ref{thm:markovian-proj}, there exists $\wtilde{\bP}\in \cP^1_{\mathcal{M}}$ such that for any $t\in[0,T]$, $X_t \# \bP = X_t \# \wtilde{\bP}$, hence in particular $\wtilde{\bP}\in \cP(\mu_0,\tau,\pi)$. Using tower property and Jensen's inequality for conditional expectation, recalling that $(\alpha,\beta)\mapsto F(\alpha,\beta)$ is convex, \begin{align*}
    \E^{\bP}\left[ \int_0^T F(t,X_t,\alpha^\bP_t,\beta^\bP_t)\, dt \right] &= \E^{\bP}\left[ \int_0^T \E^\bP \left[ F(t,X_t,\alpha^\bP_t,\beta^\bP_t) | X_t  \right] \, dt \right] \\
    &\geq \E^{\bP}\left[ \int_0^T F(t,X_t,\E^\bP \left[\alpha^\bP_t| X_t  \right],\E^\bP \left[\beta^\bP_t| X_t  \right])  \, dt \right] \\
    &\geq \E^{\wtilde{\bP}}\left[ \int_0^T F(t,X_t,\alpha^{\wtilde{\bP}}_t,\beta^{\wtilde{\bP}}_t)\, dt \right].
\end{align*}
Last inequality follows from the mimicking of the one-dimensional marginal distributions.
\end{proof}

\begin{remark}
More general payoff functions can also be considered for calibration,
including path-dependent European payoffs. The mimicking results of
\cite{brunick-shreve} apply to several such cases, for instance payoffs
depending on the running average, the running maximum, or the running
integral to date. However, this requires augmenting the state variable in order
to obtain a Markovian projection, hence increasing the dimension of the state space. Hence, in the following the dimension $d$ of the state process should be understood as possibly including additional state variables introduced to recover a Markovian formulation.
For example, consider the running integral $$A_t = A_0 + \int_0^t X_s\,ds, \qquad t\in[0,T].$$
The relevant state variable is then $Z_t=(X_t,A_t)$.
The projected characteristics are measurable functions of this
augmented state and satisfy
$$
    \alpha(t,X_t,A_t)
    =\E^\bP[\alpha_t^\bP\mid X_t,A_t],
    \qquad
    \beta(t,X_t,A_t)
    =\E^\bP[\beta_t^\bP\mid X_t,A_t].
$$
The mimicking process preserves the one-time joint marginals of
$(X,A)$. Moreover, possibly on an extension of the corresponding
filtered probability space, it admits the representation
\begin{equation*}
    \begin{split}
    X_t
    &=X_0+\int_0^t\alpha(s,X_s,A_s)\,ds
      +\int_0^t\beta(s,X_s,A_s)^{1/2}
        \,dW_s^{\widehat{\bP}},\\
    A_t
    &=A_0+\int_0^t X_s\,ds,
\end{split}
\qquad t\in[0,T].
\end{equation*}
\end{remark}

\subsection{Eulerian formulation}

For a Borel curve $(\rho_t)_{t\in[0,T]}$ of probability measures
on $\R^d$, we write $\rho_t=\rho(t,\cdot)$ and
\begin{equation*}
    \rho_t(f)=\int_{\R^d}f(x)\,\rho_t(dx).
\end{equation*}
No density with respect to Lebesgue measure is assumed.
The associated space-time measure, denoted by
$d\rho(t,x):=dt\,\rho_t(dx)$, is characterized by
\begin{equation*}
    \int_\Lambda h(t,x)\,d\rho(t,x)
    =\int_0^T\int_{\R^d}h(t,x)\,\rho_t(dx)\,dt
\end{equation*}
for every nonnegative Borel function $h$.
We denote the space of narrowly continuous probability curves by
\begin{equation*}
    \mathscr P([0,T];\R^d)
    :=\left\{(\rho_t)_{t\in[0,T]}\subseteq\mathcal P(\R^d):
        t\mapsto\rho_t(f)\text{ is continuous for every }
        f\in C_b(\R^d)\right\}.
\end{equation*}

Let $\alpha : (0,T)\times\mathbb{R}^d \to \mathbb{R}^d$ and 
$\beta : (0,T)\times\mathbb{R}^d \to \mathcal{S}_+^d$ be Borel functions. 
We define the linear differential operator $\cL=\cL(\alpha,\beta)$ by
\begin{equation*}
    \cL  = \alpha \cdot \Dx + \demi \beta : \Dxx\;.
\end{equation*}
We denote by $\cL^*$ the distributional adjoint of $\cL$, defined in the
weak sense as follows (see Definition~2.2 in \cite{trevisan16}).

\begin{definition}[Fokker--Planck Equation]
\label{def:fokker-planck}
    A Borel curve $(\nu_t)_{t\in(0,T)}\subseteq \cM(\R^d)$ is a weak solution to the Fokker--Planck~equation~\eqref{eq:fokker-planck} \begin{equation}\label{eq:fokker-planck}\tag{FP}
        \partial_t \nu_t = \cL^*_t \nu_t \quad \text{ on } (0,T)\times\R^d,
    \end{equation} if we have \begin{equation}\label{ass:integrability-fp}
        \int_\Lambda \left( |\alpha_t| + \|\beta_t\| \right)\, d|\nu_t| dt < \infty
    \end{equation} and, for any $\varphi \in C_c^{1,2}((0,T)\times\R^d)$, \begin{equation}
        \int_\Lambda [\dt \varphi(t,x)+ \cL_t \varphi(t,x)]\, d\nu_t(x) dt = 0.
    \end{equation}
\end{definition}

Following Remark~2.3 of \cite{trevisan16} (see also Remark~2.5 in
\cite{joseph-obloj-loeper}), any probability-valued weak solution
$\nu$ to~\eqref{eq:fokker-planck} admits a unique representative
$\wtilde{\nu}\in\mathscr P([0,T];\R^d)$ such that
$\nu_t=\wtilde{\nu}_t$ for $dt$-almost every $t\in(0,T)$.
In particular, its initial trace $\wtilde{\nu}_0$ is well defined.
For every $\varphi\in C_c^{1,2}(\Lambda)$ and
$0\leq t_1\leq t_2\leq T$, we have
\begin{equation}
    \wtilde{\nu}_{t_2}(\varphi_{t_2})-\wtilde{\nu}_{t_1}(\varphi_{t_1}) = \int_{t_1}^{t_2} \int_{\R^d} [\dt \varphi(t,x) + \cL_t \varphi(t,x)]d\nu_t(x)dt.
\end{equation}

\begin{theorem}\label{thm:fp-weak-sol}
    Let $\bP \in \cP^1_\mathcal{M}$, characterized by $(\alpha^\bP_t,\beta^\bP_t)$, and such that $X_0 \# \bP = \mu_0 \in \cP(\R^d)$. The one-dimensional marginal distributions $\rho_t = \rho(t,\cdot) := X_t \# \bP$, are a weak solution to the Fokker--Planck~equation~\eqref{eq:fokker-planck} on $(0,T)$ associated with the measurable functions $(\alpha_t^\bP,\beta_t^\bP)$, and $\rho_0=\mu_0$.\\
    Conversely, suppose $\nu\in\mathscr P([0,T];\R^d)$ is a weak solution to~\eqref{eq:fokker-planck} on $(0,T)$ associated with measurable functions $(\alpha_t,\beta_t)$. Then there exists $\bP\in\cP^1_{\mathcal M}$ solving the martingale problem
associated with $(\alpha,\beta)$ and satisfying $X_t\#\bP=\nu_t$ for all
$t\in[0,T]$. Equivalently, possibly after enlarging the probability space,
the process $X_.$ admits the weak SDE representation~\eqref{eq:sde-doob-decomp}.
\end{theorem}

\begin{remark}
The first statement in Theorem~\ref{thm:fp-weak-sol} follows from Itô's formula. 
The converse is given by the superposition principle, see Theorem~2.5 in 
\cite{trevisan16} (see also Theorem~2.6 in \cite{figalli07}).
\end{remark}

\begin{remark}
Assumption~\eqref{ass:integrability-fp} is equivalent to 
\eqref{eq:-char-integrability} when $\nu = \rho$.
\end{remark}

\noindent By Proposition~\ref{prop:markovian-proj} and Theorem~\ref{thm:fp-weak-sol}, the probabilistic (Lagrangian) primal problem $\cV$ admits the following Eulerian representation.
\begin{equation}\label{eq:main-primal}
    \cV = \inf_{\rho,\alpha,\beta}
    \int_0^T\int_{\R^d}
    F(t,x,\alpha(t,x),\beta(t,x))\,\rho_t(dx)\,dt,
\end{equation}
where $\rho\in\mathscr P([0,T];\R^d)$ and $\alpha,\beta$ are
Borel coefficients with values in $\R^d$ and $\mathcal S^d_+$,
respectively, satisfying
\begin{equation*}
    \int_0^T\int_{\R^d}
    \bigl(|\alpha(t,x)|+\|\beta(t,x)\|\bigr)\,\rho_t(dx)\,dt<\infty.
\end{equation*}
The constraints are
\begin{equation}\label{eq:main-primal-constr}
\begin{split}
    &\partial_t\rho_t+\nabla_x\cdot(\alpha\rho_t)
    -\frac12\sum_{i,j=1}^d\partial_{x_ix_j}(\beta_{ij}\rho_t)=0,\\
    &\rho_{\tau_p}(G_{i(p,m)})=\pi_{i(p,m)},
    \quad m=1,\ldots,M,\quad p=1,\ldots,P,
    \qquad \rho_0=\mu_0.
\end{split}
\end{equation}
The first equation is understood in the sense of
Definition~\ref{def:fokker-planck}. For almost every $t$, the
products $\alpha\rho_t$ and $\beta_{ij}\rho_t$ denote measures
obtained by weighting $\rho_t$ by the corresponding coefficients.

\begin{lemma}[Weak formulation on time intervals]\label{lem:fp-slabwise}
Let $\rho\in\mathscr{P}([0,T];\R^d)$ and let $\alpha$ and $\beta$ be
Borel coefficients satisfying
$\int_0^T\int_{\R^d}(|\alpha|+\|\beta\|)\,\rho_t(dx)\,dt<\infty$,
with $\beta$ taking values in $\mathcal{S}_+^d$.
Then~\eqref{eq:main-primal-constr} holds if and only if there exist
probability measures $\nu_0,\ldots,\nu_P\in\cP(\R^d)$ such that
\[
    \nu_0=\mu_0,
    \qquad \nu_p(G_{i(p,m)})=\pi_{i(p,m)},
    \quad p=1,\ldots,P,\quad m=1,\ldots,M,
\]
and, for every $p=1,\ldots,P$ and every
$\varphi_p\in C_c^{1,2}([\tau_{p-1},\tau_p]\times\R^d)$,
\begin{equation}\label{eq:fp-slabwise}
\begin{split}
    &\int_{\tau_{p-1}}^{\tau_p}\int_{\R^d}
    \left(\partial_t\varphi_p+\alpha\cdot\Dx\varphi_p
    +\frac12\beta:\Dxx\varphi_p\right)\rho_t(dx)\,dt\\
    &\qquad =\nu_p\bigl(\varphi_p(\tau_p,\cdot)\bigr)
    -\nu_{p-1}\bigl(\varphi_p(\tau_{p-1},\cdot)\bigr).
\end{split}
\end{equation}
Here the test functions have compact support in space and need not vanish
at either endpoint of the time interval; they may be chosen independently
on different intervals. The measures $\nu_p$ are uniquely determined by
$\rho$, with $\nu_p=\rho_{\tau_p}$ for $p=0,\ldots,P$.
\end{lemma}

\begin{proof}
If~\eqref{eq:main-primal-constr} holds, the integration-by-parts identity
for narrowly continuous Fokker--Planck solutions stated above yields
\eqref{eq:fp-slabwise} with $\nu_p=\rho_{\tau_p}$.
Conversely, apply~\eqref{eq:fp-slabwise} to the restrictions of a single
test function in $C_c^{1,2}((0,T)\times\R^d)$ and sum over $p$.
The terms at intermediate times cancel, so $\rho$ solves the global
Fokker--Planck equation in the distributional sense.
Its narrow continuity then gives the same integration-by-parts identity
on each interval with endpoint measures $\rho_{\tau_{p-1}}$ and
$\rho_{\tau_p}$. Comparing this identity with~\eqref{eq:fp-slabwise},
and choosing $\varphi_p(t,x)=\eta(t)f(x)$ with
$f\in C_c^\infty(\R^d)$ and $\eta\in C^1([\tau_{p-1},\tau_p])$
having endpoint values $(0,1)$ or $(1,0)$, identifies both endpoint
measures. Thus $\nu_p=\rho_{\tau_p}$ for every $p$, which gives the
initial condition and the price constraints.
\end{proof}

This places our problem within the so-called Benamou--Brenier framework, introduced in~\cite{BB99} as a dynamical formulation of classical optimal transport. In our setting, it corresponds to a stochastic optimal transport problem with semimartingale dynamics. The Lagrangian formulation given by the ODE associated with the velocity field is replaced by the SDE~\eqref{eq:sde-doob-decomp}, while
in the Eulerian formulation the continuity equation is replaced by the Fokker--Planck equation~\eqref{eq:fokker-planck}.

The rest of the paper studies the well-posedness of this Eulerian calibration
problem, namely existence, uniqueness, and stability of its solutions. In the
next section, we recall the corresponding strong duality and existence results
under general assumptions on the cost function $F$. Boundedness of the
effective domain of $F$, as assumed in the duality results below, provides
uniform upper bounds on the drift and covariance coefficients. Uniform
ellipticity requires an additional strictly positive lower bound on the
covariance matrices. For the subsequent uniqueness and stability analysis,
we will impose this condition together with additional regularity assumptions.
In local volatility calibration, uniform ellipticity is natural
when the state variable is the log-price and the local volatility is assumed to
be bounded away from zero.

\section{Duality and optimality}\label{sec:duality}
This section develops the dual formulation of the calibration
problem and its consequences for existence and optimality.
We first introduce a regular dual problem, establish weak
duality, and state the strong duality theorem, which yields
existence of primal optimizers whenever the value is finite.
Under additional PDE assumptions, we then identify the dual
value through a cascade of Hamilton--Jacobi--Bellman equations.
Next, we establish dual attainment for prices in the relative
interior of the admissible price set and derive optimality
conditions linking primal optimizers to the dual HJB solutions.

Our proof of strong duality follows the convex-analytic approach
used in semimartingale optimal transport calibration
\cite{guo2022lsv-calibration}, with an Eulerian formulation
in the spirit of \cite{huesmann-trevisan}.
It relies on the Fenchel--Rockafellar theorem and is presented
in Subsection~\ref{subsec:strong-duality-proof}.

\subsection{Dual formulations and strong duality}

We first introduce a class of regular dual potentials on each time interval.
For $p=1,\ldots,P$, set
$I_p=[\tau_{p-1},\tau_p]$ and $Q_p=I_p\times\R^d$.
Let $\mathcal U_p=C_b^{1,2}(Q_p)$ denote the space of functions
$u_p$ for which $u_p$, $\partial_tu_p$, $\Dx u_p$, and $\Dxx u_p$
are continuous and bounded on $Q_p$, with one-sided time derivatives
at the endpoints. For $\lambda\in\R^{MP}$, write
\begin{equation*}
    g_p^\lambda(x)=\sum_{m=1}^M\lambda_{i(p,m)}G_{i(p,m)}(x),
    \qquad u_{P+1}(\tau_P,\cdot)=0.
\end{equation*}
Define the regular dual value by
\begin{equation}\label{eq:dual-regular-slabwise}
    D_{\mathrm{reg}}(\pi)
    :=\sup_{\lambda\in\R^{MP},\ u_p\in\mathcal U_p}
    \bigl\{\lambda\cdot\pi-\mu_0(u_1(0,\cdot))\bigr\},
\end{equation}
where the supremum is restricted to potentials satisfying, for every $p$,
\begin{align}
    \partial_tu_p+
    F^*\!\left(t,x,\Dx u_p,\tfrac12\Dxx u_p\right)
    &\leq0 &&\text{on }Q_p,
    \label{eq:dual-regular-hjb}\\
    u_p(\tau_p,\cdot)
    &\geq u_{p+1}(\tau_p,\cdot)+g_p^\lambda
    &&\text{on }\R^d.
    \label{eq:dual-regular-interface}
\end{align}
In particular, the last condition is $u_P(T,\cdot)\geq g_P^\lambda$.
No equality of the traces of adjacent potentials is imposed.

\begin{prop}[Weak duality for regular potentials]\label{prop:weak-duality-slabwise}
For every admissible pair $(u,\lambda)$ in
\eqref{eq:dual-regular-slabwise} and every admissible Eulerian triplet
$(\rho,\alpha,\beta)$ satisfying~\eqref{eq:main-primal-constr},
\begin{equation}\label{eq:weak-duality-slabwise}
    \lambda\cdot\pi-\mu_0(u_1(0,\cdot))
    \leq\int_0^T\int_{\R^d}
    F(t,x,\alpha(t,x),\beta(t,x))\,\rho_t(dx)\,dt.
\end{equation}
Consequently, $D_{\mathrm{reg}}(\pi)\leq\mathcal V(\pi)$.
\end{prop}

\begin{proof}
It suffices to consider an admissible triplet of finite cost.
Fenchel's inequality and~\eqref{eq:dual-regular-hjb} imply, on each interval,
\begin{equation*}
    \partial_tu_p+\alpha\cdot\Dx u_p+\tfrac12\beta:\Dxx u_p
    \leq F(t,x,\alpha,\beta)
    \qquad dt\,\rho_t(dx)\text{-a.e.}
\end{equation*}
We apply the weak identity in Lemma~\ref{lem:fp-slabwise} with endpoint
measures $\nu_p=\rho_{\tau_p}$. To justify the use of $u_p$, which need
not have compact spatial support, choose $\chi_R\in C_c^\infty(\R^d)$
such that $0\leq\chi_R\leq1$, $\chi_R\to1$ pointwise,
$\|\Dx\chi_R\|_\infty\leq c/R$, and
$\|\Dxx\chi_R\|_\infty\leq c/R^2$, and first test with $u_p\chi_R$.
The terms in which a derivative falls on $\chi_R$ are bounded in absolute
value by
\begin{equation*}
    C_p(R^{-1}+R^{-2})
    \int_{I_p}\int_{\R^d}(|\alpha|+\|\beta\|)\,\rho_t(dx)\,dt,
\end{equation*}
where $C_p$ depends only on the cutoff bounds and the bounded derivatives
and values of $u_p$. These terms vanish as $R\to\infty$.
All remaining terms, including the endpoint terms, converge by dominated
convergence. Thus
\begin{equation*}
\begin{split}
    &\int_{I_p}\int_{\R^d}
    \left(\partial_tu_p+\alpha\cdot\Dx u_p
    +\tfrac12\beta:\Dxx u_p\right)\rho_t(dx)\,dt\\
    &\qquad=\rho_{\tau_p}(u_p(\tau_p,\cdot))
    -\rho_{\tau_{p-1}}(u_p(\tau_{p-1},\cdot)).
\end{split}
\end{equation*}
Integrating Fenchel's inequality, summing over the intervals, and using
$\rho_0=\mu_0$ gives
\begin{align*}
    &\int_0^T\int_{\R^d}F(t,x,\alpha,\beta)\,\rho_t(dx)\,dt\\
    &\quad\geq\sum_{p=1}^P
    \left[\rho_{\tau_p}(u_p(\tau_p,\cdot))
    -\rho_{\tau_{p-1}}(u_p(\tau_{p-1},\cdot))\right]\\
    &\quad=-\mu_0(u_1(0,\cdot))
    +\sum_{p=1}^P\rho_{\tau_p}
    \bigl(u_p(\tau_p,\cdot)-u_{p+1}(\tau_p,\cdot)\bigr)\\
    &\quad\geq-\mu_0(u_1(0,\cdot))
    +\sum_{p=1}^P\rho_{\tau_p}(g_p^\lambda)\\
    &\quad=\lambda\cdot\pi-\mu_0(u_1(0,\cdot)).
\end{align*}
The last inequality uses~\eqref{eq:dual-regular-interface}, and the last
equality uses the calibration constraints. Taking the infimum over
admissible Eulerian triplets and the supremum over admissible pairs
$(u,\lambda)$ proves the claim.
\end{proof}

To establish the reverse inequality and primal attainment, we impose the
following assumptions on the running cost $F$ and its Hamiltonian $H=F^*$.

\begin{assumption}[Convex-duality assumptions]\label{ass:convex-duality}
Recall that $\Lambda=[0,T]\times\R^d$.
\begin{enumerate}[label=(D\arabic*),ref=D\arabic*]
\item\label{ass:convex-duality-d1} For every $(t,x)\in\Lambda$, the function
$F(t,x,\cdot,\cdot):\R^d\times\mathcal S^d\to[0,+\infty]$
is proper, convex, and lower semicontinuous. There is a constant
$C<\infty$, independent of $(t,x)$, such that
\begin{equation*}
    \operatorname{dom}F(t,x,\cdot,\cdot)
    \subseteq\{(\alpha,\beta):
    |\alpha|\leq C,\ \|\beta\|\leq C,\ \beta\succeq0\}.
\end{equation*}
\item\label{ass:convex-duality-d2} The Hamiltonian $H$ is jointly continuous on
$\Lambda\times\R^d\times\mathcal S^d$, and
\begin{equation*}
    \sup_{(t,x)\in\Lambda}|H(t,x,0,0)|<\infty.
\end{equation*}
\end{enumerate}
\end{assumption}

\begin{remark}[Consequences of the convex duality assumptions]
\label{rk:hamiltonian-duality-properties}
Condition~\eqref{ass:convex-duality-d1} implies that $H$ is finite everywhere and satisfies
\begin{equation}\label{eq:hamiltonian-duality-bounds}
\begin{split}
    H(t,x,0,0)&\leq0,\\
    |H(t,x,a,b)-H(t,x,a',b')|
    &\leq C\bigl(|a-a'|+\|b-b'\|\bigr).
\end{split}
\end{equation}
Together with~\eqref{ass:convex-duality-d2}, this ensures that $H(t,x,a(t,x),b(t,x))$ is continuous
and bounded for continuous bounded fields $a,b$. Moreover, $F=H^*$:
taking the supremum over a fixed countable dense set of dual vectors shows
that $F$ is jointly lower semicontinuous, hence Borel measurable.
\end{remark}

\begin{theorem}[Regular strong duality and primal attainment]
\label{thm:regular-strong-duality}
Assume Assumption~\ref{ass:convex-duality}. Let $d\geq1$,
$0=\tau_0<\cdots<\tau_P=T$, $\mu_0\in\mathcal P(\R^d)$,
and $G_{i(p,m)}\in C_b(\R^d)$.
Then, for every $\pi\in\R^{MP}$,
\begin{equation}\label{eq:regular-strong-duality}
    \mathcal V(\pi)=D_{\mathrm{reg}}(\pi),
\end{equation}
where the equality holds in $[0,+\infty]$ and
$D_{\mathrm{reg}}$ is defined by
\eqref{eq:dual-regular-slabwise}--\eqref{eq:dual-regular-interface}.
If this common value is finite, the Eulerian primal
problem~\eqref{eq:main-primal} subject to
\eqref{eq:main-primal-constr} admits a minimizer of finite cost.
\end{theorem}

The proof is given in Subsection~\ref{subsec:strong-duality-proof}.
The theorem uses convexity of $F$, positive semidefinite covariances, and
bounded continuous payoffs. Strong convexity of $F$, uniform ellipticity,
and linear independence of the payoffs modulo constants are not required
for this result. The existence of a primal minimizer follows from
attainment in the abstract Fenchel--Rockafellar dual after its variables
are identified with Eulerian measures. Neither uniqueness of this
minimizer nor attainment of the supremum over $(u,\lambda)$ is asserted.

\begin{remark}[Standing finite-cost feasibility assumption]
\label{rk:finite-cost-feasibility}
Throughout the remainder of the paper, we assume that
$\mathcal V(\pi)<+\infty$. Equivalently, there exists at least one
admissible Eulerian triplet $(\rho,\alpha,\beta)$ satisfying
\eqref{eq:main-primal-constr} with finite cost.
\end{remark}

The primal attainment assertion of
Theorem~\ref{thm:regular-strong-duality}, together with the superposition
principle, yields existence in both primal formulations.

\begin{corol}[Existence of a primal minimizer]
\label{coro:primal-minimum}
Under the hypotheses of Theorem~\ref{thm:regular-strong-duality},
assume in addition that $\mathcal V(\pi)<+\infty$.
Then the primal calibration problem~\eqref{eq:main-primal} admits
a minimizer of finite cost. In particular,
\begin{equation*}
    \mathcal V(\pi)
    =
    \min_{\substack{(\rho,\alpha,\beta)\\
          \text{satisfying }\eqref{eq:main-primal-constr}}}
    \int_0^T\int_{\R^d}
    F(t,x,\alpha(t,x),\beta(t,x))\,\rho_t(dx)\,dt
    <+\infty.
\end{equation*}
The probabilistic primal problem also admits an optimal law
$\mathbb P^\star\in\mathcal P(\mu_0,\tau,\pi)$, and
\begin{equation*}
    \mathcal V(\pi)
    =\min_{\mathbb P\in\mathcal P(\mu_0,\tau,\pi)}
    \mathbb E^{\mathbb P}\!\left[
        \int_0^T F(t,X_t,\alpha_t^{\mathbb P},\beta_t^{\mathbb P})\,dt
    \right].
\end{equation*}
\end{corol}

\begin{proof}
Theorem~\ref{thm:regular-strong-duality} provides an Eulerian minimizer
$(\rho^\star,\alpha^\star,\beta^\star)$ of finite cost.
By~\eqref{ass:convex-duality-d1}, its coefficients are bounded
$dt\,\rho_t^\star(dx)$-almost everywhere, so the integrability
condition~\eqref{ass:integrability-fp} holds.
The superposition principle in Theorem~\ref{thm:fp-weak-sol} gives
a law $\mathbb P^\star$ with marginals $\rho_t^\star$ and characteristics
$(\alpha^\star(t,X_t),\beta^\star(t,X_t))$.
It satisfies the initial and calibration constraints, and its cost is
\begin{equation*}
    \mathbb E^{\mathbb P^\star}\!\left[
        \int_0^T F(t,X_t,\alpha^\star(t,X_t),\beta^\star(t,X_t))\,dt
    \right]
    =\int_0^T\int_{\R^d}
        F(t,x,\alpha^\star(t,x),\beta^\star(t,x))\,\rho_t^\star(dx)\,dt
    =\mathcal V(\pi).
\end{equation*}
Thus $\mathbb P^\star$ attains the probabilistic primal value.
\end{proof}

This establishes the existence part of the well-posedness
of the primal calibration problem. To investigate uniqueness
and stability of its minimizers, we study the dual problem,
beginning with the existence of optimal Lagrange multipliers
and subsequently examining their uniqueness and stability
with respect to the calibration prices.

\begin{lemma}[A sufficient cost condition for \eqref{ass:convex-duality-d2}]
\label{lem:cost-hamiltonian-continuity}
Suppose that $F$ satisfies~\eqref{ass:convex-duality-d1} of Assumption~\ref{ass:convex-duality}.
Write $s=(t,x)$ and $\xi=(\alpha,\beta)$. Assume that the effective domain
$D=\operatorname{dom}F(s,\cdot)$ is independent of $s$, that
\begin{equation}\label{eq:cost-spacetime-continuity}
    \lim_{s\to s_0}\sup_{\xi\in D}
    |F(s,\xi)-F(s_0,\xi)|=0
    \qquad\text{for every }s_0\in\Lambda,
\end{equation}
and that there is $C_0<\infty$ such that
\begin{equation}\label{eq:cost-minimum-bound}
    \inf_{\xi\in D}F(s,\xi)\leq C_0
    \qquad\text{for every }s\in\Lambda.
\end{equation}
Then $H$ satisfies~\eqref{ass:convex-duality-d2}.
\end{lemma}

\begin{proof}
Since the effective domain is the same at $s$ and $s_0$, comparison of
the suprema defining $H$ gives
\begin{equation*}
    |H(s,a,b)-H(s_0,a,b)|
    \leq\sup_{\xi\in D}|F(s,\xi)-F(s_0,\xi)|.
\end{equation*}
Combining this inequality with~\eqref{eq:hamiltonian-duality-bounds},
we obtain
\begin{equation*}
\begin{split}
    |H(s,a,b)-H(s_0,a_0,b_0)|
    &\leq\sup_{\xi\in D}|F(s,\xi)-F(s_0,\xi)|\\
    &\quad+C\bigl(|a-a_0|+\|b-b_0\|\bigr).
\end{split}
\end{equation*}
Condition~\eqref{eq:cost-spacetime-continuity} therefore yields joint
continuity. Finally, nonnegativity of $F$ and
\eqref{eq:cost-minimum-bound} imply
\begin{equation*}
    -C_0\leq H(s,0,0)=-\inf_{\xi\in D}F(s,\xi)\leq0.
\end{equation*}
\end{proof}

\begin{remark}
The conditions in Lemma~\ref{lem:cost-hamiltonian-continuity} are
sufficient for~\eqref{ass:convex-duality-d2}; the theorem itself allows effective domains depending
on $(t,x)$. Condition~\eqref{eq:cost-spacetime-continuity} controls the
variation of the cost in space and time, uniformly over $D$, and allows
$F$ to be unbounded on $D$.
For instance, it holds for $F(s,\xi)=F_0(\xi)+f(s,\xi)$ on $D$, with
$F_0$ of bounded effective domain $D$ and $f$ finite and continuous on
$\Lambda\times\overline D$; the remaining assumptions on the total cost
must still be verified. A sufficient condition for
\eqref{eq:cost-minimum-bound} is the existence of a fixed
$\bar\xi\in D$ with $\sup_{s\in\Lambda}F(s,\bar\xi)<\infty$.
\end{remark}

We next identify the regular dual value with the value of a
dual problem defined by a cascade of HJB equations.
For each $\lambda\in\R^{MP}$, this cascade takes the form
\begin{equation}\label{eq:hjb-cascade-duality}
\begin{cases}
\partial_t\phi_p^\lambda+
 H(t,x,\Dx\phi_p^\lambda,\tfrac12\Dxx\phi_p^\lambda)=0,
 & (t,x)\in(\tau_{p-1},\tau_p)\times\R^d,\\
\phi_p^\lambda(\tau_p,\cdot)
 =\phi_{p+1}^\lambda(\tau_p,\cdot)+g_p^\lambda,
\end{cases}
\end{equation}
for $p\in\{1,\ldots,P\}$, with
$\phi_{P+1}^\lambda(\tau_P,\cdot)=0$ and $g_p^\lambda$ as above.

We assume that the payoffs $G_{i(p,m)}$ belong to
$\BUC(\R^d)$. The following assumption ensures that the cascade
can be solved uniquely by backward induction and provides the
regularity needed for the dual identification.

\begin{assumption}[PDE properties on each time interval]
\label{ass:hjb-slabwise}
For every $p\in\{1,\ldots,P\}$, the following properties hold.
\begin{enumerate}[label=(H\arabic*),ref=H\arabic*]
\item\label{ass:hjb-slabwise-h1} For every $h\in\BUC(\R^d)$, the terminal-value problem
\begin{equation}\label{eq:hjb-slab-terminal}
    \partial_t v+H(t,x,\Dx v,\tfrac12\Dxx v)=0
    \quad\text{on }(\tau_{p-1},\tau_p)\times\R^d,
    \qquad v(\tau_p,\cdot)=h,
\end{equation}
admits a solution
$v\in C_b([\tau_{p-1},\tau_p]\times\R^d)
\cap C^{1,2}((\tau_{p-1},\tau_p)\times\R^d)$.
For every $0<\varepsilon<(\tau_p-\tau_{p-1})/2$, its derivatives
$\partial_t v$, $\Dx v$, and $\Dxx v$ are bounded on
$[\tau_{p-1}+\varepsilon,\tau_p-\varepsilon]\times\R^d$.
Moreover, $v(\tau_{p-1},\cdot)\in\BUC(\R^d)$.
\item\label{ass:hjb-slabwise-h2} Comparison holds in this bounded classical class: if
$v,w\in C_b([\tau_{p-1},\tau_p]\times\R^d)
\cap C^{1,2}((\tau_{p-1},\tau_p)\times\R^d)$ satisfy
\begin{equation*}
    \partial_t v+H(t,x,\Dx v,\tfrac12\Dxx v)\geq0,
    \qquad
    \partial_t w+H(t,x,\Dx w,\tfrac12\Dxx w)\leq0,
\end{equation*}
and $v(\tau_p,\cdot)\leq w(\tau_p,\cdot)$, then $v\leq w$ on
$[\tau_{p-1},\tau_p]\times\R^d$.
\end{enumerate}
\end{assumption}

Under Assumption~\ref{ass:hjb-slabwise}, the cascade~\eqref{eq:hjb-cascade-duality}
admits a unique solution in the stated bounded classical class on each
time interval. Indeed, $g_P^\lambda\in\BUC(\R^d)$, and
\ref{ass:hjb-slabwise-h1} allows us to solve the last terminal-value
problem. At each subsequent step of the backward induction, the terminal
datum $\phi_{p+1}^\lambda(\tau_p,\cdot)+g_p^\lambda$ remains in
$\BUC(\R^d)$ by the trace property in~\ref{ass:hjb-slabwise-h1}.
Uniqueness follows on each interval from the comparison
principle~\ref{ass:hjb-slabwise-h2}.

\begin{remark}[Sufficient hypotheses from the companion PDE article]
\label{rk:hjb-pde-inputs}
Assumption~\ref{ass:hjb-slabwise} is an explicit statement of the PDE
properties needed below. A sufficient structural condition is that $H$
satisfy Assumption~1 of~\cite{HJB-property}, on each interval after time
translation. This requires membership in the uniformly elliptic operator
class $\mathcal H(\mathfrak v,Q)$ defined there, bounded and globally
Lipschitz second derivatives in the gradient and Hessian variables,
parabolic Lipschitz continuity of $(t,x)\mapsto D_{(a,b)}H(t,x,0,0)$,
and a uniform bound on $H(t,x,0,0)$, with the constants specified in that
assumption. Here the parabolic metric on space-time is
$|t-s|^{1/2}+|x-y|$.
Under these hypotheses, Proposition~3.1 of~\cite{HJB-property} gives
comparison, while Corollary~3.2 and Proposition~3.3 give boundedness,
continuity up to the time endpoints, and the required interior derivative
bounds. The initial trace on each time interval is bounded and Lipschitz, hence BUC,
which allows the cascade to be constructed by backward induction, as
detailed in Section~6 of the companion PDE article.
We use these results in dimension $d\geq1$.
No assumption concerning strict convexity of the HJB solution map or
non-degeneracy of the payoff family is needed for this identification.
\end{remark}

We define the associated HJB dual value by
\begin{equation}\label{eq:dual-hjb-cascade}
    D_{\mathrm{HJB}}(\pi)
    :=\sup_{\lambda\in\R^{MP}}
    \bigl\{\lambda\cdot\pi-
    \mu_0(\phi_1^\lambda(0,\cdot))\bigr\}.
\end{equation}

\begin{prop}[Identification of the HJB dual value]
\label{prop:hjb-dual-identification}
Under the hypotheses of Theorem~\ref{thm:regular-strong-duality},
suppose in addition that $G_{i(p,m)}\in\BUC(\R^d)$ and that
Assumption~\ref{ass:hjb-slabwise} holds. Then, for every $\pi\in\R^{MP}$,
\begin{equation}\label{eq:hjb-regular-primal-equality}
    \mathcal V(\pi)=D_{\mathrm{reg}}(\pi)=D_{\mathrm{HJB}}(\pi).
\end{equation}
The equality holds in $[0,+\infty]$; the standing finite-cost
feasibility assumption of Remark~\ref{rk:finite-cost-feasibility}
is not imposed for this result.
\end{prop}

\begin{proof}
Fix an arbitrary $\pi\in\R^{MP}$.
First consider a regular admissible pair $(u,\lambda)$ in
\eqref{eq:dual-regular-slabwise}--\eqref{eq:dual-regular-interface}.
On the last interval, $u_P(T,\cdot)\geq g_P^\lambda
=\phi_P^\lambda(T,\cdot)$, and~\eqref{ass:hjb-slabwise-h2} gives $u_P\geq\phi_P^\lambda$.
If $u_{p+1}\geq\phi_{p+1}^\lambda$, then
\begin{equation*}
    u_p(\tau_p,\cdot)
    \geq u_{p+1}(\tau_p,\cdot)+g_p^\lambda
    \geq\phi_{p+1}^\lambda(\tau_p,\cdot)+g_p^\lambda
    =\phi_p^\lambda(\tau_p,\cdot).
\end{equation*}
Comparison on the $p$-th interval therefore gives
$u_p\geq\phi_p^\lambda$. By backward induction,
\begin{equation*}
    \lambda\cdot\pi-\mu_0(u_1(0,\cdot))
    \leq\lambda\cdot\pi-\mu_0(\phi_1^\lambda(0,\cdot)).
\end{equation*}
Taking suprema yields $D_{\mathrm{reg}}(\pi)\leq D_{\mathrm{HJB}}(\pi)$.

If $\mathcal V(\pi)=+\infty$, Theorem~\ref{thm:regular-strong-duality}
gives $D_{\mathrm{reg}}(\pi)=+\infty$, and the preceding inequality
implies $D_{\mathrm{HJB}}(\pi)=+\infty$.
This proves the claim in this case. For the remainder of the proof,
we assume that $\mathcal V(\pi)<+\infty$.

Next fix $\lambda$ and an admissible Eulerian triplet
$(\rho,\alpha,\beta)$ of finite cost, using its narrowly continuous
representative. By~\eqref{ass:convex-duality-d1}, $\alpha$ and $\beta$ are bounded
$dt\,\rho_t(dx)$-almost everywhere. Fenchel's inequality and
\eqref{eq:hjb-cascade-duality} imply
\begin{equation*}
    \partial_t\phi_p^\lambda
    +\alpha\cdot\Dx\phi_p^\lambda
    +\tfrac12\beta:\Dxx\phi_p^\lambda
    \leq F(t,x,\alpha,\beta)
    \qquad dt\,\rho_t(dx)\text{-a.e.}
\end{equation*}
Choose $0<\varepsilon<\tfrac12\min_p(\tau_p-\tau_{p-1})$.
On $[\tau_{p-1}+\varepsilon,\tau_p-\varepsilon]$, \eqref{ass:hjb-slabwise-h1} gives the
bounded derivatives needed to test the Fokker--Planck equation with
$\phi_p^\lambda$. More precisely, the time-localized weak identity
of Lemma~\ref{lem:fp-slabwise} applies on this closed subinterval;
spatial cutoffs can then be removed by the argument in
Proposition~\ref{prop:weak-duality-slabwise}. Consequently,
\begin{equation}\label{eq:hjb-duality-truncated}
\begin{split}
    &\rho_{\tau_p-\varepsilon}
      (\phi_p^\lambda(\tau_p-\varepsilon,\cdot))
     -\rho_{\tau_{p-1}+\varepsilon}
      (\phi_p^\lambda(\tau_{p-1}+\varepsilon,\cdot))\\
    &\qquad\leq
    \int_{\tau_{p-1}+\varepsilon}^{\tau_p-\varepsilon}
    \int_{\R^d}F(t,x,\alpha,\beta)\,\rho_t(dx)\,dt.
\end{split}
\end{equation}

We justify the limits of the endpoint pairings. For
$t_0\in\{\tau_{p-1},\tau_p\}$ and $t\to t_0$ within the $p$th interval,
narrow continuity gives $\rho_t\to\rho_{t_0}$ and tightness of these
measures. For every compact $K\subset\R^d$,
\begin{align*}
    &\big|\rho_t(\phi_p^\lambda(t,\cdot))
           -\rho_{t_0}(\phi_p^\lambda(t_0,\cdot))\big|\\
    &\quad\leq
    \sup_{x\in K}|\phi_p^\lambda(t,x)-\phi_p^\lambda(t_0,x)|
    +2\|\phi_p^\lambda\|_\infty\rho_t(K^c)\\
    &\qquad\quad+
    \big|(\rho_t-\rho_{t_0})(\phi_p^\lambda(t_0,\cdot))\big|.
\end{align*}
Joint continuity makes the first term tend to zero for fixed $K$;
narrow convergence does the same for the last term. Tightness makes
the middle term uniformly small by choosing $K$ sufficiently large.
Thus both endpoint pairings in~\eqref{eq:hjb-duality-truncated}
converge. Since the cost is integrable, letting $\varepsilon\downarrow0$
and summing over $p$ gives
\begin{align*}
    &\int_0^T\int_{\R^d}F(t,x,\alpha,\beta)\,\rho_t(dx)\,dt\\
    &\quad\geq
    \sum_{p=1}^P\left[
    \rho_{\tau_p}(\phi_p^\lambda(\tau_p,\cdot))
    -\rho_{\tau_{p-1}}(\phi_p^\lambda(\tau_{p-1},\cdot))\right]\\
    &\quad=-\mu_0(\phi_1^\lambda(0,\cdot))
      +\sum_{p=1}^P\rho_{\tau_p}(g_p^\lambda)\\
    &\quad=\lambda\cdot\pi-\mu_0(\phi_1^\lambda(0,\cdot)),
\end{align*}
where the last two equalities use the interface conditions and calibration.
Taking the infimum over finite-cost admissible triplets and then the
supremum over $\lambda$ proves $D_{\mathrm{HJB}}(\pi)\leq\mathcal V(\pi)$.
Theorem~\ref{thm:regular-strong-duality} now yields
\begin{equation*}
    \mathcal V(\pi)=D_{\mathrm{reg}}(\pi)
    \leq D_{\mathrm{HJB}}(\pi)\leq\mathcal V(\pi). \qedhere
\end{equation*}
\end{proof}

The argument requires regularity only on closed subintervals of the
interior of each time interval. It does not require
$\phi_p^\lambda\in\mathcal U_p$, nor a global approximation of the
cascade by regular admissible potentials. Equality of the dual values
does not assert attainment of the supremum over $\lambda$.

\subsection{Dual attainment and optimality conditions}

Throughout this subsection, we work under the hypotheses of
Proposition~\ref{prop:hjb-dual-identification}.
For each $\lambda\in\R^{MP}$, let
$(\phi_p^\lambda)_{p=1}^P$ denote the solution of the HJB
cascade~\eqref{eq:hjb-cascade-duality}. Define
\begin{equation}\label{eq:dual-attainment-value-map}
    \Phi_0:\R^{MP}\longrightarrow\R,
    \qquad
    \Phi_0(\lambda)
    :=
    \int_{\R^d}\phi_1^\lambda(0,x)\,\mu_0(dx).
\end{equation}
This quantity is finite since $\phi_1^\lambda$ is bounded
and $\mu_0$ is a probability measure. In the particular case
$\mu_0=\delta_{x_0}$, it reduces to
$\Phi_0(\lambda)=\phi_1^\lambda(0,x_0)$.

By Proposition~\ref{prop:hjb-dual-identification},
\begin{equation}\label{eq:dual-attainment-supremum}
    \mathcal V(\pi)
    =
    \sup_{\lambda\in\R^{MP}}
    \bigl\{\lambda\cdot\pi-\Phi_0(\lambda)\bigr\}
    =
    \Phi_0^*(\pi),
\end{equation}
where $\Phi_0^*$ denotes the convex conjugate of $\Phi_0$.
We now seek a sufficient condition on the price vector $\pi$
under which this supremum is attained by a multiplier
$\lambda^\star\in\R^{MP}$. The argument will rely on the
convexity of $\Phi_0$ and on the geometry of the set of prices
that can be calibrated with finite cost.

\begin{prop}[Convexity of the dual value map]
\label{prop:convexity-dual-value-map}
Under the hypotheses of
Proposition~\ref{prop:hjb-dual-identification},
the map $\Phi_0$ defined in~\eqref{eq:dual-attainment-value-map}
is finite, convex, and continuous on $\R^{MP}$.
In particular, it is a proper closed convex function.
\end{prop}

\begin{proof}
Since $H=F^*$, the Hamiltonian is convex in its last two
variables. The backward comparison argument of
Corollary~6.2 in~\cite{HJB-property} applies under
Assumption~\ref{ass:hjb-slabwise}: convex combinations of
solutions are supersolutions on each interval, and the
linearity of $g_p^\lambda$ in $\lambda$ propagates the
terminal inequalities backward through the cascade.
Consequently, for every $\lambda,\mu\in\R^{MP}$,
$\theta\in[0,1]$, and $x\in\R^d$,
\begin{equation*}
    \phi_1^{\theta\lambda+(1-\theta)\mu}(0,x)
    \leq
    \theta\phi_1^\lambda(0,x)
    +(1-\theta)\phi_1^\mu(0,x).
\end{equation*}
Integrating against $\mu_0$ gives
\begin{equation*}
    \Phi_0(\theta\lambda+(1-\theta)\mu)
    \leq
    \theta\Phi_0(\lambda)+(1-\theta)\Phi_0(\mu),
\end{equation*}
which proves convexity.
Since $\Phi_0$ is finite on all of $\R^{MP}$, it is continuous
there. It is therefore proper and lower semicontinuous,
hence closed.
\end{proof}

\begin{definition}[Admissible prices]
\label{def:dual-attainment-admissible-prices}
The set of admissible prices is
\begin{equation*}
    \mathfrak P
    :=\{\pi\in\R^{MP}:\mathcal V(\pi)<+\infty\}.
\end{equation*}
\end{definition}

\begin{lemma}[Domain of the dual conjugate]
\label{lem:dual-attainment-domain}
Under the standing assumptions, the admissible price set satisfies
\begin{equation*}
    \mathfrak P=\operatorname{dom}\Phi_0^*.
\end{equation*}
\end{lemma}

\begin{proof}
By Proposition~\ref{prop:hjb-dual-identification},
$\mathcal V(\pi)=\Phi_0^*(\pi)$ for every $\pi\in\R^{MP}$,
with equality understood in the extended sense. The claimed identity follows from the definition of $\mathfrak P$
and of the effective domain of $\Phi_0^*$.
\end{proof}

By Lemma~\ref{lem:dual-attainment-domain}, $\mathfrak P$ is convex,
being the effective domain of the convex function $\Phi_0^*$.
It is nonempty by the standing finite-cost feasibility assumption
of Remark~\ref{rk:finite-cost-feasibility}.
We denote by $\operatorname{ri}(\mathfrak P)$ the relative
interior of $\mathfrak P$, that is, its interior in its
affine hull endowed with the topology induced by $\R^{MP}$.
We will establish dual attainment for price vectors satisfying
\begin{equation}\label{eq:dual-attainment-relative-interior}
    \pi\in\operatorname{ri}(\mathfrak P).
\end{equation}

\begin{prop}[Dual attainment]\label{prop:dual-attainement}
Under the hypotheses of
Proposition~\ref{prop:hjb-dual-identification}, suppose that
$\pi\in\operatorname{ri}(\mathfrak P)$. Then the HJB dual
problem admits an optimal multiplier $\lambda^\star\in\R^{MP}$,
and
\begin{equation}\label{eq:dual-attainment-maximum}
    \mathcal V(\pi)
    =\max_{\lambda\in\R^{MP}}
    \bigl\{\lambda\cdot\pi-\Phi_0(\lambda)\bigr\}
    <+\infty.
\end{equation}
The set of optimal multipliers is characterized by
\begin{equation}\label{eq:dual-attainment-subdifferential}
\begin{split}
    \argmax_{\lambda\in\R^{MP}}
    \bigl\{\lambda\cdot\pi-\Phi_0(\lambda)\bigr\}
    &=\partial\Phi_0^*(\pi)\\
    &=\{\lambda\in\R^{MP}:\pi\in\partial\Phi_0(\lambda)\},
\end{split}
\end{equation}
where $\partial$ denotes the convex subdifferential.
\end{prop}

\begin{proof}
By Proposition~\ref{prop:convexity-dual-value-map}, $\Phi_0$
is proper, closed, and convex. Its conjugate $\Phi_0^*$ is
therefore also proper, closed, and convex
by~\cite[Theorem~12.2]{rockafellar-1970a}.
Lemma~\ref{lem:dual-attainment-domain} gives
\begin{equation*}
    \pi\in\operatorname{ri}(\mathfrak P)
    =\operatorname{ri}(\operatorname{dom}\Phi_0^*).
\end{equation*}
Thus $\partial\Phi_0^*(\pi)\neq\varnothing$
by~\cite[Theorem~23.4]{rockafellar-1970a}.
Choose $\lambda^\star\in\partial\Phi_0^*(\pi)$.
Since $\Phi_0$ is proper, closed, and convex,
\cite[Theorem~23.5]{rockafellar-1970a} gives
\begin{equation*}
    \Phi_0^*(\pi)+\Phi_0(\lambda^\star)
    =\lambda^\star\cdot\pi.
\end{equation*}
Since $\mathcal V(\pi)=\Phi_0^*(\pi)<+\infty$,
this proves~\eqref{eq:dual-attainment-maximum}.
More generally, a multiplier $\lambda\in\R^{MP}$ is optimal
if and only if
$\Phi_0^*(\pi)+\Phi_0(\lambda)=\lambda\cdot\pi$.
By the same equality criterion, this is equivalent to
$\lambda\in\partial\Phi_0^*(\pi)$ and to
$\pi\in\partial\Phi_0(\lambda)$, proving
\eqref{eq:dual-attainment-subdifferential}.
\end{proof}

\begin{remark}[Scope of the relative-interior condition]
\label{rk:dual-attainment-interiority}
The condition $\pi\in\operatorname{ri}(\mathfrak P)$ means
that every sufficiently small perturbation of $\pi$ within
the affine hull of $\mathfrak P$ remains feasible with finite
cost. It accommodates affine relations between the calibration
constraints and does not require $\mathfrak P$ to have nonempty
interior in $\R^{MP}$. If $\operatorname{int}(\mathfrak P)
\neq\varnothing$, its affine hull is all of $\R^{MP}$,
and relative interior and interior coincide.
For any $\pi\in\mathfrak P$, the Fenchel equality criterion
used in the preceding proof shows that dual attainment is
equivalent to $\partial\Phi_0^*(\pi)\neq\varnothing$.
Relative interior is therefore a sufficient condition;
attainment may also hold at boundary points.
Neither uniqueness of the optimal multiplier nor compactness
of the set of optimal multipliers follows from the proposition.
\end{remark}

We next relate optimal primal coefficients to the optimal HJB cascade.
The following conditions hold whenever the HJB dual is attained,
in particular under the hypotheses of
Proposition~\ref{prop:dual-attainement}.

\begin{prop}[Optimality conditions for the coefficients]
\label{prop:optimality-coefficients}
Under the hypotheses of
Proposition~\ref{prop:hjb-dual-identification}, let $\pi\in\mathfrak P$.
Let $\lambda^\star$ be an optimal multiplier for
\eqref{eq:dual-attainment-supremum}, and let
$(\rho^\star,\alpha^\star,\beta^\star)$ be a primal minimizer.
Set $\phi_p^\star:=\phi_p^{\lambda^\star}$ for $p=1,\ldots,P$.
Then, on each $(\tau_{p-1},\tau_p)\times\R^d$,
\begin{equation}\label{eq:optimality-subgradient-coeff}
    (\alpha^\star(t,x),\beta^\star(t,x))
    \in\partial_{(a,b)}H\!\left(
        t,x,\Dx\phi_p^\star(t,x),
        \tfrac12\Dxx\phi_p^\star(t,x)\right)
\end{equation}
for $dt\,\rho_t^\star(dx)$-almost every $(t,x)$, where the
subdifferential is taken with respect to the last two variables.
If, in addition, $H(t,x,\cdot,\cdot)$ is differentiable for every
$(t,x)$, then
\begin{equation}\label{eq:optimality-eq-coeff}
    (\alpha^\star(t,x),\beta^\star(t,x))
    =\nabla_{(a,b)}H\!\left(
        t,x,\Dx\phi_p^\star(t,x),
        \tfrac12\Dxx\phi_p^\star(t,x)\right)
\end{equation}
with the same almost-everywhere interpretation on each interval.
\end{prop}

\begin{proof}
For $\tau_{p-1}<t<\tau_p$, define the nonnegative Fenchel defect
\begin{equation*}
\begin{split}
    R_p(t,x):={}&F(t,x,\alpha^\star(t,x),\beta^\star(t,x))
       +H\!\left(t,x,\Dx\phi_p^\star(t,x),
                      \tfrac12\Dxx\phi_p^\star(t,x)\right)\\
       &-\alpha^\star(t,x)\cdot\Dx\phi_p^\star(t,x)
        -\tfrac12\beta^\star(t,x):\Dxx\phi_p^\star(t,x)
       \geq0.
\end{split}
\end{equation*}
The primal cost is finite, and~\eqref{ass:convex-duality-d1}
makes $\alpha^\star$ and $\beta^\star$ bounded
$dt\,\rho_t^\star(dx)$-almost everywhere.
Choose $0<\varepsilon<\tfrac12\min_p(\tau_p-\tau_{p-1})$.
On each closed interval
$[\tau_{p-1}+\varepsilon,\tau_p-\varepsilon]$,
the derivative bounds in~\eqref{ass:hjb-slabwise-h1} allow us
to test the Fokker--Planck equation with $\phi_p^\star$,
using Lemma~\ref{lem:fp-slabwise} and the spatial-cutoff argument
of Proposition~\ref{prop:weak-duality-slabwise}.
Since $\phi_p^\star$ solves~\eqref{eq:hjb-cascade-duality},
we obtain
\begin{equation}\label{eq:optimality-truncated-defect}
\begin{split}
    &\int_{\tau_{p-1}+\varepsilon}^{\tau_p-\varepsilon}
      \int_{\R^d}R_p(t,x)\,\rho_t^\star(dx)\,dt\\
    &\quad=\int_{\tau_{p-1}+\varepsilon}^{\tau_p-\varepsilon}
      \int_{\R^d}F(t,x,\alpha^\star(t,x),\beta^\star(t,x))
      \,\rho_t^\star(dx)\,dt\\
    &\qquad-\rho_{\tau_p-\varepsilon}^\star
       (\phi_p^\star(\tau_p-\varepsilon,\cdot))
       +\rho_{\tau_{p-1}+\varepsilon}^\star
       (\phi_p^\star(\tau_{p-1}+\varepsilon,\cdot)).
\end{split}
\end{equation}
All integrals in this identity are finite on the truncated interval.

We justify the convergence of the endpoint pairings.
By narrow continuity, the family
$\{\rho_t^\star:t\in[\tau_{p-1},\tau_p]\}$ is compact
for the topology of narrow convergence, hence uniformly
tight by Prokhorov's theorem.
Fix $t_0\in\{\tau_{p-1},\tau_p\}$ and let $t\to t_0$
within the $p$th interval. For every compact
$K\subset\R^d$,
\begin{align*}
    &\big|\rho_t^\star(\phi_p^\star(t,\cdot))
      -\rho_{t_0}^\star(\phi_p^\star(t_0,\cdot))\big|\\
    &\quad\leq
      \sup_{x\in K}
      |\phi_p^\star(t,x)-\phi_p^\star(t_0,x)|
      +2\|\phi_p^\star\|_\infty\,\rho_t^\star(K^c)\\
    &\qquad+
      \big|(\rho_t^\star-\rho_{t_0}^\star)
      (\phi_p^\star(t_0,\cdot))\big|.
\end{align*}
Uniform tightness makes the second term uniformly small
by choosing $K$ sufficiently large. For this fixed compact,
joint continuity of $\phi_p^\star$ makes the first term
tend to zero, while narrow continuity makes the last term
tend to zero. Thus both endpoint pairings in
\eqref{eq:optimality-truncated-defect} converge.
The cost integrals converge by integrability.
Summing~\eqref{eq:optimality-truncated-defect} over $p$ and
applying monotone convergence to the nonnegative defects gives
\begin{equation}\label{eq:optimality-zero-defect}
\begin{split}
    &\sum_{p=1}^P\int_{\tau_{p-1}}^{\tau_p}
      \int_{\R^d}R_p(t,x)\,\rho_t^\star(dx)\,dt\\
    &\quad=\mathcal V(\pi)
      -\sum_{p=1}^P\left[
        \rho_{\tau_p}^\star(\phi_p^\star(\tau_p,\cdot))
        -\rho_{\tau_{p-1}}^\star
          (\phi_p^\star(\tau_{p-1},\cdot))\right]\\
    &\quad=\mathcal V(\pi)+\Phi_0(\lambda^\star)
       -\lambda^\star\cdot\pi
      =0.
\end{split}
\end{equation}
Here the second equality follows from the cascade interface
conditions, the initial condition, and calibration; the last
uses optimality of $\lambda^\star$ and
\eqref{eq:dual-attainment-supremum}.
This limiting argument does not require separate integrability
of the terms involving derivatives of $\phi_p^\star$ up to
the maturities.

Nonnegativity now implies $R_p=0$
$dt\,\rho_t^\star(dx)$-almost everywhere on each interval.
Since $F(t,x,\cdot,\cdot)$ is proper, closed, and convex and
$H=F^*$, the equality case in Fenchel's inequality
\cite[Theorem~23.5]{rockafellar-1970a} gives
\eqref{eq:optimality-subgradient-coeff}.
When $H(t,x,\cdot,\cdot)$ is differentiable, its
subdifferential consists of its gradient
\cite[Theorem~25.1]{rockafellar-1970a}, which yields
\eqref{eq:optimality-eq-coeff}.
\end{proof}

The preceding argument also applies to optimal laws of the
probabilistic primal problem, without assuming that their characteristics
depend only on the current time and state. In this setting, It\^o's
formula replaces the use of the weak Fokker--Planck equation.

\begin{prop}[Optimality conditions for optimal laws]
\label{prop:optimality-laws}
Under the hypotheses of
Proposition~\ref{prop:hjb-dual-identification}, let $\pi\in\mathfrak P$.
Let $\lambda^\star$ be a dual optimizer for
\eqref{eq:dual-attainment-supremum}, and let
$\mathbb P\in\mathcal P(\mu_0,\tau,\pi)$ be an optimal law for
the probabilistic primal problem.
Set $\phi_p^\star:=\phi_p^{\lambda^\star}$ for $p=1,\ldots,P$.
Then, on each interval $(\tau_{p-1},\tau_p)$,
\begin{equation}\label{eq:optimality-law-subgradient}
    (\alpha_t^{\mathbb P},\beta_t^{\mathbb P})
    \in\partial_{(a,b)}H\!\left(
        t,X_t,\Dx\phi_p^\star(t,X_t),
        \tfrac12\Dxx\phi_p^\star(t,X_t)\right)
\end{equation}
$dt\otimes\mathbb P$-almost everywhere, where the subdifferential
is taken with respect to the last two variables.
If, in addition, $H(t,x,\cdot,\cdot)$ is differentiable for every
$(t,x)$, then
\begin{equation}\label{eq:optimality-law-coefficients}
    (\alpha_t^{\mathbb P},\beta_t^{\mathbb P})
    =\nabla_{(a,b)}H\!\left(
        t,X_t,\Dx\phi_p^\star(t,X_t),
        \tfrac12\Dxx\phi_p^\star(t,X_t)\right)
\end{equation}
with the same almost-everywhere interpretation on each interval.
\end{prop}

\begin{proof}
Fix an optimal law $\mathbb P$. Its cost equals
$\mathcal V(\pi)<+\infty$, so~\eqref{ass:convex-duality-d1}
implies that $\alpha^{\mathbb P}$ and $\beta^{\mathbb P}$ are bounded
$dt\otimes\mathbb P$-almost everywhere.
For each $p=1,\ldots,P$ and $t\in(\tau_{p-1},\tau_p)$, define
the Fenchel defect along the canonical process by
\begin{equation}\label{eq:optimality-law-defect}
\begin{split}
    R_p^{\mathbb P}(t):={}&
       F(t,X_t,\alpha_t^{\mathbb P},\beta_t^{\mathbb P})
       +H\!\left(t,X_t,\Dx\phi_p^\star(t,X_t),
                         \tfrac12\Dxx\phi_p^\star(t,X_t)\right)\\
       &-\alpha_t^{\mathbb P}\cdot\Dx\phi_p^\star(t,X_t)
        -\tfrac12\beta_t^{\mathbb P}:\Dxx\phi_p^\star(t,X_t).
\end{split}
\end{equation}
Since $H=F^*$, Fenchel's inequality gives
$R_p^{\mathbb P}(t)\geq0$. This inequality applies directly to
the progressively measurable characteristics of $\mathbb P$;
no Markovian representation is needed.

Fix $p\in\{1,\ldots,P\}$ and
$\tau_{p-1}<s<u<\tau_p$.
By~\eqref{ass:hjb-slabwise-h1}, the derivatives
$\partial_t\phi_p^\star$, $\Dx\phi_p^\star$ and $\Dxx\phi_p^\star$
are bounded on $[s,u]\times\R^d$.
The stochastic integral in It\^o's formula for
$\phi_p^\star(t,X_t)$ on $[s,u]$ is a square-integrable martingale,
since the boundedness of $\Dx\phi_p^\star$ and $\beta^{\mathbb P}$ gives
\begin{equation*}
    \mathbb E^{\mathbb P}\!\left[
        \int_s^u
        \Dx\phi_p^\star(t,X_t)^\top\beta_t^{\mathbb P}
        \Dx\phi_p^\star(t,X_t)\,dt
    \right]<+\infty.
\end{equation*}
Its expectation is therefore zero.
Taking expectations in It\^o's formula and using the HJB
equation~\eqref{eq:hjb-cascade-duality}, together with
\eqref{eq:optimality-law-defect}, yields
\begin{equation}\label{eq:optimality-law-truncated-defect}
\begin{split}
    \mathbb E^{\mathbb P}\!\left[\int_s^u R_p^{\mathbb P}(t)\,dt\right]
    ={}&\mathbb E^{\mathbb P}\!\left[
        \int_s^u F(t,X_t,\alpha_t^{\mathbb P},\beta_t^{\mathbb P})\,dt
    \right]\\
    &-\mathbb E^{\mathbb P}[\phi_p^\star(u,X_u)]
     +\mathbb E^{\mathbb P}[\phi_p^\star(s,X_s)].
\end{split}
\end{equation}
All terms in this identity are finite, by the finite cost of
$\mathbb P$, the boundedness of $\phi_p^\star$, and the preceding
interior derivative bounds.

We now let $s\downarrow\tau_{p-1}$ and $u\uparrow\tau_p$.
For $t_0\in\{\tau_{p-1},\tau_p\}$, continuity of the canonical
paths and joint continuity of $\phi_p^\star$ on $[\tau_{p-1},\tau_p]\times\R^d$ give
\begin{equation*}
    \phi_p^\star(t,X_t)\longrightarrow
    \phi_p^\star(t_0,X_{t_0})
    \qquad\mathbb P\text{-almost surely}
\end{equation*}
as $t\to t_0$ within $[\tau_{p-1},\tau_p]$. Since $\phi_p^\star$ is bounded,
dominated convergence yields convergence of the endpoint expectations
in~\eqref{eq:optimality-law-truncated-defect}.
The cost integrals converge by integrability, and monotone convergence
applies to the nonnegative defects on the increasing intervals $(s,u)$.
Summing the resulting identities over $p$ gives
\begin{equation}\label{eq:optimality-law-zero-defect}
\begin{split}
    &\sum_{p=1}^P\mathbb E^{\mathbb P}\!\left[
        \int_{\tau_{p-1}}^{\tau_p}R_p^{\mathbb P}(t)\,dt\right]\\
    &\quad=\mathcal V(\pi)
       -\sum_{p=1}^P\mathbb E^{\mathbb P}\!\left[
           \phi_p^\star(\tau_p,X_{\tau_p})
           -\phi_p^\star(\tau_{p-1},X_{\tau_{p-1}})\right]\\
    &\quad=\mathcal V(\pi)+\Phi_0(\lambda^\star)
       -\lambda^\star\cdot\pi=0.
\end{split}
\end{equation}
Indeed, the interface conditions in~\eqref{eq:hjb-cascade-duality}
make the sum of the endpoint terms telescope, leaving the initial
term $\mathbb E^{\mathbb P}[\phi_1^\star(0,X_0)]
=\Phi_0(\lambda^\star)$ and the payoff terms
$\sum_{p=1}^P\mathbb E^{\mathbb P}[g_p^{\lambda^\star}(X_{\tau_p})]
=\lambda^\star\cdot\pi$ by calibration.
The last equality in~\eqref{eq:optimality-law-zero-defect} follows
from optimality of $\lambda^\star$ in~\eqref{eq:dual-attainment-supremum}.

Nonnegativity implies that $R_p^{\mathbb P}=0$
$dt\otimes\mathbb P$-almost everywhere on each time interval.
The equality case in Fenchel's inequality
\cite[Theorem~23.5]{rockafellar-1970a} therefore gives
\eqref{eq:optimality-law-subgradient}.
If $H(t,x,\cdot,\cdot)$ is differentiable, its subdifferential
consists of its gradient \cite[Theorem~25.1]{rockafellar-1970a},
which yields~\eqref{eq:optimality-law-coefficients}.
\end{proof}

\begin{remark}[Strict convexity and the gradient formula]
\label{rk:optimality-strict-convexity}
Under the hypotheses of Propositions~\ref{prop:optimality-coefficients}
and~\ref{prop:optimality-laws},
Assumption~\ref{ass:strict-convexity-cost} is sufficient for the
differentiability required in their gradient assertions.
Indeed, for each $(t,x)$, the function $F(t,x,\cdot,\cdot)$ is
proper, closed, and strictly convex, hence essentially strictly convex.
By~\cite[Theorem~26.3]{rockafellar-1970a}, its conjugate
$H(t,x,\cdot,\cdot)$ is therefore essentially smooth and, in particular,
differentiable on the interior of its effective domain.
Since $H(t,x,\cdot,\cdot)$ is finite on all of
$\R^d\times\mathcal S^d$ under~\eqref{ass:convex-duality-d1},
it is differentiable throughout this space.
Consequently, under Assumption~\ref{ass:strict-convexity-cost},
the optimal Eulerian coefficients and the characteristics of any
optimal law satisfy the gradient identities
\eqref{eq:optimality-eq-coeff} and~\eqref{eq:optimality-law-coefficients},
respectively.
\end{remark}

\begin{remark}[The HJB--Fokker--Planck optimality system]
\label{rk:optimality-hjb-fp}
The preceding optimality conditions reflect a standard
primal--dual structure in variational formulations of optimal
control and dynamic transport: the constraint governing the
evolution of the marginal flow is paired with a Hamilton--Jacobi
equation for the dual potential. In the deterministic setting,
this evolution is described by the continuity equation;
in the present diffusive setting, it is described by the
Fokker--Planck equation.

When $H$ is differentiable in its last two variables, the HJB
solution determines the drift and covariance
through~\eqref{eq:optimality-eq-coeff}, and these coefficients
govern the forward evolution of the marginal flow.
The system is completed by the initial law, the interface
conditions of the HJB cascade, and the calibration constraints
at the prescribed maturities. The dual variables entering
the interface conditions are chosen so that the resulting
marginal flow satisfies these constraints.
\end{remark}

\subsection{Proof of strong duality}\label{subsec:strong-duality-proof}

We prove Theorem~\ref{thm:regular-strong-duality}
under its structural assumptions, for an arbitrary price vector
$\pi\in\R^{MP}$, without imposing the standing finite-cost
feasibility assumption of Remark~\ref{rk:finite-cost-feasibility}.
The proof proceeds by applying Fenchel--Rockafellar duality to
the regular dual problem. The resulting optimisation problem
is initially formulated over continuous linear functionals.
We show that its feasible functionals with finite conjugate value
are represented by measures, from which we reconstruct an
admissible marginal flow and its drift and covariance coefficients.
We then identify the conjugate functional with the primal cost.
The optimiser provided by Fenchel--Rockafellar duality therefore
determines an admissible Eulerian triplet whose cost equals the
regular dual value. Together with weak duality, this proves
equality of the primal and dual values and existence of a primal
minimiser when the value is finite.

Since $H(t,x,0,0)\leq0$, the pair $(u,\lambda)=(0,0)$
is admissible in the regular dual, so $D_{\mathrm{reg}}(\pi)\geq0$.
If $D_{\mathrm{reg}}(\pi)=+\infty$, weak duality
(Proposition~\ref{prop:weak-duality-slabwise}) gives
$\mathcal V(\pi)=+\infty$, and the asserted equality follows.
For the remainder of the proof, we therefore assume that
$D_{\mathrm{reg}}(\pi)<+\infty$.

\subsubsection{Functional setting and Fenchel--Rockafellar duality}

We first formulate the regular dual problem on Banach spaces
so that Fenchel--Rockafellar duality applies.

Equip each space
$\mathcal U_p$ with the norm
\begin{equation}\label{eq:strong-duality-potential-norm}
    \|u_p\|_{\mathcal U_p}
    :=\max\bigl\{\|u_p\|_\infty,
    \|\partial_tu_p\|_\infty,\|\Dx u_p\|_\infty,
    \|\Dxx u_p\|_\infty\bigr\},
\end{equation}
where the suprema are taken over $Q_p$. This is a Banach space:
uniform convergence of a Cauchy sequence and of its indicated derivatives
preserves the derivative identities, including the one-sided time
derivatives at the endpoints. Consider the Banach spaces
\begin{equation}\label{eq:strong-duality-banach-spaces}
\begin{split}
    \mathbb X
    &:=\left(\prod_{p=1}^P\mathcal U_p\right)\times\R^{MP},\\
    \mathbb E
    &:=\left(\prod_{p=1}^P
    C_b(Q_p;\R\times\R^d\times\mathcal S^d)\right)
    \times\prod_{p=1}^P C_b(\R^d).
\end{split}
\end{equation}
We use the maximum product norms, with the Euclidean norm on $\R^{MP}$
and the uniform norms on the spaces of bounded continuous functions.
More explicitly, for
$z=((r_p,a_p,b_p)_{p=1}^P,(s_p)_{p=1}^P)\in\mathbb E$, set
\begin{equation*}
    \|z\|_{\mathbb E}
    :=\max_{1\leq p\leq P}
    \max\bigl\{\|r_p\|_\infty,\|a_p\|_\infty,
    \|b_p\|_\infty,\|s_p\|_\infty\bigr\}.
\end{equation*}
The coordinates $s_p$ will encode the interface inequalities.

Define the linear operator $L:\mathbb X\to\mathbb E$ by
\begin{equation}\label{eq:strong-duality-operator}
    L(u,\lambda)
    :=\left(
    (\partial_tu_p,\Dx u_p,\tfrac12\Dxx u_p)_{p=1}^P,
    (g_p^\lambda+u_{p+1}(\tau_p,\cdot)
                     -u_p(\tau_p,\cdot))_{p=1}^P
    \right),
\end{equation}
with the convention $u_{P+1}(\tau_P,\cdot)=0$.
The derivative and trace maps are bounded for the chosen norms, and
$\lambda\mapsto g_p^\lambda$ is continuous from $\R^{MP}$ to
$C_b(\R^d)$ because the payoff family is finite and bounded continuous.
Thus $L$ is well defined and continuous.

Let $\ell\in\mathbb X^*$ be the continuous linear functional
\begin{equation}\label{eq:strong-duality-linear-functional}
    \ell(u,\lambda):=\mu_0(u_1(0,\cdot))-\lambda\cdot\pi,
\end{equation}
and define $\Phi:\mathbb E\to[0,+\infty]$ by
\begin{equation}\label{eq:strong-duality-indicator}
    \Phi(z):=
    \begin{cases}
    0,&\text{if, } r_p(t,x)+H(t,x,a_p(t,x),b_p(t,x))\leq0
        \quad\text{on }Q_p, \\[-2pt]
      &s_p(x)\leq0\quad\text{on }\R^d, \text{ for every p,}\\
    +\infty,&\text{otherwise}.
    \end{cases}
\end{equation}
The function $H(t,x,\cdot,\cdot)$ is convex as a convex conjugate.
Consequently, the set on which $\Phi$ vanishes is convex and contains
the origin since $H(t,x,0,0)\leq0$, so $\Phi$ is proper and convex.
Moreover, the bound~\eqref{eq:hamiltonian-duality-bounds} ensures uniform
continuity of $H$ in its last two variables, uniformly in $(t,x)$, making
the constraint set closed in $\mathbb E$ and hence $\Phi$ lower semicontinuous.

By construction, $\Phi(L(u,\lambda))=0$ precisely when $(u,\lambda)$
satisfies~\eqref{eq:dual-regular-hjb}--\eqref{eq:dual-regular-interface}.
The definition of the regular dual value therefore gives
\begin{equation}\label{eq:strong-duality-convex-reformulation}
    -D_{\mathrm{reg}}(\pi)
    =\inf_{(u,\lambda)\in\mathbb X}
    \bigl\{\ell(u,\lambda)+\Phi(L(u,\lambda))\bigr\}.
\end{equation}
We next verify the qualification condition. Set $\lambda^\circ=0$ and
\begin{equation}\label{eq:strong-duality-strict-candidate}
    u_p^\circ(t,x):=T-t+P-p+1,
    \qquad (t,x)\in Q_p,\quad p=1,\ldots,P.
\end{equation}
Then $(u^\circ,0)\in\mathbb X$ and
\begin{equation*}
    z^\circ:=L(u^\circ,0)
    =\bigl((-1,0,0)_{p=1}^P,(-1)_{p=1}^P\bigr).
\end{equation*}
In particular, the last interface coordinate is also $-1$, since
$u_P^\circ(T,\cdot)=1$ and $u_{P+1}^\circ(T,\cdot)=0$ by convention.
Let $C$ be the constant in~\eqref{eq:hamiltonian-duality-bounds} and
choose $0<\delta<(1+2C)^{-1}$. If
$z=((r_p,a_p,b_p)_p,(s_p)_p)$ satisfies
$\|z-z^\circ\|_{\mathbb E}<\delta$, then, for every $p$,
\begin{align*}
    r_p(t,x)+H(t,x,a_p(t,x),b_p(t,x))
    &\leq -1+\delta+H(t,x,0,0)+2C\delta\\
    &\leq -1+(1+2C)\delta<0
    \qquad\text{on }Q_p,\\
    s_p(x)&\leq-1+\delta<0
    \qquad\text{on }\R^d.
\end{align*}
Thus $\Phi$ vanishes on an open ball about $z^\circ$ and is continuous
at $L(u^\circ,0)$, while $\ell$ is finite at $(u^\circ,0)$.

The Fenchel--Rockafellar theorem in its operator form
(a consequence of~\cite[Theorem~1.12]{brezis2011}) therefore yields
\begin{equation}\label{eq:strong-duality-fenchel-rockafellar}
    -D_{\mathrm{reg}}(\pi)
    =\max_{Y\in\mathbb E^*}
    \bigl\{-\ell^*(-L^*Y)-\Phi^*(Y)\bigr\},
\end{equation}
where $L^*:\mathbb E^*\to\mathbb X^*$ is the adjoint of $L$.
Since $D_{\mathrm{reg}}(\pi)<+\infty$ by the reduction
at the beginning of the proof and $D_{\mathrm{reg}}(\pi)\geq0$,
the value in~\eqref{eq:strong-duality-fenchel-rockafellar}
is finite.

Since $\ell$ is continuous and linear, its convex conjugate satisfies
\begin{equation*}
    \ell^*(\xi)=
    \begin{cases}
    0,&\xi=\ell,\\
    +\infty,&\xi\neq\ell,
    \end{cases}
    \qquad \xi\in\mathbb X^*.
\end{equation*}
Consequently,~\eqref{eq:strong-duality-fenchel-rockafellar} becomes
\begin{equation}\label{eq:strong-duality-abstract-dual}
    -D_{\mathrm{reg}}(\pi)
    =\max_{\substack{Y\in\mathbb E^*\\L^*Y=-\ell}}
    \bigl\{-\Phi^*(Y)\bigr\}.
\end{equation}
The constraint $L^*Y=-\ell$ is an equality in $\mathbb X^*$; explicitly,
\begin{equation}\label{eq:strong-duality-adjoint-constraint}
    \langle Y,L(u,\lambda)\rangle
    =-\mu_0(u_1(0,\cdot))+\lambda\cdot\pi
    \qquad\text{for every }(u,\lambda)\in\mathbb X.
\end{equation}
The maximum in~\eqref{eq:strong-duality-abstract-dual} is attained by
some $Y\in\mathbb E^*$ with $\Phi^*(Y)<+\infty$.
We next establish a measure representation for every such feasible $Y$.

\subsubsection{Representation of the dual variables by measures}

Fix $Y\in\mathbb E^*$ such that $L^*Y=-\ell$ and
$\Phi^*(Y)<+\infty$, and write
\begin{equation}\label{eq:strong-duality-functional-components}
    Y=\bigl((\rho^p,A^p,B^p)_{p=1}^P,(\nu_p)_{p=1}^P\bigr).
\end{equation}
Here, $\rho^p\in C_b(Q_p)^*,$ $
    A^p\in C_b(Q_p;\R^d)^*,$ $
    B^p\in C_b(Q_p;\mathcal S^d)^*,$ and $
    \nu_p\in C_b(\R^d)^*$ for any $p$.
Thus, for $z=((r_p,a_p,b_p)_p,(s_p)_p)\in\mathbb E$,
\begin{equation*}
    \langle Y,z\rangle
    =\sum_{p=1}^P\bigl[\rho^p(r_p)+A^p(a_p)+B^p(b_p)
                           +\nu_p(s_p)\bigr].
\end{equation*}
At this stage, these objects are continuous linear functionals,
not yet measures. We first prove that $\rho^p$ and $\nu_p$ are positive
and establish the domination inequalities
$|A^p(a)|\leq C\rho^p(|a|)$ and
$|B^p(b)|\leq C\rho^p(\|b\|)$ for bounded continuous fields $a$ and $b$.
We then use the adjoint constraint~\eqref{eq:strong-duality-adjoint-constraint}
to establish tightness estimates,
which allow us to represent $\rho^p$ and $\nu_p$ by finite Radon measures
on $Q_p$ and $\R^d$, respectively. Finally, the domination inequalities
yield measures representing $A^p$ and $B^p$ that are absolutely continuous
with respect to $\rho^p$, with bounded densities $\alpha^p$ and $\beta^p$.
We also prove that $\beta^p$ is positive semidefinite
$\rho^p$-almost everywhere.

To establish positivity and the domination inequalities, we use
\begin{equation}\label{eq:strong-duality-support-functional}
    \Phi^*(Y)=\sup_{z\in\operatorname{dom}\Phi}\langle Y,z\rangle
    <+\infty.
\end{equation}
In each of the following tests, all unspecified coordinates of $z$
are set to zero; these coordinates satisfy the constraints because
$H(t,x,0,0)\leq0$.

Fix $p$. For $f\in C_b(Q_p)$ with $f\geq0$ and $n\geq1$, the choice
$(r_p,a_p,b_p)=(-nf,0,0)$ belongs to $\operatorname{dom}\Phi$.
Hence $-n\rho^p(f)\leq\Phi^*(Y)$ for every $n$, which implies
$\rho^p(f)\geq0$. Likewise, taking $s_p=-nh$ with
$h\in C_b(\R^d)$ nonnegative shows that $\nu_p(h)\geq0$.
Thus $\rho^p$ and $\nu_p$ are positive functionals.
For a field $\Sigma\in C_b(Q_p;\mathcal S^d)$ with
$\Sigma(t,x)\succeq0$, condition~\eqref{ass:convex-duality-d1} gives
\begin{equation*}
    H(t,x,0,-n\Sigma(t,x))
    =\sup_{(\alpha,\beta)\in\operatorname{dom}F(t,x,\cdot,\cdot)}
    \bigl\{-n\beta:\Sigma(t,x)-F(t,x,\alpha,\beta)\bigr\}
    \leq0.
\end{equation*}
The test $(r_p,a_p,b_p)=(0,0,-n\Sigma)$ is therefore admissible.
Letting $n\to\infty$ in $-nB^p(\Sigma)\leq\Phi^*(Y)$ yields
$B^p(\Sigma)\geq0$.

Next, let $a\in C_b(Q_p;\R^d)$ and let $C$ be the constant
in~\eqref{eq:hamiltonian-duality-bounds}. The test
$(r_p,a_p,b_p)=(-nC|a|,na,0)$ is admissible, since
\begin{equation*}
    -nC|a(t,x)|+H(t,x,na(t,x),0)
    \leq H(t,x,0,0)\leq0.
\end{equation*}
It follows that
$n[A^p(a)-C\rho^p(|a|)]\leq\Phi^*(Y)$.
Dividing by $n$, letting $n\to\infty$, and repeating the argument
with $-a$ gives the first inequality below. Applying the same argument
to $(-nC\|b\|,0,nb)$ and $(-nC\|b\|,0,-nb)$ gives the second:
\begin{equation}\label{eq:strong-duality-flux-domination}
    |A^p(a)|\leq C\rho^p(|a|),\qquad
    |B^p(b)|\leq C\rho^p(\|b\|),
    \quad
    \begin{aligned}
    &a\in C_b(Q_p;\R^d),\\[-2pt]
    &b\in C_b(Q_p;\mathcal S^d).
    \end{aligned}
\end{equation}

We now use the adjoint constraint to identify the linear equations
satisfied by these functionals. Taking $u=0$ and varying each component
of $\lambda$ separately in~\eqref{eq:strong-duality-adjoint-constraint}
gives
\begin{equation}\label{eq:strong-duality-functional-calibration}
    \nu_p(G_{i(p,m)})=\pi_{i(p,m)},
    \qquad p=1,\ldots,P,\quad m=1,\ldots,M.
\end{equation}
Set $\nu_0:=\mu_0$, viewed as a functional on $C_b(\R^d)$.
Taking $\lambda=0$ and allowing only the $p$th potential to be nonzero
in the same identity yields
\begin{equation}\label{eq:strong-duality-functional-fp}
    \rho^p(\partial_tu)+A^p(\Dx u)+B^p(\tfrac12\Dxx u)
    =\nu_p(u(\tau_p,\cdot))
     -\nu_{p-1}(u(\tau_{p-1},\cdot)),
    \quad u\in\mathcal U_p.
\end{equation}
Testing with $u\equiv1$ shows that $\nu_p(1)=\nu_{p-1}(1)$,
and hence $\nu_p(1)=1$ for every $p$. Testing next with
$u(t,x)=t-\tau_{p-1}$ and using $\nu_p(1)=1$ gives
\begin{equation}\label{eq:strong-duality-functional-masses}
    \rho^p(1)=\Delta_p:=\tau_p-\tau_{p-1},
    \qquad p=1,\ldots,P.
\end{equation}
These identities are established at the level of continuous linear
functionals. To represent them by Radon measures on $Q_p$ and $\R^d$,
we next prove tightness using~\eqref{eq:strong-duality-functional-fp}
and~\eqref{eq:strong-duality-flux-domination}.

For $R\geq1$, choose $\theta_R\in C_b^\infty(\R^d)$ such that
$0\leq\theta_R\leq1$, $\theta_R=0$ on the ball $B_R(0)$,
$\theta_R=1$ outside $B_{2R}(0)$, and
\begin{equation*}
    \|\Dx\theta_R\|_\infty\leq cR^{-1},\qquad
    \|\Dxx\theta_R\|_\infty\leq cR^{-2},
\end{equation*}
where $c$ is independent of $R$. We also regard $\theta_R$ as a
time-independent function on each $Q_p$.
Testing~\eqref{eq:strong-duality-functional-fp} with $u(t,x)=\theta_R(x)$
and using positivity of $\rho^p$, the flux domination, and
$\rho^p(1)=\Delta_p$, we obtain
\begin{equation}\label{eq:strong-duality-interface-tightness}
    \nu_p(\theta_R)
    \leq\nu_{p-1}(\theta_R)
        +cC\Delta_p(R^{-1}+R^{-2}).
\end{equation}
Since $\nu_0=\mu_0$ is a probability measure and $\theta_R\to0$
pointwise, dominated convergence gives $\nu_0(\theta_R)\to0$.
Induction in $p$ therefore yields $\nu_p(\theta_R)\to0$ for every $p$.
Next, the test $u(t,x)=(\tau_p-t)\theta_R(x)$ gives
\begin{equation*}
\begin{split}
    \rho^p(\theta_R)
    &=\Delta_p\nu_{p-1}(\theta_R)
      +A^p((\tau_p-t)\Dx\theta_R)\\
    &\quad+B^p(\tfrac12(\tau_p-t)\Dxx\theta_R).
\end{split}
\end{equation*}
Applying~\eqref{eq:strong-duality-flux-domination} to the two flux terms,
using $0\leq\tau_p-t\leq\Delta_p$ on $Q_p$ and
$\rho^p(1)=\Delta_p$ from~\eqref{eq:strong-duality-functional-masses}, gives
\begin{equation}\label{eq:strong-duality-slab-tightness}
    0\leq\rho^p(\theta_R)
    \leq\Delta_p\nu_{p-1}(\theta_R)
        +cC\Delta_p^2(R^{-1}+R^{-2})
    \longrightarrow0.
\end{equation}
The convergence follows from $\nu_{p-1}(\theta_R)\to0$, established
above using~\eqref{eq:strong-duality-interface-tightness} and
$\nu_0=\mu_0$.

We now show that these tightness estimates yield Radon representations
on the full spaces of bounded continuous functions. Let $\mathcal T$
denote one of the positive functionals $\rho^p$ or $\nu_p$, and let
$S=Q_p$ or $S=\R^d$, respectively. Since $S$ is locally compact,
the Riesz--Markov representation theorem applied to the restriction of
$\mathcal T$ to $C_0(S)$ gives a finite positive Radon measure $m$ with
\begin{equation*}
    \mathcal T(h)=\int_S h\,dm,\qquad h\in C_0(S),
\end{equation*}
where $C_0(S)$ denotes the continuous functions vanishing at infinity.
Set $\chi_R=1-\theta_R$. For every $f\in C_b(S)$, one has
$f\chi_R\in C_c(S)$; when $S=Q_p$, this uses the compactness of the
time interval $[\tau_{p-1},\tau_p]$. Therefore
\begin{equation*}
    \mathcal T(f\chi_R)=\int_S f\chi_R\,dm.
\end{equation*}
Positivity and the tightness estimates above give
\begin{equation*}
    |\mathcal T(f)-\mathcal T(f\chi_R)|
    \leq\|f\|_\infty\mathcal T(\theta_R)\longrightarrow0.
\end{equation*}
On the measure side, dominated convergence gives
$\int_S f\chi_R\,dm\to\int_S f\,dm$. Hence
$\mathcal T(f)=\int_S f\,dm$ for every $f\in C_b(S)$.
We henceforth identify $\rho^p$ and $\nu_p$ with their representing
Radon measures. Their previously established normalizations become
\begin{equation}\label{eq:strong-duality-radon-masses}
    \nu_p\in\mathcal P(\R^d),\qquad
    \rho^p(Q_p)=\Delta_p.
\end{equation}

Having represented $\rho^p$ and $\nu_p$ by Radon measures,
we now turn to $A^p$ and $B^p$. We show that these functionals
are given by integration against bounded densities with respect
to $\rho^p$.
The domination estimates~\eqref{eq:strong-duality-flux-domination}
give
\begin{equation*}
    |A^p(a)|\leq C\int_{Q_p}|a|\,d\rho^p,\qquad
    |B^p(b)|\leq C\int_{Q_p}\|b\|\,d\rho^p.
\end{equation*}
These bounds imply that each functional depends only on the
$\rho^p$-almost everywhere equivalence class of its argument and is
continuous for the corresponding $L^1(\rho^p)$ norm.
Since compactly supported continuous fields are dense in these $L^1$
spaces, $A^p$ and $B^p$ extend uniquely to continuous linear functionals
on $L^1(\rho^p;\R^d)$ and $L^1(\rho^p;\mathcal S^d)$, respectively.
The representation of their duals by $L^\infty$ gives densities
\begin{equation*}
    \alpha^p\in L^\infty(\rho^p;\R^d),\qquad
    \beta^p\in L^\infty(\rho^p;\mathcal S^d),\qquad
    |\alpha^p|,\|\beta^p\|\leq C\quad\rho^p\text{-a.e.},
\end{equation*}
such that
\begin{equation}\label{eq:strong-duality-flux-densities}
    A^p(a)=\int_{Q_p}a\cdot\alpha^p\,d\rho^p,\qquad
    B^p(b)=\int_{Q_p}b:\beta^p\,d\rho^p
\end{equation}
for all bounded continuous fields $a$ and $b$ of the corresponding
dimensions. In particular, the fluxes are represented by finite
vector- and matrix-valued Radon measures, which we denote by
$A^p=\alpha^p\rho^p$ and $B^p=\beta^p\rho^p$.
Finally, positivity of $B^p$ gives
\begin{equation*}
    \int_{Q_p}f\,v^\top\beta^p v\,d\rho^p
    =B^p(fvv^\top)\geq0,
    \qquad f\in C_c(Q_p),\quad f\geq0,\quad v\in\mathbb Q^d.
\end{equation*}
Thus $v^\top\beta^p v\geq0$ almost everywhere for each rational vector
$v$. Taking a common set of full measure for this countable family and
using continuity of the quadratic form in $v$, we conclude that
$\beta^p\succeq0$ $\rho^p$-almost everywhere.

\subsubsection{Construction of the marginal flow}

We now construct a narrowly continuous family of probability
measures from the space-time measures $\rho^p$. We first use
\eqref{eq:strong-duality-functional-fp} to show that the time marginal
of $\rho^p$ is Lebesgue measure on $[\tau_{p-1},\tau_p]$. Disintegration then gives
$\rho^p(dt,dx)=dt\,\rho_t^p(dx)$, with $\rho_t^p$ a probability measure
for almost every $t$. We next prove that this family admits a narrowly
continuous representative on $[\tau_{p-1},\tau_p]$, with endpoint values $\nu_{p-1}$
and $\nu_p$. The representatives on adjacent intervals therefore agree
at each maturity and define a single narrowly continuous family
$(\rho_t)_{t\in[0,T]}$. Together with the coefficients constructed above,
this family satisfies the constraints~\eqref{eq:main-primal-constr}.
To establish this, we use Lemma~\ref{lem:fp-slabwise}
from Section~\ref{sec:framework}, which characterises these
constraints through the weak Fokker--Planck
identity~\eqref{eq:fp-slabwise} on each time interval
and the prescribed initial and calibration conditions.

Fix $p$ and let $\vartheta_p$ be the time marginal of $\rho^p$ on
$[\tau_{p-1},\tau_p]$.
Taking $u(t,x)=\eta(t)$, with $\eta\in C^1([\tau_{p-1},\tau_p])$, in
\eqref{eq:strong-duality-functional-fp} gives
\begin{equation*}
    \int_{[\tau_{p-1},\tau_p]}\eta'(t)\,\vartheta_p(dt)
    =\eta(\tau_p)-\eta(\tau_{p-1})
    =\int_{\tau_{p-1}}^{\tau_p}\eta'(t)\,dt.
\end{equation*}
Every continuous function on $[\tau_{p-1},\tau_p]$ is the derivative of a $C^1$ function,
so $\vartheta_p=dt|_{[\tau_{p-1},\tau_p]}$. In particular, $\rho^p$ has no mass on the
endpoint slices. Disintegration with respect to time therefore gives
\begin{equation}\label{eq:strong-duality-time-disintegration}
    \rho^p(dt,dx)=dt\,\rho_t^p(dx),\qquad
    \rho_t^p\in\mathcal P(\R^d)\quad\text{for a.e. }t\in[\tau_{p-1},\tau_p].
\end{equation}

Let $f\in C_b^2(\R^d)$, with bounded first and second derivatives, and set
\begin{equation*}
    k_f^p(t):=\int_{\R^d}
    \left(\alpha^p(t,x)\cdot\Dx f(x)
    +\tfrac12\beta^p(t,x):\Dxx f(x)\right)\rho_t^p(dx).
\end{equation*}
The coefficient bounds give $k_f^p\in L^\infty([\tau_{p-1},\tau_p])$.
Testing~\eqref{eq:strong-duality-functional-fp} with $u(t,x)=\eta(t)f(x)$
and $\eta\in C_c^1((\tau_{p-1},\tau_p))$ shows that
$t\mapsto\rho_t^p(f)$ has distributional derivative $k_f^p$.
Since $k_f^p\in L^\infty([\tau_{p-1},\tau_p])$, the function $t\mapsto\rho_t^p(f)$
belongs to $W^{1,\infty}((\tau_{p-1},\tau_p))$ and therefore admits a
Lipschitz, hence absolutely continuous, representative $h_f^p$ on $[\tau_{p-1},\tau_p]$.
Testing~\eqref{eq:strong-duality-functional-fp} with
$u(t,x)=\eta(t)f(x)$, for arbitrary $\eta\in C^1([\tau_{p-1},\tau_p])$, and
integrating by parts in time gives
\begin{equation*}
    \eta(\tau_p)h_f^p(\tau_p)
    -\eta(\tau_{p-1})h_f^p(\tau_{p-1})
    =\eta(\tau_p)\nu_p(f)-\eta(\tau_{p-1})\nu_{p-1}(f).
\end{equation*}
Choosing the endpoint values of $\eta$ independently identifies both
traces. Thus
\begin{equation}\label{eq:strong-duality-scalar-traces}
    h_f^p(t)=\nu_{p-1}(f)+\int_{\tau_{p-1}}^t k_f^p(s)\,ds,
    \qquad h_f^p(\tau_p)=\nu_p(f).
\end{equation}

We have obtained a continuous representative for each scalar integral
$t\mapsto\rho_t^p(f)$. We must still construct a probability measure
at every time, agreeing almost everywhere with the disintegration, whose
integrals realize these representatives. Uniform tightness in time will
provide probability limits at the exceptional times, and a countable
determining family of test functions will ensure their uniqueness.
Indeed, applying~\eqref{eq:strong-duality-scalar-traces} to $f=\theta_R$
and using the derivative bounds for $\theta_R$ gives
\begin{equation}\label{eq:strong-duality-uniform-time-tightness}
    \rho_t^p(\theta_R)
    \leq\nu_{p-1}(\theta_R)
        +cC\Delta_p(R^{-1}+R^{-2})
    \quad\text{for a.e. }t\in[\tau_{p-1},\tau_p].
\end{equation}
Choose a countable family $(f_n)_{n\geq1}\subset C_c^\infty(\R^d)$
dense in $C_0(\R^d)$ for the uniform norm; in particular, it determines
finite Radon measures. There exists a set $J_p\subset[\tau_{p-1},\tau_p]$ of full
Lebesgue measure on which $\rho_t^p$ is a probability measure,
$\rho_t^p(f_n)=h_{f_n}^p(t)$ for every $n$, and
\eqref{eq:strong-duality-uniform-time-tightness} holds for every integer
$R\geq1$. Since $\theta_R=1$ outside $B_{2R}(0)$, we have
\begin{equation*}
    \sup_{t\in J_p}\rho_t^p(\R^d\setminus\overline B_{2R}(0))
    \leq\nu_{p-1}(\theta_R)+cC\Delta_p(R^{-1}+R^{-2})
    \longrightarrow0.
\end{equation*}
Hence $(\rho_t^p)_{t\in J_p}$ is uniformly tight.

For any $t\in[\tau_{p-1},\tau_p]$, choose $t_j\in J_p$ with $t_j\to t$, which is
possible because $J_p$ is dense in $[\tau_{p-1},\tau_p]$. By tightness and Prokhorov's
theorem, a subsequence of $(\rho_{t_j}^p)_j$ converges narrowly to a
probability measure $\widetilde\rho_t^p$. For every $n$,
\begin{equation*}
    \widetilde\rho_t^p(f_n)=\lim_j h_{f_n}^p(t_j)=h_{f_n}^p(t),
\end{equation*}
where the limit is taken along that subsequence. These identities
uniquely determine $\widetilde\rho_t^p$, independently of the approximating
times and the subsequence. For $t\in J_p$, it coincides with $\rho_t^p$.

Since $\theta_R$ is bounded continuous, the bound in
\eqref{eq:strong-duality-uniform-time-tightness} passes to the limit
and holds for $\widetilde\rho_t^p$ at every $t\in[\tau_{p-1},\tau_p]$.
Thus the extended family is uniformly tight as well.
For any sequence $s_j\to t$ in $[\tau_{p-1},\tau_p]$, every subsequential narrow limit
of $\widetilde\rho_{s_j}^p$ integrates each $f_n$ to $h_{f_n}^p(t)$,
by continuity of $h_{f_n}^p$. It must therefore equal
$\widetilde\rho_t^p$. Relative compactness, by Prokhorov's theorem,
and uniqueness of the limit give
$\widetilde\rho_{s_j}^p\to\widetilde\rho_t^p$ narrowly, proving
narrow continuity on $[\tau_{p-1},\tau_p]$.
The scalar trace identities~\eqref{eq:strong-duality-scalar-traces}
then imply
\begin{equation}\label{eq:strong-duality-measure-traces}
    \widetilde\rho_{\tau_{p-1}}^p=\nu_{p-1},\qquad
    \widetilde\rho_{\tau_p}^p=\nu_p.
\end{equation}

The curves on adjacent intervals have the same trace at their common
endpoint. They therefore define a narrowly continuous curve
$\rho\in\mathscr P([0,T];\R^d)$ satisfying
$\rho_0=\mu_0$, $\rho_{\tau_p}=\nu_p$, and
$\rho^p(dt,dx)=dt\,\rho_t(dx)$ on each $Q_p$.
Choose bounded Borel versions of $\alpha^p$ and $\beta^p$, with
$\beta^p\succeq0$, and define $(\alpha,\beta)=(\alpha^p,\beta^p)$ on
each open time interval, up to $dt\,\rho_t(dx)$-null sets.
They are Borel and satisfy
\begin{equation*}
    \int_0^T\int_{\R^d}(|\alpha|+\|\beta\|)\,\rho_t(dx)\,dt
    \leq2CT.
\end{equation*}
With these identifications,
\eqref{eq:strong-duality-functional-fp} is the weak Fokker--Planck
identity on each interval, and
\eqref{eq:strong-duality-functional-calibration} gives the price
constraints at the actual marginal measures $\rho_{\tau_p}$.
Lemma~\ref{lem:fp-slabwise} therefore shows that $(\rho,\alpha,\beta)$
satisfies~\eqref{eq:main-primal-constr}.

\subsubsection{Identification of the cost and conclusion}

We now show that $\Phi^*(Y)$ equals the cost of the
Eulerian triplet constructed above. We first express $\Phi^*(Y)$
as a sum of suprema over bounded continuous fields on each $Q_p$.
We then identify each supremum with the integral of $F$
over the corresponding interval.
Applying the resulting cost identity to an optimiser provided by
Fenchel--Rockafellar duality gives an admissible Eulerian triplet
whose cost equals $D_{\mathrm{reg}}(\pi)$. Weak duality then yields
equality of the primal and regular dual values and existence of a
primal minimiser in the finite-value case. Finally, we show that
every finite-cost admissible Eulerian triplet defines a feasible
functional in the abstract problem with the same cost.

We compute $\Phi^*(Y)$ using its support representation
\eqref{eq:strong-duality-support-functional}.
Each interface coordinate contributes zero, since
$\sup_{s_p\leq0}\nu_p(s_p)=0$ by positivity of $\nu_p$.
For fixed bounded continuous fields $a,b$ on $Q_p$, the function
$H(\cdot,a,b)$ is bounded continuous by
\eqref{ass:convex-duality-d2} and~\eqref{eq:hamiltonian-duality-bounds}.
Positivity of $\rho^p$ therefore allows the supremum over $r_p$
subject to $r_p+H(\cdot,a,b)\leq0$ to be attained at
$r_p=-H(\cdot,a,b)$.
As the coordinates of different intervals are independent, it follows that
\begin{equation}\label{eq:strong-duality-conjugate-slab-sum}
    \Phi^*(Y)=\sum_{p=1}^P S_p,
\end{equation}
where
\begin{equation}\label{eq:strong-duality-slab-support}
    S_p:=\sup_{(a,b)\in C_b(Q_p;\R^d\times\mathcal S^d)}
    \int_{Q_p}\bigl[a\cdot\alpha^p+b:\beta^p-H(t,x,a,b)\bigr]
    \,d\rho^p.
\end{equation}

Fix $p$, write $s=(t,x)$, and set
\begin{equation*}
    q(s,z):=a\cdot\alpha^p(s)+b:\beta^p(s)-H(s,a,b),
    \qquad z=(a,b)\in\R^d\times\mathcal S^d.
\end{equation*}
Pointwise convex duality gives
\begin{equation}\label{eq:strong-duality-pointwise-cost}
    F(s,\alpha^p(s),\beta^p(s))
    =\sup_{z\in\R^d\times\mathcal S^d}q(s,z).
\end{equation}
To identify $S_p$ with the integral of $F$, we justify interchanging
the supremum and the integral. In $S_p$, the supremum is taken over
bounded continuous fields, whereas the supremum
in~\eqref{eq:strong-duality-pointwise-cost} is taken separately at each
point. The upper bound follows from Fenchel's inequality:
\begin{equation*}
    S_p\leq\int_{Q_p}F(s,\alpha^p(s),\beta^p(s))\,d\rho^p(s).
\end{equation*}
For the reverse bound, we construct bounded continuous fields whose
integral values approach the integral of the pointwise supremum.

Choose a countable dense family $(\zeta_k)_{k\geq1}$ in
$\R^d\times\mathcal S^d$ with
$\zeta_1=0$. Since $q(s,\cdot)$ is continuous, the supremum in
\eqref{eq:strong-duality-pointwise-cost} can be taken over this family.
Consequently, the bounded Borel functions
\begin{equation*}
    M_n(s):=\max_{1\leq k\leq n}q(s,\zeta_k)
\end{equation*}
satisfy
\begin{equation}\label{eq:strong-duality-finite-maxima}
    0\leq M_n(s)\uparrow F(s,\alpha^p(s),\beta^p(s)).
\end{equation}
Here nonnegativity follows from
$M_n(s)\geq q(s,0)=-H(s,0,0)\geq0$.
For each $n$, define
\begin{equation*}
    k_n(s):=\min\{k\leq n:q(s,\zeta_k)=M_n(s)\},
    \qquad z^{(n)}(s):=\zeta_{k_n(s)}.
\end{equation*}
The sets on which $k_n$ takes a given value are Borel, being determined
by finitely many comparisons of Borel functions. Thus $z^{(n)}$ is a
bounded Borel field with $q(s,z^{(n)}(s))=M_n(s)$.
It need not be continuous and hence cannot yet be used in the
supremum~\eqref{eq:strong-duality-slab-support}.

Since $\rho^p$ is a finite Radon measure, choose
$z^{(n,j)}\in C_c(Q_p;\R^d\times\mathcal S^d)$ such that
\begin{equation*}
    \int_{Q_p}|z^{(n,j)}-z^{(n)}|\,d\rho^p\longrightarrow0
    \qquad (j\to\infty).
\end{equation*}
The bounds on $\alpha^p,\beta^p$ and
\eqref{eq:hamiltonian-duality-bounds} give a constant $K<\infty$,
independent of $s,z,z'$, such that
\begin{equation*}
    |q(s,z)-q(s,z')|\leq K|z-z'|
    \quad\text{for }\rho^p\text{-a.e. }s.
\end{equation*}
Therefore, for each fixed $n$,
\begin{equation*}
\begin{split}
    &\left|\int_{Q_p}q(s,z^{(n,j)}(s))\,d\rho^p(s)
                    -\int_{Q_p}M_n(s)\,d\rho^p(s)\right|\\
    &\qquad\leq K\int_{Q_p}|z^{(n,j)}-z^{(n)}|\,d\rho^p
    \longrightarrow0.
\end{split}
\end{equation*}
These integrals are finite because the fields are bounded,
$H(\cdot,0,0)$ is bounded, and $\rho^p$ is finite.
Each $z^{(n,j)}$ is admissible in the supremum defining $S_p$, so
letting $j\to\infty$ gives $S_p\geq\int_{Q_p}M_n\,d\rho^p$.
Letting $n\to\infty$ and applying monotone convergence to
\eqref{eq:strong-duality-finite-maxima} yields
\begin{equation*}
    S_p\geq\int_{Q_p}F(s,\alpha^p(s),\beta^p(s))\,d\rho^p(s),
\end{equation*}
including when this integral is infinite. Combining both inequalities
and summing over $p$, we obtain
\begin{equation}\label{eq:strong-duality-conjugate-action}
\begin{split}
    \Phi^*(Y)
    &=\sum_{p=1}^P\int_{Q_p}
      F(t,x,\alpha^p(t,x),\beta^p(t,x))\,d\rho^p(t,x)\\
    &=\int_0^T\int_{\R^d}
      F(t,x,\alpha(t,x),\beta(t,x))\,\rho_t(dx)\,dt.
\end{split}
\end{equation}
In particular, the reconstructed triplet has finite cost whenever
$\Phi^*(Y)<+\infty$.

Applied to an abstract optimizer $Y^\star$ provided by
Fenchel--Rockafellar duality, the preceding construction yields
an admissible Eulerian triplet
$(\rho^\star,\alpha^\star,\beta^\star)$ whose cost satisfies
\begin{equation*}
    \int_0^T\int_{\R^d}
    F(t,x,\alpha^\star(t,x),\beta^\star(t,x))
    \,\rho_t^\star(dx)\,dt
    =\Phi^*(Y^\star)
    =D_{\mathrm{reg}}(\pi)<+\infty.
\end{equation*}
Together with weak duality, this gives
\begin{equation*}
    D_{\mathrm{reg}}(\pi)
    \leq \mathcal V(\pi)
    \leq \int_0^T\int_{\R^d}
    F(t,x,\alpha^\star(t,x),\beta^\star(t,x))
    \,\rho_t^\star(dx)\,dt
    =D_{\mathrm{reg}}(\pi).
\end{equation*}
Thus $\mathcal V(\pi)=D_{\mathrm{reg}}(\pi)<+\infty$,
and the reconstructed triplet attains the primal infimum. The same argument applies to every abstract optimizer. The converse construction below explains how an arbitrary
finite-cost admissible Eulerian triplet is represented in
the abstract formulation.

Let $(\rho,\alpha,\beta)$ be any admissible Eulerian triplet of
finite cost, and define $Y$ by its interval measures
$dt\,\rho_t(dx)$, its fluxes, and the marginals
$\nu_p=\rho_{\tau_p}$.
Condition~\eqref{ass:convex-duality-d1} ensures that its coefficients
are bounded almost everywhere. The identity in
Lemma~\ref{lem:fp-slabwise}, extended to bounded regular tests by the
spatial-cutoff argument in Proposition~\ref{prop:weak-duality-slabwise},
together with calibration gives $L^*Y=-\ell$.
The preceding conjugate calculation applies to these measures and
bounded coefficients, so~\eqref{eq:strong-duality-conjugate-action}
holds for this $Y$ as well. Thus the finite-cost feasible objects in the
abstract dual and the Eulerian primal have the same objective values,
up to the minus sign in~\eqref{eq:strong-duality-abstract-dual}.

Together with the case $D_{\mathrm{reg}}(\pi)=+\infty$
treated at the beginning of the proof, this establishes
Theorem~\ref{thm:regular-strong-duality}.

\section{Uniqueness and stability}\label{sec:stability}

\subsection{Uniqueness of primal optimizers}\label{subsec:primal-uniqueness}

We use the optimality conditions of
Propositions~\ref{prop:optimality-coefficients} and~\ref{prop:optimality-laws}
to identify common
dynamics for all primal minimizers and deduce their uniqueness from
uniqueness in law of the associated SDE. A single optimizer
$\lambda^\star$ of the HJB dual problem suffices for this argument:
the optimality conditions apply to every primal minimizer paired with
that same dual optimizer. Uniqueness of the dual optimizer is therefore
not required. We first establish uniqueness in the Eulerian and
probabilistic formulations under an explicit uniqueness-in-law
assumption, then verify this assumption using the companion PDE results.

Throughout this subsection, we work under the hypotheses of
Proposition~\ref{prop:hjb-dual-identification}.
We assume that $\mathcal V(\pi)<+\infty$ and that the HJB dual
problem~\eqref{eq:dual-attainment-supremum} is attained.
By Proposition~\ref{prop:dual-attainement}, these two conditions
hold whenever $\pi\in\operatorname{ri}(\mathfrak P)$.
We fix one dual optimizer $\lambda^\star\in\R^{MP}$.
We further assume that
$H(t,x,\cdot,\cdot)$ is differentiable for every $(t,x)\in\Lambda$.
By Remark~\ref{rk:optimality-strict-convexity},
Assumption~\ref{ass:strict-convexity-cost} is sufficient for this
differentiability.

Set $\phi_p^\star:=\phi_p^{\lambda^\star}$ for $p=1,\ldots,P$.
On each $(\tau_{p-1},\tau_p)\times\R^d$, define the feedback coefficients
\begin{equation}\label{eq:primal-uniqueness-feedback}
    (\widehat\alpha(t,x),\widehat\beta(t,x))
    :=\nabla_{(a,b)}H\!\left(
        t,x,\Dx\phi_p^\star(t,x),
        \tfrac12\Dxx\phi_p^\star(t,x)\right).
\end{equation}
Since $H$ is jointly continuous and $H(t,x,\cdot,\cdot)$ is convex
and differentiable, $\nabla_{(a,b)}H$ is jointly continuous.
Together with the interior $C^{1,2}$ regularity of $\phi_p^\star$
provided by~\eqref{ass:hjb-slabwise-h1}, this implies that the feedback
coefficients are continuous on each $(\tau_{p-1},\tau_p)\times\R^d$, hence Borel measurable.
By conjugacy, the gradient in~\eqref{eq:primal-uniqueness-feedback}
belongs to $\operatorname{dom}F(t,x,\cdot,\cdot)$.
Consequently,~\eqref{ass:convex-duality-d1} ensures that
$\widehat\alpha$ and $\widehat\beta$ are uniformly bounded and
$\widehat\beta\succeq0$.

\begin{assumption}[Uniqueness in law]
\label{ass:primal-uniqueness-in-law}
For the coefficients defined in~\eqref{eq:primal-uniqueness-feedback},
the SDE
\begin{equation}\label{eq:primal-uniqueness-sde}
    dX_t=\widehat\alpha(t,X_t)\,dt
        +\widehat\beta(t,X_t)^{1/2}\,dW_t,
    \qquad X_0\sim\mu_0,
\end{equation}
where $W$ is a $d$-dimensional Brownian motion, admits at most one
law of a weak solution on $\Omega=C([0,T];\R^d)$.
Thus any two weak solutions with initial law $\mu_0$, possibly defined
on different filtered probability spaces, have the same law on $\Omega$.
\end{assumption}

Weak existence is not part of Assumption~\ref{ass:primal-uniqueness-in-law}.
It will follow by identifying the coefficients of a primal minimizer
through Proposition~\ref{prop:optimality-coefficients} and applying
the superposition principle in Theorem~\ref{thm:fp-weak-sol}.

\begin{prop}[Uniqueness of the Eulerian minimizer]
\label{prop:primal-uniqueness-eulerian}
Under the standing hypotheses of this subsection, suppose that
Assumption~\ref{ass:primal-uniqueness-in-law} holds.
Then the Eulerian primal problem~\eqref{eq:main-primal} subject to
\eqref{eq:main-primal-constr} admits a unique minimizer in the following
sense: if $(\rho^1,\alpha^1,\beta^1)$ and $(\rho^2,\alpha^2,\beta^2)$
are two minimizers, then
\begin{equation*}
    \rho_t^1=\rho_t^2=:\rho_t
    \qquad\text{for every }t\in[0,T],
\end{equation*}
and
\begin{equation*}
    (\alpha^1,\beta^1)=(\alpha^2,\beta^2)
    \qquad dt\,\rho_t(dx)\text{-almost everywhere}.
\end{equation*}
\end{prop}

\begin{proof}
Since $\pi\in\mathfrak P$, the primal value is finite, and
Corollary~\ref{coro:primal-minimum} provides a minimizer of finite cost.
Let $(\rho^1,\alpha^1,\beta^1)$ and $(\rho^2,\alpha^2,\beta^2)$
be two minimizers of~\eqref{eq:main-primal} subject to
\eqref{eq:main-primal-constr}. The fixed dual optimizer $\lambda^\star$
can be paired with either minimizer, so
Proposition~\ref{prop:optimality-coefficients} applies to both pairs.
By~\eqref{eq:optimality-eq-coeff} and the definition
\eqref{eq:primal-uniqueness-feedback}, for $i=1,2$ we obtain
\begin{equation}\label{eq:primal-uniqueness-coefficient-identification}
    (\alpha^i(t,x),\beta^i(t,x))
    =(\widehat\alpha(t,x),\widehat\beta(t,x))
    \qquad dt\,\rho_t^i(dx)\text{-almost everywhere}
\end{equation}
on each $(\tau_{p-1},\tau_p)\times\R^d$.
Thus the same feedback coefficients represent both minimizers, with
each identity understood relative to its own occupation measure.

Since the maturity times have zero Lebesgue measure,
\eqref{eq:primal-uniqueness-coefficient-identification} holds
$dt\,\rho_t^i(dx)$-almost everywhere on $(0,T)\times\R^d$.
For each $i=1,2$, we may therefore replace $(\alpha^i,\beta^i)$
by $(\widehat\alpha,\widehat\beta)$ in the weak Fokker--Planck
formulation satisfied by $\rho^i$. Both flows are thus narrowly
continuous probability-valued solutions of the same equation
\begin{equation}\label{eq:primal-uniqueness-common-fp}
    \partial_t\rho
    +\nabla_x\cdot(\widehat\alpha\rho)
    -\tfrac12\sum_{j,k=1}^d
        \partial_{x_jx_k}(\widehat\beta_{jk}\rho)=0,
    \qquad \rho_0=\mu_0,
\end{equation}
on $(0,T)\times\R^d$, in the sense of
Definition~\ref{def:fokker-planck}.

The boundedness of $(\widehat\alpha,\widehat\beta)$ ensures
the integrability condition~\eqref{ass:integrability-fp} for each flow.
Applying Theorem~\ref{thm:fp-weak-sol}, we obtain
$\mathbb P^1,\mathbb P^2\in\mathcal P^1_{\mathcal M}$ solving the
martingale problem associated with these coefficients and satisfying
\begin{equation*}
    X_t\#\mathbb P^i=\rho_t^i
    \qquad\text{for every }t\in[0,T],\quad i=1,2.
\end{equation*}
Under each $\mathbb P^i$, the canonical process admits a weak solution
representation of~\eqref{eq:primal-uniqueness-sde}, possibly after
enlarging the probability space. Both initial laws are $\mu_0$.
Assumption~\ref{ass:primal-uniqueness-in-law} therefore gives
$\mathbb P^1=\mathbb P^2$ as probability measures on $\Omega$.

Taking the marginals of these equal laws yields
\begin{equation*}
    \rho_t^1=X_t\#\mathbb P^1
    =X_t\#\mathbb P^2=\rho_t^2=:\rho_t
    \qquad\text{for every }t\in[0,T].
\end{equation*}
The two identities in~\eqref{eq:primal-uniqueness-coefficient-identification}
are therefore valid with respect to the same occupation measure
$dt\,\rho_t(dx)$. Consequently,
\begin{equation*}
    (\alpha^1,\beta^1)
    =(\widehat\alpha,\widehat\beta)
    =(\alpha^2,\beta^2)
    \qquad dt\,\rho_t(dx)\text{-almost everywhere},
\end{equation*}
which proves the claimed uniqueness.
\end{proof}

We can now apply Proposition~\ref{prop:optimality-laws} to obtain
uniqueness among all optimal semimartingale laws.

\begin{prop}[Uniqueness of the optimal law]
\label{prop:primal-uniqueness-law}
Under the standing hypotheses of this subsection, suppose that
Assumption~\ref{ass:primal-uniqueness-in-law} holds.
Then the probabilistic primal problem admits a unique optimal law
in $\mathcal P(\mu_0,\tau,\pi)$.
\end{prop}

\begin{proof}
Existence of an optimal law follows from
Corollary~\ref{coro:primal-minimum}.

For any optimal law $\mathbb P$, Proposition~\ref{prop:optimality-laws},
applied with the fixed dual optimizer $\lambda^\star$, gives
\begin{equation*}
    (\alpha_t^{\mathbb P},\beta_t^{\mathbb P})
    =(\widehat\alpha(t,X_t),\widehat\beta(t,X_t))
    \qquad dt\otimes\mathbb P\text{-almost everywhere}.
\end{equation*}
Thus, possibly after enlarging the probability space, the canonical
process under $\mathbb P$ is a weak solution
of~\eqref{eq:primal-uniqueness-sde} with initial law $\mu_0$.
Assumption~\ref{ass:primal-uniqueness-in-law} implies that all such
optimal laws coincide on $\Omega$.
\end{proof}

We now verify Assumption~\ref{ass:primal-uniqueness-in-law} under the
structural hypotheses of the companion PDE article~\cite{HJB-property}, using its pathwise
uniqueness result on each interval of the cascade.

\begin{corol}[Primal uniqueness under the PDE hypotheses]
\label{coro:primal-uniqueness-pde}
Under the standing hypotheses of this subsection, suppose that $H$
satisfies Assumption~1 of~\cite{HJB-property} on each time interval
after time translation. Then
Assumption~\ref{ass:primal-uniqueness-in-law} holds.
Consequently, the Eulerian primal problem admits a unique minimizer
in the sense of Proposition~\ref{prop:primal-uniqueness-eulerian},
and the probabilistic primal problem admits a unique optimal law
in $\mathcal P(\mu_0,\tau,\pi)$.
\end{corol}

\begin{proof}
Each component $\phi_p^\star$ has bounded uniformly continuous terminal
data by the cascade construction; see also Section~6 of~\cite{HJB-property}.
After translating time, Lemma~3.1 of~\cite{HJB-property} therefore
applies to its associated SDE. In particular, pathwise uniqueness holds
on the whole closed interval $[\tau_{p-1},\tau_p]$ for every
deterministic initial point.

This uniqueness also holds for a common random initial value at
$\tau_{p-1}$. Indeed, condition the joint law of two candidate solutions
and their driving Brownian increments on this initial value.
Since Brownian increments after $\tau_{p-1}$ are independent of
the sigma-field at $\tau_{p-1}$, the conditional laws still describe
two solutions driven by the same Brownian motion, now with a fixed
initial point. The preceding pathwise uniqueness result applies
for almost every initial value.

Consider two solutions of~\eqref{eq:primal-uniqueness-sde} on the
same filtered probability space, driven by the same Brownian motion
and with the same initial variable. They coincide on $[0,\tau_1]$,
including at $\tau_1$. Applying the preceding argument successively
on $[\tau_1,\tau_2],\ldots,[\tau_{P-1},T]$ shows that they coincide
on $[0,T]$. Thus pathwise uniqueness holds for the full cascade.

The Yamada--Watanabe theorem
\cite[Proposition~1]{yamada-watanabe1971} gives uniqueness in law
for each deterministic initial point. Conditioning on $X_0$ and
integrating against $\mu_0$ then gives uniqueness in law for the
prescribed initial distribution $\mu_0$, which is
Assumption~\ref{ass:primal-uniqueness-in-law}.
Propositions~\ref{prop:primal-uniqueness-eulerian}
and~\ref{prop:primal-uniqueness-law} complete the proof.
\end{proof}

We now consider the feedback diffusion associated with an arbitrary
parameter $\lambda\in\R^{MP}$ and identify its model prices through
the gradient of $\Phi_0$. For the following proposition and remark, we do not
impose the standing finite-cost feasibility assumption of
Remark~\ref{rk:finite-cost-feasibility} or assume dual attainment;
the required hypotheses are stated explicitly below.

\begin{prop}[Dual gradient and exact calibration]
\label{prop:dual-gradient-calibration}
Under the hypotheses of
Proposition~\ref{prop:hjb-dual-identification},
assume that, on each time interval after time translation,
$H$ satisfies Assumption~1 of~\cite{HJB-property},
whose conditions are recalled in
Remark~\ref{rk:hjb-pde-inputs}.

For each $\lambda\in\R^{MP}$, define
$(\alpha^\lambda,\beta^\lambda)$ pointwise on each $(\tau_{p-1},\tau_p)\times\R^d$
by the right-hand side of~\eqref{eq:optimality-eq-coeff},
with $\phi_p^\star$ replaced by $\phi_p^\lambda$.
Their values at the maturity times may be chosen arbitrarily,
with $\beta^\lambda$ positive semidefinite.
Let $\mathbb P^\lambda$ denote the unique law on $\Omega$
of the solution to
\begin{equation}\label{eq:dual-gradient-diffusion}
    dX_t
    =\alpha^\lambda(t,X_t)\,dt
     +\bigl(\beta^\lambda(t,X_t)\bigr)^{1/2}\,dW_t,
    \qquad \operatorname{Law}(X_0)=\mu_0.
\end{equation}

The function $\Phi_0$ defined
in~\eqref{eq:dual-attainment-value-map}
is continuously differentiable on $\R^{MP}$.
For $i=i(p,m)$ and $q\in\{1,\ldots,P\}$, its associated
first variations
\[
    u_{q,i}^\lambda
    :=\partial_{\lambda_i}\phi_q^\lambda
\]
are the unique bounded solutions, continuous on
$[\tau_{q-1},\tau_q]\times\R^d$ and classical in its interior, of
\begin{equation}\label{eq:dual-gradient-linearized-cascade}
\begin{cases}
    \partial_tu_{q,i}^\lambda
    +\alpha^\lambda\cdot\Dx u_{q,i}^\lambda
    +\tfrac12\beta^\lambda:\Dxx u_{q,i}^\lambda=0,
    & t\in(\tau_{q-1},\tau_q),\\
    u_{q,i}^\lambda(\tau_q,\cdot)
    =u_{q+1,i}^\lambda(\tau_q,\cdot)
     +\mathbf 1_{\{q=p\}}G_i,
\end{cases}
\end{equation}
with $u_{P+1,i}^\lambda(\tau_P,\cdot)=0$.
Moreover,
\begin{equation}\label{eq:dual-gradient-model-prices}
    \partial_{\lambda_i}\Phi_0(\lambda)
    =
    \int_{\R^d}u_{1,i}^\lambda(0,x)\,\mu_0(dx)
    =
    \mathbb E^{\mathbb P^\lambda}
        [G_i(X_{\tau_p})],
    \qquad i=i(p,m).
\end{equation}

Consequently, for any $\pi\in\R^{MP}$,
a vector $\lambda^\star\in\R^{MP}$ is a dual optimizer
in~\eqref{eq:dual-attainment-supremum} if and only if
\begin{equation}\label{eq:dual-gradient-first-order}
    \nabla\Phi_0(\lambda^\star)=\pi.
\end{equation}
Equivalently, the associated law
$\mathbb P^{\lambda^\star}$ satisfies all the calibration
constraints:
\begin{equation}\label{eq:dual-gradient-exact-calibration}
    \mathbb E^{\mathbb P^{\lambda^\star}}
        [G_{i(p,m)}(X_{\tau_p})]
    =\pi_{i(p,m)},
    \qquad
    1\leq p\leq P,\quad 1\leq m\leq M.
\end{equation}
\end{prop}
\begin{proof}
By Proposition~6.1 of~\cite{HJB-property}, for each
$q\in\{1,\ldots,P\}$, the map
\[
    \lambda\longmapsto\phi_q^\lambda
    \quad\text{from }\R^{MP}\text{ to }
    C_b([\tau_{q-1},\tau_q]\times\R^d)
\]
is continuously differentiable for the supremum norm.
Here the solution map is viewed as a function of the full
vector $\lambda$, independent of the components associated
with maturities preceding $\tau_q$.

Since $\mu_0$ is a probability measure, the map
\[
    v\longmapsto\int_{\R^d}v(0,x)\,\mu_0(dx)
\]
is a bounded linear functional on
$C_b([0,\tau_1]\times\R^d)$.
Composing this functional with $\lambda\mapsto\phi_1^\lambda$
shows that $\Phi_0$ is continuously differentiable and that
\begin{equation*}
    \partial_{\lambda_i}\Phi_0(\lambda)
    =\int_{\R^d}\partial_{\lambda_i}\phi_1^\lambda(0,x)\,\mu_0(dx)
    =\int_{\R^d}u_{1,i}^\lambda(0,x)\,\mu_0(dx),
    \qquad i=1,\ldots,MP.
\end{equation*}
This proves the first equality
in~\eqref{eq:dual-gradient-model-prices}.

Fix $i=i(p,m)$. Proposition~6.1 of~\cite{HJB-property}
also identifies $u_{q,i}^\lambda$ with the solution of the
linearized equation on each time interval. By the definition of
$(\alpha^\lambda,\beta^\lambda)$, its spatial operator is
\begin{equation*}
    \mathcal L_t^\lambda v
    :=\alpha^\lambda(t,x)\cdot\Dx v
      +\tfrac12\beta^\lambda(t,x):\Dxx v.
\end{equation*}
Thus $\partial_tu_{q,i}^\lambda
+\mathcal L_t^\lambda u_{q,i}^\lambda=0$ on
$(\tau_{q-1},\tau_q)\times\R^d$. Since evaluation at a time endpoint is
continuous for the supremum norm, differentiating the
interface conditions in~\eqref{eq:hjb-cascade-duality} gives
\begin{equation*}
    u_{q,i}^\lambda(\tau_q,\cdot)
    =u_{q+1,i}^\lambda(\tau_q,\cdot)
      +\mathbf 1_{\{q=p\}}G_i,
\end{equation*}
where we used
$\partial_{\lambda_i}g_q^\lambda
=\mathbf 1_{\{q=p\}}G_i$.
This yields~\eqref{eq:dual-gradient-linearized-cascade}.
Uniqueness in the stated class follows from the uniqueness
of the linearized problem on each time interval
\cite[Corollary~3.7]{HJB-property} and backward induction.
In particular, $u_{q,i}^\lambda=0$ for $q>p$,
$u_{p,i}^\lambda(\tau_p,\cdot)=G_i$, and the traces agree
at every earlier maturity.

For each $\lambda\in\R^{MP}$, Lemma~3.1
of~\cite{HJB-property}, applied on each time interval after time
translation, gives strong existence and pathwise uniqueness.
Concatenating these solutions and applying the uniqueness
argument in the proof of
Corollary~\ref{coro:primal-uniqueness-pde} yields a strong
solution on $[0,T]$ with initial law $\mu_0$, unique in law.
That argument uses only the PDE hypotheses and does not
require $\lambda$ to be a dual optimizer.
Thus~\eqref{eq:dual-gradient-diffusion} defines a unique
law $\mathbb P^\lambda$ on $\Omega$.

Fix $i=i(p,m)$. On each time interval $[\tau_{q-1},\tau_q]$ with $q\leq p$,
$u_{q,i}^\lambda$ is a bounded solution of
\eqref{eq:dual-gradient-linearized-cascade}, classical in
the interior and continuous up to the time endpoints.
For $\tau_{q-1}<s<r<\tau_q$, the coefficients are continuous
and bounded on $[s,r]\times\R^d$, and the associated SDE
is well posed. The Feynman--Kac representation
\cite[Theorem~5.7.6]{karatzas-shreve1991}, together with
the Markov property, gives
\begin{equation*}
    u_{q,i}^\lambda(s,X_s)
    =\mathbb E^{\mathbb P^\lambda}\!\left[
        u_{q,i}^\lambda(r,X_r)\mid\mathcal F_s\right].
\end{equation*}
As $r\uparrow\tau_q$, continuity of $u_{q,i}^\lambda$
on $[\tau_{q-1},\tau_q]\times\R^d$ and of the trajectories gives
\begin{equation*}
    u_{q,i}^\lambda(r,X_r)
    \longrightarrow u_{q,i}^\lambda(\tau_q,X_{\tau_q})
    \qquad\mathbb P^\lambda\text{-almost surely}.
\end{equation*}
Since $u_{q,i}^\lambda$ is bounded, the dominated convergence
theorem yields convergence in $L^1(\mathbb P^\lambda)$.
Conditional expectation is a contraction in $L^1$, so
\begin{equation*}
    u_{q,i}^\lambda(s,X_s)
    =\mathbb E^{\mathbb P^\lambda}\!\left[
        u_{q,i}^\lambda(\tau_q,X_{\tau_q})
        \mid\mathcal F_s\right],
    \qquad \tau_{q-1}<s<\tau_q.
\end{equation*}
To reach the initial endpoint, take the conditional expectation
of this identity with respect to $\mathcal F_{\tau_{q-1}}$
and let $s\downarrow\tau_{q-1}$. The dominated convergence
theorem and the same $L^1$ contraction give
\begin{equation*}
    u_{q,i}^\lambda(\tau_{q-1},X_{\tau_{q-1}})
    =\mathbb E^{\mathbb P^\lambda}\!\left[
        u_{q,i}^\lambda(\tau_q,X_{\tau_q})
        \mid\mathcal F_{\tau_{q-1}}\right]
    \qquad\mathbb P^\lambda\text{-almost surely}.
\end{equation*}
For $q<p$, the interface condition identifies the terminal
value inside the conditional expectation with
$u_{q+1,i}^\lambda(\tau_q,X_{\tau_q})$.
At $\tau_p$, the terminal value is $G_i(X_{\tau_p})$.
Successive application of these representations and the
tower property thus yields
\begin{equation*}
    u_{1,i}^\lambda(0,X_0)
    =\mathbb E^{\mathbb P^\lambda}
        [G_i(X_{\tau_p})\mid\mathcal F_0].
\end{equation*}
Taking expectations and using the initial law $\mu_0$, we obtain
\begin{equation*}
    \int_{\R^d}u_{1,i}^\lambda(0,x)\,\mu_0(dx)
    =\mathbb E^{\mathbb P^\lambda}
        [u_{1,i}^\lambda(0,X_0)]
    =\mathbb E^{\mathbb P^\lambda}[G_i(X_{\tau_p})].
\end{equation*}
Together with the first step, this proves
\eqref{eq:dual-gradient-model-prices}.

Finally, fix $\pi\in\R^{MP}$.
By Proposition~\ref{prop:convexity-dual-value-map}, $\Phi_0$
is convex. Hence the differentiable function
$\lambda\mapsto\lambda\cdot\pi-\Phi_0(\lambda)$ is concave
on $\R^{MP}$. If $\lambda^\star$ is a dual optimizer,
the first-order necessary condition for this unconstrained
problem gives
\begin{equation*}
    0=\pi-\nabla\Phi_0(\lambda^\star).
\end{equation*}
Conversely, if $\nabla\Phi_0(\lambda^\star)=\pi$,
convexity yields, for every $\lambda\in\R^{MP}$,
\begin{equation*}
    \Phi_0(\lambda)
    \geq\Phi_0(\lambda^\star)
       +\pi\cdot(\lambda-\lambda^\star),
\end{equation*}
and therefore
$\lambda\cdot\pi-\Phi_0(\lambda)
\leq\lambda^\star\cdot\pi-\Phi_0(\lambda^\star)$.
Thus $\lambda^\star$ is a dual optimizer if and only if
\eqref{eq:dual-gradient-first-order} holds.
By~\eqref{eq:dual-gradient-model-prices}, this condition
is equivalent, component by component, to the calibration
constraints~\eqref{eq:dual-gradient-exact-calibration}.
\end{proof}

\begin{remark}[Recovery of the primal optimizer]
\label{rk:dual-gradient-primal-recovery}
Under the hypotheses of
Proposition~\ref{prop:dual-gradient-calibration},
suppose that $\nabla\Phi_0(\lambda^\star)=\pi$.
Then $\lambda^\star$ is a dual optimizer, and strong duality yields
\begin{equation*}
    \mathcal V(\pi)
    =\lambda^\star\cdot\pi-\Phi_0(\lambda^\star)<+\infty.
\end{equation*}
Corollary~\ref{coro:primal-uniqueness-pde} therefore guarantees
the existence of a unique primal optimal law.
By Proposition~\ref{prop:optimality-laws}, this law solves
\eqref{eq:dual-gradient-diffusion} with parameter $\lambda^\star$.
Uniqueness in law identifies it with $\mathbb P^{\lambda^\star}$.
Thus solving the dual first-order condition and constructing
the associated diffusion recovers the unique optimizer
of the probabilistic primal problem.

The gradient representation also provides a basis for numerical
methods: the gradient of the finite-dimensional dual objective
is $\pi-\nabla\Phi_0(\lambda)$, which can be evaluated through
the linearized pricing equations above. This enables gradient-based
optimization of the dual problem; once a dual optimizer is obtained,
the associated feedback diffusion reconstructs the primal optimal law.
\end{remark}

\subsection{Uniqueness and stability of dual optimizers}\label{subsec:dual-stability}

Throughout this subsection, we work under the hypotheses of
Proposition~\ref{prop:hjb-dual-identification}. The running cost $F$,
the initial distribution $\mu_0$, the maturity times, and the payoff
family $G$ are fixed, while the price vector $\pi$ varies in
$\mathfrak P$. Under the hypotheses introduced below, we will show
that $\mathfrak P$ has nonempty interior and that, on this interior,
the dual problem admits a unique optimizer which depends locally
Lipschitz continuously on $\pi$.
The argument relies on local strong convexity of the value map
$\Phi_0$ defined in~\eqref{eq:dual-attainment-value-map},
which we obtain from the solution-map results of the companion
PDE article~\cite{HJB-property}. We formulate its non-degeneracy
condition along the linearized directions of the cascade.

\begin{assumption}[Hamiltonian hypotheses for dual stability]
\label{ass:dual-stability-hamiltonian}
\begin{enumerate}[label=(\roman*)]
\item\label{ass:dual-stability-regularity}
On each time interval $[\tau_{p-1},\tau_p]$, after time
translation, $H$ satisfies Assumption~1 in~\cite{HJB-property}.
Its structural and regularity conditions are recalled in
Remark~\ref{rk:hjb-pde-inputs}.
In particular, $H$ is twice continuously differentiable in its
last two variables, with bounded and globally Lipschitz Hessian.
\item\label{ass:dual-stability-nondegeneracy}
For $\lambda,\eta\in\R^{MP}$, denote by
\begin{equation*}
    u_p^{\lambda,\eta}:=D_\lambda\phi_p^\lambda[\eta],
    \qquad
    Z_p^{\lambda,\eta}
    :=\left(\Dx u_p^{\lambda,\eta},\tfrac12\Dxx u_p^{\lambda,\eta}\right)
\end{equation*}
the first variation on the $p$-th interval and its spatial derivatives.
These variations are well defined under~\ref{ass:dual-stability-regularity}
by Proposition~6.1 of~\cite{HJB-property}.
For every $\lambda\in\R^{MP}$ and every $\eta\in\R^{MP}$ with $|\eta|=1$,
there exist $p\in\{1,\ldots,P\}$ and a nonempty parabolic cylinder
$Q$ relatively compact in
$(\tau_{p-1},\tau_p)\times\R^d$, such that
\begin{equation}\label{eq:dual-stability-directional-nondegeneracy}
    \Hessian H\!\left(t,x,\Dx\phi_p^\lambda,
                         \tfrac12\Dxx\phi_p^\lambda\right)
       [Z_p^{\lambda,\eta},Z_p^{\lambda,\eta}]>0
    \qquad\text{on }Q.
\end{equation}
Here $\Hessian H:=D^2_{(a,b)}H$ is the Hessian with respect to
the last two variables, viewed as a symmetric bilinear form on
$\R^d\times\mathcal S^d$.
\end{enumerate}
\end{assumption}

The identity $H=F^*$ ensures convexity in the last two variables,
while the bounded effective domain in~\eqref{ass:convex-duality-d1}
yields global Lipschitz continuity in these variables.
The quadratic form
in~\eqref{eq:dual-stability-directional-nondegeneracy} is therefore
nonnegative on every time interval.
Condition~\ref{ass:dual-stability-nondegeneracy}
requires a strictly positive contribution on at least one interval
for each parameter and each nonzero direction. It is the cascade
version of Assumption~4 in~\cite{HJB-property}.
The additional smoothness and uniform ellipticity required
in~\ref{ass:dual-stability-regularity} are structural restrictions
on the running cost through its conjugate; they do not follow
from convex conjugacy alone.

We assume
Assumption~\ref{ass:dual-stability-hamiltonian} throughout this
subsection. We also impose Assumption~\ref{hyp:G}: at each maturity,
the payoffs are bounded and uniformly continuous, and are linearly
independent modulo constants. Explicitly, for every
$p\in\{1,\ldots,P\}$ and $\eta\in\R^M$,
\begin{equation*}
    \sum_{m=1}^M \eta_m G_{i(p,m)}(\cdot)
    \text{ is constant on }\R^d
    \quad\Longrightarrow\quad \eta=0.
\end{equation*}
This is Assumption~5 of~\cite{HJB-property} for the cascade.

\begin{remark}[A sufficient condition on the Hamiltonian]
\label{rk:dual-stability-positive-hessian}
Under~\ref{ass:dual-stability-regularity} and Assumption~\ref{hyp:G},
a sufficient condition for~\ref{ass:dual-stability-nondegeneracy}
is the existence of $\Delta>0$ such that
\begin{equation*}
    \Hessian H(t,x,a,b)\succ0
    \qquad\text{whenever }|(a,b)|<\Delta,
\end{equation*}
for every $(t,x)\in(\tau_{p-1},\tau_p)\times\R^d$ and every $p$.
Together with convexity of $H$, this is Assumption~2
of~\cite{HJB-property}. Indeed, for a nonzero direction $\eta$, choose
the last maturity $p$ for which $(\eta_{i(p,m)})_{m=1}^M\neq0$.
For every $q>p$, the solution $\phi_q^\lambda$ depends only on the
multiplier blocks associated with maturities $q,\ldots,P$, which
are unchanged along the direction $\eta$. Hence
$u_q^{\lambda,\eta}=0$ on every subsequent interval.
Differentiating the terminal condition on the $p$-th interval gives
\begin{equation*}
    u_p^{\lambda,\eta}(\tau_p,\cdot)
    =\sum_{m=1}^M \eta_{i(p,m)}G_{i(p,m)}(\cdot).
\end{equation*}
This terminal datum is nonconstant by Assumption~\ref{hyp:G}.
The one-interval localization argument in Step~2 of the proof
of Theorem~5.1 in~\cite{HJB-property} then gives a cylinder
on this interval where
\eqref{eq:dual-stability-directional-nondegeneracy} holds.
The directional condition thus allows for Hamiltonians whose
Hessian is not positive definite in all directions.
\end{remark}

\begin{remark}[The log-martingale constraint]
\label{rk:dual-stability-log-martingale}
In dimension one, for a running cost independent of $(t,x)$,
the constraint $\alpha=-\beta/2$ leads to Hamiltonians of the form
$H(a,b)=h(b-a/2)$. Suppose that $h$ is
smooth and convex and that the resulting Hamiltonian satisfies
\ref{ass:dual-stability-regularity}. Its Hessian has a nontrivial
kernel, whereas, for any smooth function $u$,
\begin{equation*}
    \Hessian H(a,b)
       \left[\left(\Dx u,\tfrac12\Dxx u\right),
             \left(\Dx u,\tfrac12\Dxx u\right)\right]
    =\tfrac14 h''(b-a/2)(\Dxx u-\Dx u)^2.
\end{equation*}
Consequently, the directional condition only requires that, for
each $\lambda$ and nonzero $\eta$, the quantities
$h''(\tfrac12\Dxx\phi_p^\lambda-\tfrac12\Dx\phi_p^\lambda)$
and $(\Dxx u_p^{\lambda,\eta}-\Dx u_p^{\lambda,\eta})^2$ be strictly
positive on some cylinder of one interval.
This is the mechanism used in the log-martingale example
of~\cite[Section~5.2]{HJB-property}, with a smooth convex
regularization of the Hamiltonian and log-put payoffs
with distinct positive strikes.
For the cascade, the same argument applies on the last interval
with a nonzero payoff direction, whose first variation has the
terminal datum displayed in
Remark~\ref{rk:dual-stability-positive-hessian}.
Thus degeneracy of the Hamiltonian Hessian does not itself
exclude local volatility calibration under the log-martingale
constraint from our hypotheses.
\end{remark}

\begin{remark}[Payoff independence and dual uniqueness]
\label{rk:dual-stability-payoff-independence}
To explain the role of Assumption~\ref{hyp:G}, suppose instead
that there exist a maturity $p$, a nonzero vector
$\eta\in\R^{MP}$ supported on its multiplier block, and a constant
$c\in\R$ such that
\begin{equation*}
    \sum_{m=1}^M\eta_{i(p,m)}G_{i(p,m)}(\cdot)=c.
\end{equation*}
Replacing $\lambda$ by $\lambda+s\eta$ adds the constant $sc$
to the terminal datum on the $p$-th interval. Since the Hamiltonian
does not depend on the value of the solution, this constant is
preserved on that interval and on every preceding interval.
Consequently, for every $\lambda\in\R^{MP}$ and $s\in\R$,
\begin{equation*}
    \Phi_0(\lambda+s\eta)=\Phi_0(\lambda)+sc.
\end{equation*}
For any admissible price vector $\pi\in\mathfrak P$, calibration
implies $\pi\cdot\eta=c$. Hence
\begin{equation*}
    \pi\cdot(\lambda+s\eta)-\Phi_0(\lambda+s\eta)
    =\pi\cdot\lambda-\Phi_0(\lambda).
\end{equation*}
Thus, whenever a dual optimizer exists, every translate along
$\eta$ is also optimal. Assumption~\ref{hyp:G} excludes these
directions of invariance, which prevent uniqueness in the full
multiplier space. For admissible prices, redundant payoffs can be
removed at each maturity without changing the feasible set or the
primal optimization problem. Linear independence modulo constants
therefore amounts to choosing a nonredundant representation of the
calibration constraints.
\end{remark}

Recall that $\Phi_0(\lambda)$ is the initial value of the HJB
cascade, averaged against the initial law:
\begin{equation*}
    \Phi_0(\lambda)
    =\int_{\R^d}\phi_1^\lambda(0,x)\,\mu_0(dx).
\end{equation*}
By~\eqref{eq:dual-attainment-supremum}, the dual value is its
convex conjugate $\Phi_0^*(\pi)=\mathcal V(\pi)$.
To establish uniqueness and stability of the maximizer in this
representation, we use the convexity results for the HJB solution
map from~\cite{HJB-property} to prove local strong convexity of
$\Phi_0$.

For $n\geq1$, recall that a function $f:\R^n\to\R$ is locally strongly convex
if, for every nonempty compact convex set $K\subset\R^n$,
there exists $m_K>0$ such that, for all $\lambda_1,\lambda_2\in K$
and $\theta\in[0,1]$,
\begin{equation}\label{eq:dual-value-local-strong-convexity}
    f((1-\theta)\lambda_1+\theta\lambda_2)
    \leq(1-\theta)f(\lambda_1)+\theta f(\lambda_2)-\frac{m_K}{2}\theta(1-\theta)|\lambda_2-\lambda_1|^2.
\end{equation}

\begin{theorem}[Local strong convexity of the dual value map]
\label{prop:dual-value-local-strong-convexity}
Under the standing hypotheses of this subsection, the map
$\Phi_0$ defined in~\eqref{eq:dual-attainment-value-map} is locally
strongly convex on $\R^{MP}$.
In particular, it is strictly convex.
\end{theorem}

\begin{proof}
Fix a nonempty compact convex set $K\subset\R^{MP}$ and a closed
ball $B\subset\R^{MP}$ of positive radius containing $K$.
We use Proposition~6.3 of~\cite{HJB-property} in its formulation
under the directional non-degeneracy condition
\ref{ass:dual-stability-nondegeneracy}.
For every $x\in\R^d$, this result gives local strong convexity of
$\lambda\mapsto\phi_1^\lambda(0,x)$.
Consequently, there exists $c_B(x)>0$ such that, for all
$\lambda_1,\lambda_2\in B$ and $\theta\in[0,1]$,
\begin{equation}\label{eq:dual-value-pointwise-strong-convexity}
\begin{split}
    \phi_1^{(1-\theta)\lambda_1+\theta\lambda_2}(0,x)
    &\leq(1-\theta)\phi_1^{\lambda_1}(0,x)
          +\theta\phi_1^{\lambda_2}(0,x)\\
    &\quad-\frac{c_B(x)}{2}\theta(1-\theta)
                |\lambda_2-\lambda_1|^2.
\end{split}
\end{equation}
If $\mu_0=\delta_{x_0}$, then
$\Phi_0(\lambda)=\phi_1^\lambda(0,x_0)$, so this inequality
proves strong convexity on $K$ with modulus $c_B(x_0)$.
For a general initial law, we will choose a bounded measurable
positive modulus and integrate the pointwise inequality against
$\mu_0$.

Set
\begin{equation*}
    \mathcal T_B
    :=\{(\lambda_1,\lambda_2,\theta)\in B\times B\times(0,1):
           \lambda_1\neq\lambda_2\}.
\end{equation*}
For $x\in\R^d$ and $(\lambda_1,\lambda_2,\theta)\in\mathcal T_B$,
define the normalized convexity defect
\begin{equation*}
    R_B(x;\lambda_1,\lambda_2,\theta)
    :=\frac{2\bigl[(1-\theta)\phi_1^{\lambda_1}(0,x)
                  +\theta\phi_1^{\lambda_2}(0,x)
                  -\phi_1^{(1-\theta)\lambda_1+\theta\lambda_2}(0,x)
             \bigr]}
            {\theta(1-\theta)|\lambda_2-\lambda_1|^2},
\end{equation*}
and put
\begin{equation}\label{eq:dual-value-measurable-modulus}
    m_B(x):=\min\left\{1,
        \inf_{(\lambda_1,\lambda_2,\theta)\in\mathcal T_B}
        R_B(x;\lambda_1,\lambda_2,\theta)\right\}.
\end{equation}
The set $\mathcal T_B$ is nonempty since $B$ has positive radius.
By~\eqref{eq:dual-value-pointwise-strong-convexity},
\begin{equation*}
    0<\min\{1,c_B(x)\}\leq m_B(x)\leq1
    \qquad\text{for every }x\in\R^d.
\end{equation*}

For each fixed $x$, the map $\lambda\mapsto\phi_1^\lambda(0,x)$
is continuous, so $R_B(x;\cdot)$ is continuous on $\mathcal T_B$.
Choose a countable dense subset $\mathcal D_B\subset\mathcal T_B$.
The infimum in~\eqref{eq:dual-value-measurable-modulus} is therefore
unchanged when restricted to $\mathcal D_B$.
For each fixed triple in $\mathcal D_B$, the function
$x\mapsto R_B(x;\lambda_1,\lambda_2,\theta)$ is continuous,
since the initial traces of the HJB solutions are continuous.
Thus $m_B$ is Borel measurable, being the minimum of $1$ and a
countable infimum of continuous functions.
By its definition, inequality~\eqref{eq:dual-value-pointwise-strong-convexity}
holds with $m_B(x)$ in place of $c_B(x)$ for every
$\lambda_1,\lambda_2\in B$ and $\theta\in[0,1]$; the cases
$\lambda_1=\lambda_2$ or $\theta\in\{0,1\}$ are equalities.

Define
\begin{equation*}
    m_K:=\int_{\R^d}m_B(x)\,\mu_0(dx).
\end{equation*}
Since $m_B$ is Borel measurable, strictly positive, and bounded
by $1$, and $\mu_0$ is a probability measure, we have
$0<m_K\leq1$.
For fixed $\lambda_1,\lambda_2\in K$ and $\theta\in[0,1]$,
the three initial traces in
\eqref{eq:dual-value-pointwise-strong-convexity} are bounded
and hence integrable against $\mu_0$.
Integrating that inequality with $m_B(x)$ in place of $c_B(x)$
and using~\eqref{eq:dual-attainment-value-map} therefore yields
\begin{equation}\label{eq:dual-value-integrated-strong-convexity}
\begin{split}
    \Phi_0((1-\theta)\lambda_1+\theta\lambda_2)
    &\leq(1-\theta)\Phi_0(\lambda_1)+\theta\Phi_0(\lambda_2)\\
    &\quad-\frac{m_K}{2}\theta(1-\theta)|\lambda_2-\lambda_1|^2.
\end{split}
\end{equation}
Thus $\Phi_0$ is strongly convex on $K$, with a modulus
independent of $\lambda_1$, $\lambda_2$, and $\theta$.
Since $K$ was arbitrary, $\Phi_0$ is locally strongly convex on
$\R^{MP}$. Finally, any two distinct points of $\R^{MP}$ belong
to a common compact convex set. Applying the strong convexity
inequality on that set with $\theta\in(0,1)$ yields a strict
convexity inequality, proving that $\Phi_0$ is strictly convex.
\end{proof}

We now apply convex conjugacy to relate these properties of $\Phi_0$
to the geometry of the admissible price set and the regularity of
the value function.

\begin{corol}[Interior of the admissible price set and differentiability of the value]
\label{coro:dual-conjugate-regularity}
Under the standing hypotheses of this subsection, it follows that
the admissible price set $\mathfrak P$ has nonempty interior
and that its relative interior coincides with its interior:
\begin{equation}\label{eq:dual-price-interior}
    \operatorname{int}(\mathfrak P)\neq\varnothing,
    \qquad
    \operatorname{ri}(\mathfrak P)=\operatorname{int}(\mathfrak P).
\end{equation}
Moreover, the convex conjugate $\Phi_0^*$ belongs to
$C^1(\operatorname{int}(\mathfrak P))$.
Since $\mathcal V=\Phi_0^*$, the same regularity holds for
the primal value function $\mathcal V$ on this set.
\end{corol}

\begin{proof}
By Proposition~\ref{prop:convexity-dual-value-map}, $\Phi_0$ is
finite on $\R^{MP}$, and
Theorem~\ref{prop:dual-value-local-strong-convexity} gives its
strict convexity. The first part of
Lemma~\ref{lem:local-lipschitz-conjugate}, applied with $f=\Phi_0$,
therefore yields
\begin{equation*}
    \operatorname{int}(\operatorname{dom}\Phi_0^*)\neq\varnothing,
    \qquad
    \Phi_0^*\in C^1(\operatorname{int}(\operatorname{dom}\Phi_0^*)).
\end{equation*}
Lemma~\ref{lem:dual-attainment-domain} identifies
$\operatorname{dom}\Phi_0^*$ with $\mathfrak P$.
Since this convex set has nonempty interior in $\R^{MP}$,
its affine hull is $\R^{MP}$, so its relative interior coincides
with its interior. Finally,
\eqref{eq:dual-attainment-supremum} gives $\mathcal V=\Phi_0^*$.
\end{proof}

\begin{prop}[Uniqueness and local Lipschitz stability of dual optimizers]
\label{prop:main-stability-result}
Under the standing hypotheses of this subsection, for every
$\pi\in\operatorname{int}(\mathfrak P)$, the HJB dual
problem~\eqref{eq:dual-attainment-supremum} admits a unique
optimizer $\lambda^\star(\pi)\in\R^{MP}$, characterized by
\begin{equation}\label{eq:dual-optimizer-value-gradient}
    \lambda^\star(\pi)
    =\nabla\Phi_0^*(\pi)
    =\nabla\mathcal V(\pi).
\end{equation}
Moreover, the map
\begin{equation*}
    \lambda^\star:\operatorname{int}(\mathfrak P)\longrightarrow\R^{MP}
\end{equation*}
is locally Lipschitz continuous.
\end{prop}

\begin{proof}
Fix $\pi\in\operatorname{int}(\mathfrak P)$.
By Corollary~\ref{coro:dual-conjugate-regularity},
$\operatorname{int}(\mathfrak P)=\operatorname{ri}(\mathfrak P)$.
Proposition~\ref{prop:dual-attainement} therefore ensures that
the HJB dual problem admits an optimizer.
By Theorem~\ref{prop:dual-value-local-strong-convexity}, $\Phi_0$
is strictly convex on $\R^{MP}$. Consequently, the objective
\begin{equation*}
    \lambda\longmapsto\pi\cdot\lambda-\Phi_0(\lambda)
\end{equation*}
is strictly concave and has at most one maximizer.
The optimizer is therefore unique; we denote it by
$\lambda^\star(\pi)$.

By~\eqref{eq:dual-attainment-subdifferential},
$\lambda^\star(\pi)\in\partial\Phi_0^*(\pi)$.
Corollary~\ref{coro:dual-conjugate-regularity} gives the
differentiability of $\Phi_0^*$ on $\operatorname{int}(\mathfrak P)$,
so its subdifferential at $\pi$ is the singleton
$\{\nabla\Phi_0^*(\pi)\}$.
Together with $\mathcal V=\Phi_0^*$, this yields
\begin{equation*}
    \lambda^\star(\pi)=\nabla\Phi_0^*(\pi)=\nabla\mathcal V(\pi),
\end{equation*}
which proves~\eqref{eq:dual-optimizer-value-gradient}.

Finally, $\Phi_0$ is finite on $\R^{MP}$ and locally strongly
convex by Theorem~\ref{prop:dual-value-local-strong-convexity}.
The second part of Lemma~\ref{lem:local-lipschitz-conjugate},
applied with $f=\Phi_0$, therefore implies that $\nabla\Phi_0^*$
is locally Lipschitz continuous on
$\operatorname{int}(\operatorname{dom}\Phi_0^*)$.
By Lemma~\ref{lem:dual-attainment-domain}, this set is
$\operatorname{int}(\mathfrak P)$.
The identification~\eqref{eq:dual-optimizer-value-gradient}
then gives the claimed local Lipschitz continuity of
$\pi\mapsto\lambda^\star(\pi)$.
\end{proof}

\begin{remark}[Sensitivity of the primal value to prices]
\label{rk:primal-value-sensitivity}
Proposition~\ref{prop:main-stability-result} shows that the primal
value function $\mathcal V$ is continuously differentiable on
$\operatorname{int}(\mathfrak P)$, with locally Lipschitz gradient
$\nabla\mathcal V=\lambda^\star$.
For every $\pi_0\in\operatorname{int}(\mathfrak P)$, choose an open
ball $U$ containing $\pi_0$ and contained in
$\operatorname{int}(\mathfrak P)$ on which $\lambda^\star$ is
Lipschitz with constant $L_U$. For all $\pi,\pi+h\in U$,
\begin{equation*}
    0\leq\mathcal V(\pi+h)-\mathcal V(\pi)
          -\lambda^\star(\pi)\cdot h
    \leq\frac{L_U}{2}|h|^2.
\end{equation*}
The lower bound follows from convexity of $\mathcal V$.
For the upper bound, the segment from $\pi$ to $\pi+h$ lies in $U$,
and the fundamental theorem of calculus gives
\begin{equation*}
\begin{split}
    &\mathcal V(\pi+h)-\mathcal V(\pi)
       -\lambda^\star(\pi)\cdot h\\
    &\qquad=\int_0^1
       \bigl(\lambda^\star(\pi+th)-\lambda^\star(\pi)\bigr)\cdot h\,dt
       \leq\int_0^1 L_U t|h|^2\,dt
       =\frac{L_U}{2}|h|^2.
\end{split}
\end{equation*}
Thus the dual optimizer gives the first-order sensitivity of the
minimal calibration cost to changes in the quoted prices, with a
quadratic remainder.
\end{remark}

We next study how the optimal calibrated law varies with the price
vector. We will combine the local Lipschitz stability of the dual
optimizer with stability estimates for the HJB feedback coefficients
and the associated SDEs to establish continuity of the optimal laws
in the adapted Wasserstein distance.

\subsection{Stability of optimal laws}\label{subsec:primal-stability}

Throughout this subsection, we work under the hypotheses of
Proposition~\ref{prop:hjb-dual-identification} and
Assumption~\ref{ass:dual-stability-hamiltonian}.
The running cost $F$, the Hamiltonian $H=F^*$, the payoff family
$G$, the maturity times, and the initial law $\mu_0$ are fixed,
while $\pi$ varies in $\operatorname{int}(\mathfrak P)$.
This set is nonempty by Corollary~\ref{coro:dual-conjugate-regularity}.
For each such $\pi$, Proposition~\ref{prop:main-stability-result}
provides a unique dual optimizer $\lambda^\star(\pi)$, which depends
locally Lipschitz continuously on $\pi$.
Corollary~\ref{coro:primal-uniqueness-pde} also gives a unique
optimal law for the probabilistic primal problem, denoted by
$\mathbb P^{\star,\pi}$.
We study the continuity of $\pi\mapsto\mathbb P^{\star,\pi}$
in the adapted Wasserstein distance of order two.

Here $\mathbb P^{\star,\pi}$ is the optimizer over the full
probabilistic admissible class, whose characteristics may
a priori depend on the entire past trajectory.
Proposition~\ref{prop:optimality-laws} forces the characteristics
of every optimizer to coincide with the feedback coefficients
in~\eqref{eq:primal-uniqueness-feedback}, evaluated at
$\lambda^\star(\pi)$. Uniqueness in law of the associated SDE
then identifies its solution law with $\mathbb P^{\star,\pi}$.
Thus the feedback representation follows from optimality and
allows us to establish stability of the full optimal
trajectory law through stability estimates for these SDEs.

For $\lambda\in\R^{MP}$, let $(\phi_p^\lambda)_{p=1}^P$ be the
HJB cascade defined in~\eqref{eq:hjb-cascade-duality}. On each
$(\tau_{p-1},\tau_p)\times\R^d$, set
\begin{equation}\label{eq:primal-stability-feedback}
\begin{split}
    (\alpha^\lambda,\beta^\lambda)(t,x)
    &:=\nabla_{(a,b)}H\!\left(
        t,x,\Dx\phi_p^\lambda(t,x),
        \tfrac12\Dxx\phi_p^\lambda(t,x)\right),\\
    \sigma^\lambda(t,x)&:=\bigl(\beta^\lambda(t,x)\bigr)^{1/2}.
\end{split}
\end{equation}
Here $\sigma^\lambda$ denotes the symmetric positive definite
square root of the covariance matrix $\beta^\lambda$;
positive definiteness follows from the uniform ellipticity
in Assumption~\ref{ass:dual-stability-regularity}.
At the finitely many times $\tau_0,\ldots,\tau_P$, set
$\alpha^\lambda=0$ and $\beta^\lambda=\sigma^\lambda=I_d$.
These values are chosen only to define the coefficients at these
finitely many times. They need not belong to the effective domain
of $F$ there. Since these times form a set of zero Lebesgue measure,
the choices affect neither the associated SDEs nor the
time-integrated cost.
For $\pi\in\operatorname{int}(\mathfrak P)$, the coefficients
$(\alpha^{\lambda^\star(\pi)},\beta^{\lambda^\star(\pi)})$
are the feedback characteristics of the optimal law
$\mathbb P^{\star,\pi}$.

We retain the deterministic initial condition
$\mu_0=\delta_{x_0}$, with the same $x_0\in\R^d$ for all prices.
By~\eqref{ass:convex-duality-d1} and the preceding identification,
the drift and covariance characteristics of the optimal laws
are uniformly bounded, independently of $\pi$.
In particular, $\mathbb P^{\star,\pi}\in\mathcal{SM}_2$ for every
$\pi\in\operatorname{int}(\mathfrak P)$.

We use the following notions of causality and adapted Wasserstein
distance, following Definitions~1.2--1.3
of~\cite{backhoff-adaptedwas}.

\begin{definition}[Bicausal couplings]\label{def:bicausal-couplings}
Let $\mathbb P,\mathbb Q$ be probability measures on $\Omega$.
On $\Omega\times\Omega$, denote the canonical processes by
$X$ and $Y$, and their raw natural filtrations by
$(\mathcal F_t^X)$ and $(\mathcal F_t^Y)$.
A coupling $\Gamma$ of $\mathbb P$ and $\mathbb Q$ is
\emph{causal from $X$ to $Y$} if, for every $t\in[0,T]$
and every bounded $\mathcal F_t^Y$-measurable random variable $Z$,
$\mathbb E_\Gamma[Z\mid\mathcal F_T^X]$ admits an
$\mathcal F_t^X$-measurable version.
It is \emph{bicausal} if the same property holds with $X$ and $Y$
interchanged. We denote the set of such couplings by
$\Pi_{\mathrm{BC}}(\mathbb P,\mathbb Q)$.
\end{definition}

For a bicausal coupling, the marginal semimartingale decompositions
remain valid in the joint filtration; see
\cite[Lemma~3.1]{backhoff-adaptedwas}, applied coordinatewise.

\begin{definition}[Adapted Wasserstein distance]\label{def:adapted-wasserstein}
Let $\mathbb P,\mathbb Q\in\mathcal{SM}_2$ have the same
deterministic initial point $x_0$.
Write $X=x_0+M^X+A^X$ and $Y=x_0+M^Y+A^Y$ for their
semimartingale decompositions, lifted to $\Omega\times\Omega$.
Their adapted Wasserstein distance of order two is defined by
\begin{equation}\label{eq:adapted-wasserstein-distance}
    \mathcal{AW}_2(\mathbb P,\mathbb Q)^2
    :=\inf_{\Gamma\in\Pi_{\mathrm{BC}}(\mathbb P,\mathbb Q)}
    \mathbb E_\Gamma\!\left[
        \operatorname{Tr}[M^X-M^Y]_T
        +\operatorname{Var}_{[0,T]}(A^X-A^Y)^2
    \right].
\end{equation}
Here $[M^X-M^Y]$ is the matrix quadratic variation,
and $\operatorname{Var}_{[0,T]}(A^X-A^Y)$ denotes the total
variation of $A^X-A^Y$ on $[0,T]$, defined pathwise
by~\eqref{eq:notation-path-total-variation}.
\end{definition}

To compare the optimal laws, we use estimates for the feedback
coefficients in~\eqref{eq:primal-stability-feedback}.
The interior estimates of the companion PDE article~\cite{HJB-property},
applied successively to the cascade, yield Lipschitz dependence
on $\lambda$ and spatial Lipschitz bounds on each truncated interval
$[\tau_{p-1},\tau_p-\varepsilon]$.
These estimates are uniform for $\lambda$ in a fixed compact set,
with constants that may depend on $\varepsilon>0$.
Their precise formulation and proof are given in
Lemma~\ref{lem:feedback-coefficient-estimates} in the appendix.

\begin{theorem}[Continuity of optimal calibrated laws]
\label{thm:stability-primal}
Under the standing hypotheses of this subsection, with
$\mu_0=\delta_{x_0}$, let $\mathbb P^{\star,\pi}$ denote the
unique optimizer of the probabilistic primal problem over
$\mathcal P(\mu_0,\tau,\pi)$ for
$\pi\in\operatorname{int}(\mathfrak P)$.
The map
\begin{equation*}
    \operatorname{int}(\mathfrak P)\longrightarrow\mathcal{SM}_2,
    \qquad \pi\longmapsto\mathbb P^{\star,\pi},
\end{equation*}
is continuous for the Euclidean topology on the price space
and the adapted Wasserstein distance $\mathcal{AW}_2$ on the
space of laws.
Equivalently, for every sequence
$(\pi_n)_{n\geq1}\subset\operatorname{int}(\mathfrak P)$
converging to $\pi\in\operatorname{int}(\mathfrak P)$,
\begin{equation}\label{eq:primal-law-adapted-stability}
    \mathcal{AW}_2\!\left(
        \mathbb P^{\star,\pi_n},\mathbb P^{\star,\pi}
    \right)\longrightarrow0.
\end{equation}
\end{theorem}

\begin{proof}
We compare the optimal laws by constructing solutions of their
associated SDEs with the same initial condition and the same Brownian
motion. We first show that their joint law is a bicausal coupling.
We then prove convergence of the solutions in the expected squared
uniform norm. On each interval between consecutive maturities, we use
the Lipschitz estimates for the feedback coefficients away from the
right endpoint and their uniform boundedness near that endpoint.
Combining these estimates and proceeding successively through the
maturity times gives convergence on the whole time interval.
Finally, we express the adapted Wasserstein cost of the coupling
in terms of the drift and diffusion differences and show that
this cost tends to zero.

Fix a sequence $(\pi_n)_{n\geq1}\subset\operatorname{int}(\mathfrak P)$
converging to $\pi\in\operatorname{int}(\mathfrak P)$, and set
\begin{equation*}
    \lambda_n:=\lambda^\star(\pi_n),
    \qquad \lambda:=\lambda^\star(\pi).
\end{equation*}
Proposition~\ref{prop:main-stability-result} gives
$\lambda_n\to\lambda$. Hence all these multipliers belong to
a common compact set $K\subset\R^{MP}$, to which
Lemma~\ref{lem:feedback-coefficient-estimates} applies.

\medskip
We first construct solutions of the SDEs associated with
$\lambda_n$ and $\lambda$, driven by the same Brownian motion,
and identify their laws with the optimal calibrated laws.
Fix a probability space
$(\overline\Omega,\overline{\mathcal F},\overline{\mathbb P})$
carrying a $d$-dimensional Brownian motion $W$, with its
completed natural filtration.
On each time interval, Lemma~3.1 of~\cite{HJB-property}, applied after
time translation, gives strong existence and pathwise uniqueness
for the SDE associated with the corresponding HJB solution.
We concatenate these strong solutions across the maturity times,
using the terminal value on each time interval as the initial value on the
next one. The construction applies to these random initial values,
which are measurable with respect to the Brownian past and
independent of the subsequent Brownian increments.
Pathwise uniqueness across the maturities follows as in the
proof of Corollary~\ref{coro:primal-uniqueness-pde}.
We thus obtain, on this same probability space and driven by
the same $W$, strong solutions $X^n$ and $X$ on $[0,T]$ of
\begin{equation}\label{eq:primal-stability-synchronous-sdes}
    \begin{aligned}
        dX_t^n&=\alpha^{\lambda_n}(t,X_t^n)\,dt
                    +\sigma^{\lambda_n}(t,X_t^n)\,dW_t,
        &X_0^n&=x_0,\\
        dX_t&=\alpha^\lambda(t,X_t)\,dt
                    +\sigma^\lambda(t,X_t)\,dW_t,
        &X_0&=x_0.
    \end{aligned}
\end{equation}

By Proposition~\ref{prop:optimality-laws}, every probabilistic
primal optimizer at price $\pi_n$, respectively $\pi$, has the
feedback characteristics associated with $\lambda_n$,
respectively $\lambda$.
The uniqueness in law established in
Corollary~\ref{coro:primal-uniqueness-pde} therefore identifies
the laws of the constructed solutions:
\begin{equation}\label{eq:primal-stability-synchronous-laws}
    \operatorname{Law}_{\overline{\mathbb P}}(X^n)
        =\mathbb P^{\star,\pi_n},
    \qquad
    \operatorname{Law}_{\overline{\mathbb P}}(X)
        =\mathbb P^{\star,\pi}.
\end{equation}

\medskip
We now show that the joint law of these solutions is
bicausal in the sense of Definition~\ref{def:bicausal-couplings},
so that it is an admissible coupling in the definition
of the adapted Wasserstein distance.
Define
\begin{equation*}
    \Gamma_n:=\operatorname{Law}_{\overline{\mathbb P}}(X^n,X).
\end{equation*}
By~\eqref{eq:primal-stability-synchronous-laws}, this is a
coupling of $\mathbb P^{\star,\pi_n}$ and $\mathbb P^{\star,\pi}$.
For a process $Y$ on the common probability space, denote by
$\overline{\mathcal F}_t^Y$ the
$\overline{\mathbb P}$-completion of
$\sigma(Y_s:0\leq s\leq t)$.
Fix $t\in[0,T]$. Strong existence and pathwise uniqueness
allow the equation for $X^n$ to be restarted at $t$: its future
path is a measurable function of $X_t^n$ and
$(W_{t+s}-W_t)_{0\leq s\leq T-t}$.
These Brownian increments are independent of
$\overline{\mathcal F}_t^W$, which contains
$\overline{\mathcal F}_t^{X^n}\vee\overline{\mathcal F}_t^X$.
Consequently, conditionally on $\overline{\mathcal F}_t^{X^n}$,
the future of $X^n$ is independent of
$\overline{\mathcal F}_t^X$. In particular, for every bounded
$\overline{\mathcal F}_t^X$-measurable random variable $Z$,
\begin{equation*}
    \overline{\mathbb E}
        [Z\mid\overline{\mathcal F}_T^{X^n}]
    =
    \overline{\mathbb E}
        [Z\mid\overline{\mathcal F}_t^{X^n}],
\end{equation*}
where $\overline{\mathbb E}$ denotes expectation under
$\overline{\mathbb P}$.
The same argument applies with $X^n$ and $X$ interchanged.
Under the joint law $\Gamma_n$, the preceding identity becomes,
for every bounded Borel function $f:C([0,t];\R^d)\to\R$,
\begin{equation*}
    \mathbb E_{\Gamma_n}
        [f(X_{\vert[0,t]})\mid\mathcal F_T^{X^n}]
    =
    \mathbb E_{\Gamma_n}
        [f(X_{\vert[0,t]})\mid\mathcal F_t^{X^n}].
\end{equation*}
Here we use the raw natural filtrations of $X^n$ and $X$; completing
these sigma-fields does not change conditional expectations
up to almost-sure equality.
Every bounded $\mathcal F_t^X$-measurable random variable
can be written as such a function of $X_{\vert[0,t]}$.
Thus $\Gamma_n$ is causal from $X^n$ to $X$ in the sense of
Definition~\ref{def:bicausal-couplings}.
Interchanging the processes gives the reverse causality,
and therefore
\begin{equation}\label{eq:primal-stability-bicausal-coupling}
    \Gamma_n\in\Pi_{\mathrm{BC}}
        \bigl(\mathbb P^{\star,\pi_n},\mathbb P^{\star,\pi}\bigr).
\end{equation}
For a characterization of bicausal couplings of scalar SDEs
through strong solutions driven by Brownian motions in a common
filtration, see \cite[Proposition~2.2]{backhoff-kallblad-robinson2025}.

\medskip
Set $D_t^n:=X_t^n-X_t$ and fix
$\varepsilon\in(0,\Delta_\tau)$, where
$\Delta_\tau=\min_{1\leq p\leq P}(\tau_p-\tau_{p-1})$.
To compare the trajectories, we first work, for each $p$,
on $[\tau_{p-1},\tau_p-\varepsilon]$, where
Lemma~\ref{lem:feedback-coefficient-estimates} provides uniform
Lipschitz bounds in the state variable and the multiplier.
We estimate the difference between the solutions in terms of
their difference at $\tau_{p-1}$ and of $|\lambda_n-\lambda|$.
Denote their coefficient differences by
\begin{equation*}
    \delta\alpha_t^n
    :=\alpha^{\lambda_n}(t,X_t^n)-\alpha^\lambda(t,X_t),
    \qquad
    \delta\sigma_t^n
    :=\sigma^{\lambda_n}(t,X_t^n)-\sigma^\lambda(t,X_t).
\end{equation*}
Splitting the drift difference as
\begin{align*}
    \delta\alpha_t^n
    &=
    \alpha^{\lambda_n}(t,X_t^n)-\alpha^\lambda(t,X_t^n)
    +\alpha^\lambda(t,X_t^n)-\alpha^\lambda(t,X_t),
\end{align*}
and using the analogous decomposition for $\delta\sigma_t^n$,
Lemma~\ref{lem:feedback-coefficient-estimates} gives
\begin{equation}\label{eq:primal-stability-coefficient-differences}
    |\delta\alpha_t^n|^2+\|\delta\sigma_t^n\|^2
    \leq C_{K,\varepsilon}
          \bigl(|\lambda_n-\lambda|^2+|D_t^n|^2\bigr),
    \qquad t\in[\tau_{p-1},\tau_p-\varepsilon].
\end{equation}
Here and below, $C_{K,\varepsilon}$ denotes a finite constant
independent of $n$ and $p$, which may increase from one
estimate to the next.

Subtracting the two equations in
\eqref{eq:primal-stability-synchronous-sdes}, and applying
Cauchy--Schwarz to the drift integral and the
Burkholder--Davis--Gundy inequality to the martingale integral,
we obtain, for $t\in[\tau_{p-1},\tau_p-\varepsilon]$,
\begin{align*}
    \overline{\mathbb E}
       \left[\sup_{\tau_{p-1}\leq s\leq t}|D_s^n|^2\right]
    &\leq 3\overline{\mathbb E}|D_{\tau_{p-1}}^n|^2
       +C\overline{\mathbb E}\int_{\tau_{p-1}}^t
           \bigl(|\delta\alpha_s^n|^2
                    +\|\delta\sigma_s^n\|^2\bigr)\,ds\\
    &\leq 3\overline{\mathbb E}|D_{\tau_{p-1}}^n|^2
       +C_{K,\varepsilon}\int_{\tau_{p-1}}^t
          \left(
             |\lambda_n-\lambda|^2
             +\overline{\mathbb E}
                \left[\sup_{\tau_{p-1}\leq r\leq s}|D_r^n|^2\right]
          \right)\,ds,
\end{align*}
where $C$ depends only on $T$ and the martingale inequality.
The expectations are finite by boundedness of the coefficients
and the common deterministic initial condition.
Gr\"onwall's inequality therefore yields
\begin{equation}\label{eq:primal-stability-truncated-estimate}
    \overline{\mathbb E}
       \left[\sup_{\tau_{p-1}\leq t\leq\tau_p-\varepsilon}
                    |D_t^n|^2\right]
    \leq C_{K,\varepsilon}
          \left(
             \overline{\mathbb E}|D_{\tau_{p-1}}^n|^2
             +|\lambda_n-\lambda|^2
          \right).
\end{equation}
The constant in this estimate may diverge as
$\varepsilon\downarrow0$.

\medskip
To extend this comparison to $\tau_p$, we control the
increments of both solutions on $[\tau_p-\varepsilon,\tau_p]$
using only the uniform coefficient bound~\eqref{eq:feedback-uniform-bounds}.
This gives an error that tends to zero with $\varepsilon$,
uniformly in $n$.
Cauchy--Schwarz and the Burkholder--Davis--Gundy inequality give
\begin{align*}
    &\overline{\mathbb E}
       \left[\sup_{\tau_p-\varepsilon\leq t\leq\tau_p}
          |X_t^n-X_{\tau_p-\varepsilon}^n|^2\right]\\
    &\quad\leq
       2\varepsilon\,\overline{\mathbb E}
          \int_{\tau_p-\varepsilon}^{\tau_p}
             |\alpha^{\lambda_n}(s,X_s^n)|^2\,ds
       +C\overline{\mathbb E}
          \int_{\tau_p-\varepsilon}^{\tau_p}
             \|\sigma^{\lambda_n}(s,X_s^n)\|^2\,ds\\
    &\quad\leq C(\varepsilon+\varepsilon^2).
\end{align*}
The same bound holds for $X$.
Here $C$ depends only on the uniform coefficient bound and
the martingale inequality, and is independent of $n$, $p$,
and $\varepsilon$.
For $t\in[\tau_p-\varepsilon,\tau_p]$, write
\begin{equation*}
    D_t^n
    =D_{\tau_p-\varepsilon}^n
     +(X_t^n-X_{\tau_p-\varepsilon}^n)
     -(X_t-X_{\tau_p-\varepsilon}).
\end{equation*}
Using $|u+v+w|^2\leq3(|u|^2+|v|^2+|w|^2)$ and
the preceding increment bounds, we obtain
\begin{equation}\label{eq:primal-stability-full-slab-estimate}
    \overline{\mathbb E}
       \left[\sup_{\tau_{p-1}\leq t\leq\tau_p}|D_t^n|^2\right]
    \leq
       3\overline{\mathbb E}
          \left[\sup_{\tau_{p-1}\leq t\leq\tau_p-\varepsilon}
                     |D_t^n|^2\right]
       +C(\varepsilon+\varepsilon^2).
\end{equation}

\medskip
We combine the preceding estimates to prove convergence
on each interval $[\tau_{p-1},\tau_p]$ by induction on $p$.
At each stage, we first choose $\varepsilon$ to control the increments
near $\tau_p$, and then take $n$ large with this $\varepsilon$ fixed.
Suppose that
$\overline{\mathbb E}|D_{\tau_{p-1}}^n|^2\to0$.
This holds for $p=1$, since $D_0^n=0$.
Let $\eta>0$. Since the constant $C$ in
\eqref{eq:primal-stability-full-slab-estimate} is independent
of $\varepsilon$, choose $\varepsilon\in(0,\Delta_\tau)$ such that
\begin{equation*}
    C(\varepsilon+\varepsilon^2)<\frac{\eta}{2}.
\end{equation*}
For this fixed $\varepsilon$, the induction hypothesis,
$\lambda_n\to\lambda$, and
\eqref{eq:primal-stability-truncated-estimate} give an integer
$N\geq1$ such that, for every $n\geq N$,
\begin{equation*}
    \overline{\mathbb E}
       \left[\sup_{\tau_{p-1}\leq t\leq\tau_p-\varepsilon}
                    |D_t^n|^2\right]<\frac{\eta}{6}.
\end{equation*}
Consequently, \eqref{eq:primal-stability-full-slab-estimate}
implies, for every $n\geq N$,
\begin{equation*}
    \overline{\mathbb E}
       \left[\sup_{\tau_{p-1}\leq t\leq\tau_p}|D_t^n|^2\right]
    <3\frac{\eta}{6}+\frac{\eta}{2}=\eta.
\end{equation*}
Since $\eta>0$ was arbitrary, this proves
\begin{equation}\label{eq:primal-stability-slab-convergence}
    \overline{\mathbb E}
       \left[\sup_{\tau_{p-1}\leq t\leq\tau_p}|D_t^n|^2\right]
    \xrightarrow[n\to\infty]{}0.
\end{equation}
In particular,
$\overline{\mathbb E}|D_{\tau_p}^n|^2\to0$, which provides
the initial convergence for the next time interval and completes
the induction.
Finally, let $\eta>0$. For each $p$, choose an integer
$N_p\geq1$ such that, for every $n\geq N_p$,
\begin{equation*}
    \overline{\mathbb E}
       \left[\sup_{\tau_{p-1}\leq t\leq\tau_p}|D_t^n|^2\right]
    <\frac{\eta}{P}.
\end{equation*}
Taking $N=\max_{1\leq p\leq P}N_p$, we obtain, for every $n\geq N$,
\begin{equation*}
    \overline{\mathbb E}
       \left[\sup_{0\leq t\leq T}|X_t^n-X_t|^2\right]
    \leq\sum_{p=1}^P\overline{\mathbb E}
       \left[\sup_{\tau_{p-1}\leq t\leq\tau_p}|D_t^n|^2\right]
    <\eta.
\end{equation*}
Thus
\begin{equation}\label{eq:primal-stability-path-convergence}
    \overline{\mathbb E}
       \left[\sup_{0\leq t\leq T}|X_t^n-X_t|^2\right]
    \longrightarrow0.
\end{equation}

\medskip
We next estimate the adapted Wasserstein cost of $\Gamma_n$.
Since both solutions are driven by the same Brownian motion,
this cost can be expressed in terms of the differences between
their drift and diffusion coefficients along the trajectories.
On the common probability space, write
$X^n=x_0+M^n+A^n$ and $X=x_0+M+A$, where
\begin{equation*}
    \begin{aligned}
        M_t^n&=\int_0^t\sigma^{\lambda_n}(s,X_s^n)\,dW_s,
        & A_t^n&=\int_0^t\alpha^{\lambda_n}(s,X_s^n)\,ds,\\
        M_t&=\int_0^t\sigma^\lambda(s,X_s)\,dW_s,
        & A_t&=\int_0^t\alpha^\lambda(s,X_s)\,ds.
    \end{aligned}
\end{equation*}
These are the canonical semimartingale decompositions of the
two solutions: the drift integrals are adapted continuous
processes of finite variation, and the remaining terms are
square-integrable martingales also in the respective natural
filtrations.
Since the same Brownian motion drives both solutions,
\begin{equation*}
    M_t^n-M_t=\int_0^t\delta\sigma_s^n\,dW_s,
    \qquad
    A_t^n-A_t=\int_0^t\delta\alpha_s^n\,ds.
\end{equation*}
Consequently, the trace of the quadratic variation and the
total variation entering the adapted cost satisfy
\begin{equation*}
    \operatorname{Tr}[M^n-M]_T
       =\int_0^T\|\delta\sigma_t^n\|^2\,dt,
    \qquad
    \operatorname{Var}_{[0,T]}(A^n-A)
       =\int_0^T|\delta\alpha_t^n|\,dt.
\end{equation*}
The bicausality established in
\eqref{eq:primal-stability-bicausal-coupling} makes $\Gamma_n$
an admissible coupling in~\eqref{eq:adapted-wasserstein-distance}.
Evaluating its cost on the common probability space therefore gives
\begin{equation}\label{eq:primal-stability-adapted-cost-bound}
    \begin{aligned}
    \mathcal{AW}_2^2\!
       \left(\mathbb P^{\star,\pi_n},\mathbb P^{\star,\pi}\right)
    &\leq
       \overline{\mathbb E}\left[\int_0^T
          \|\sigma^{\lambda_n}(t,X_t^n)-\sigma^\lambda(t,X_t)\|^2\,dt\right]\\
    &\quad+
       \overline{\mathbb E}\left[
          \left(\int_0^T
             |\alpha^{\lambda_n}(t,X_t^n)-\alpha^\lambda(t,X_t)|
             \,dt\right)^2\right].
    \end{aligned}
\end{equation}

\medskip
To show that this cost tends to zero, we prove convergence
of the drift and diffusion differences in $L^2(dt\,d\overline{\mathbb P})$.
We again separate the intervals away from the maturity times, where
we use the coefficient estimates and the convergence of the trajectories,
from the remaining intervals, where uniform boundedness suffices.
Set
\begin{equation}\label{eq:primal-stability-integrated-coefficient-error}
    J_n:=\overline{\mathbb E}\left[
       \int_0^T
          \bigl(|\delta\alpha_t^n|^2+\|\delta\sigma_t^n\|^2\bigr)
       \,dt\right].
\end{equation}
Fix $\varepsilon\in(0,\Delta_\tau)$.
On each interval $[\tau_{p-1},\tau_p-\varepsilon]$, integrating
\eqref{eq:primal-stability-coefficient-differences} gives
\begin{equation*}
    \sum_{p=1}^P\overline{\mathbb E}\left[
       \int_{\tau_{p-1}}^{\tau_p-\varepsilon}
          \bigl(|\delta\alpha_t^n|^2+\|\delta\sigma_t^n\|^2\bigr)
       \,dt\right]
    \leq C_{K,\varepsilon}
       \left(
          |\lambda_n-\lambda|^2
          +\overline{\mathbb E}
             \left[\sup_{0\leq t\leq T}|D_t^n|^2\right]
       \right),
\end{equation*}
where the time factor is absorbed into $C_{K,\varepsilon}$.
On the remaining intervals, the uniform bound
\eqref{eq:feedback-uniform-bounds} yields
\begin{equation*}
    \sum_{p=1}^P\overline{\mathbb E}\left[
       \int_{\tau_p-\varepsilon}^{\tau_p}
          \bigl(|\delta\alpha_t^n|^2+\|\delta\sigma_t^n\|^2\bigr)
       \,dt\right]
    \leq CP\varepsilon,
\end{equation*}
with $C$ independent of $n$ and $\varepsilon$.
Combining the two estimates, we obtain
\begin{equation}\label{eq:primal-stability-integrated-error-bound}
    J_n\leq C_{K,\varepsilon}
       \left(
          |\lambda_n-\lambda|^2
          +\overline{\mathbb E}
             \left[\sup_{0\leq t\leq T}|D_t^n|^2\right]
       \right)+CP\varepsilon.
\end{equation}
Let $\eta>0$ and choose $\varepsilon\in(0,\Delta_\tau)$
such that $CP\varepsilon<\eta/2$.
For this fixed $\varepsilon$, the convergence
$\lambda_n\to\lambda$ and
\eqref{eq:primal-stability-path-convergence} give an integer
$N\geq1$ such that the first term on the right-hand side of
\eqref{eq:primal-stability-integrated-error-bound} is less
than $\eta/2$ for every $n\geq N$.
Thus $J_n<\eta$ for every $n\geq N$, proving
\begin{equation}\label{eq:primal-stability-integrated-error-convergence}
    J_n\xrightarrow[n\to\infty]{}0.
\end{equation}

It remains to control the squared total variation term
in the coupling cost. Cauchy--Schwarz in time gives
\begin{equation*}
    \overline{\mathbb E}\left[
       \left(\int_0^T|\delta\alpha_t^n|\,dt\right)^2
    \right]
    \leq T\,\overline{\mathbb E}\left[
       \int_0^T|\delta\alpha_t^n|^2\,dt
    \right].
\end{equation*}
Combining this inequality with
\eqref{eq:primal-stability-adapted-cost-bound} and the definition
of $J_n$, we conclude that
\begin{equation*}
    \mathcal{AW}_2^2\!
       \left(\mathbb P^{\star,\pi_n},\mathbb P^{\star,\pi}\right)
    \leq \max\{1,T\}\,J_n
    \xrightarrow[n\to\infty]{}0.
\end{equation*}
This proves~\eqref{eq:primal-law-adapted-stability} and hence
the continuity of the optimal law with respect to the price vector.
\end{proof}

\begin{remark}[Consequences and scope of the stability result]
\label{rk:primal-stability-scope}
The proof constructs the optimal processes on a common probability
space, driven by the same Brownian motion, and establishes
convergence in the expected squared uniform norm, as stated
in~\eqref{eq:primal-stability-path-convergence}.
In particular, their time marginals converge in the usual
Wasserstein distance of order two, uniformly over $t\in[0,T]$.

This also implies continuity of model prices for European options
beyond those used as calibration constraints. In particular, in
log-price coordinates, the put payoff $x\mapsto(K-e^x)^+$ is
$K$-Lipschitz. Hence put prices converge for every strike $K>0$
and every maturity $t\in[0,T]$, uniformly in maturity and over
bounded strike ranges, whether or not these strikes and maturities
occur among the calibration instruments. When forward prices and
discount factors are fixed, the same conclusion holds for calls
by put--call parity.

The argument establishes continuity without a quantitative rate.
Indeed, the constants $C_{K,\varepsilon}$ in the interior
coefficient estimates may diverge as $\varepsilon\downarrow0$;
the proof therefore does not yield local Lipschitz continuity
of the optimal law with respect to the prices.
\end{remark}

\begin{remark}[Extension to a fixed random initial condition]
\label{rk:primal-stability-random-initial-law}
The stability argument extends to a fixed initial law
$\mu_0\in\mathcal P_2(\R^d)$, where
$\mathcal P_2(\R^d):=\{\mu\in\mathcal P(\R^d):
\int_{\R^d}|x|^2\,\mu(dx)<\infty\}$,
under the other hypotheses of the theorem.
Indeed, one can construct the solutions with
the same initial random variable $\xi\sim\mu_0$ and the same
Brownian motion, independent of $\xi$.
Since the coefficient estimates are uniform in space and
$X_0^n=X_0=\xi$ almost surely, the estimates for the difference
of the solutions remain unchanged. The bicausality argument
also applies, using the filtration generated by $\xi$ and
the Brownian motion.
For random initial conditions, the definition of
$\mathcal{AW}_2$ must additionally account for the initial
displacement, by including $|X_0-Y_0|^2$ in the coupling cost.
This term vanishes for the coupling constructed above.
The resulting extension concerns stability with respect
to the prices while keeping $\mu_0$ fixed.
\end{remark}

\appendix
\section{Convex analysis}\label{app:convex-analysis}

\begin{lemma}[Local Lipschitz continuity of the gradient of a convex conjugate]
\label{lem:local-lipschitz-conjugate}
Let $n\geq1$ and let $f:\R^n\to\R$ be strictly convex.
Then its convex conjugate $f^*$ satisfies
\begin{equation*}
    U:=\operatorname{int}(\operatorname{dom}f^*)\neq\varnothing,
    \qquad f^*\in C^1(U).
\end{equation*}
If, in addition, $f$ is locally strongly convex on $\R^n$,
in the sense of~\eqref{eq:dual-value-local-strong-convexity},
then $\nabla f^*$ is locally Lipschitz continuous on $U$.
\end{lemma}

\begin{proof}
    The first part follows from Theorem E.4.1.1 in \cite{hiriart-convex}.
    Assume now that $f$ is locally strongly convex on $\R^n$.
    Let $K^*$ be a nonempty compact subset of
    $\mathrm{int}(\dom{f^*})$ and define
    $K=\text{conv}(\nabla f^*(K^*))$.
    Since $\nabla f^*$ is continuous on $\mathrm{int}(\dom{f^*})$,
    its image $\nabla f^*(K^*)$ is compact.
    By Theorem~17.2 in \cite{rockafellar-1970a}, its convex hull $K$
    is therefore compact as well.
    Denote by $c_K>0$ a modulus of strong convexity of $f$ on $K$.
    Let $x^*_1,x^*_2$ in $K^*,$ and set $x_i = \nabla f^*(x^*_i)$ for $i=1,2.$
    Then $x_i\in K$ and $x_i\in\partial f^*(x_i^*)$.
    Since $f$ is finite and convex on $\R^n$, it is proper and closed.
    By the subdifferential inversion formula
    \cite[Corollary~23.5.1]{rockafellar-1970a},
    we obtain $x_i^*\in\partial f(x_i)$.
    Hence, the strong convexity of $f$ on $K$ implies that \begin{equation*}
        (x_1^*-x_2^*)\cdot(x_1-x_2) \geq c_K |x_1-x_2|^2.
    \end{equation*} Applying the Cauchy--Schwarz inequality, we conclude that \begin{equation*}
        |\nabla f^*(x_1^*)-\nabla f^*(x_2^*)| \leq \frac{1}{c_K} |x_1^*-x_2^*|. \qedhere
    \end{equation*}
\end{proof}

\section{Estimates for the HJB feedback coefficients}
\label{app:feedback-coefficient-estimates}

\begin{lemma}[Bounds and stability of the feedback coefficients]
\label{lem:feedback-coefficient-estimates}
Under the hypotheses of Proposition~\ref{prop:hjb-dual-identification},
assume in addition that condition~\ref{ass:dual-stability-regularity}
holds. Let $(\alpha^\lambda,\sigma^\lambda)$ be
defined by~\eqref{eq:primal-stability-feedback}, with the endpoint
convention specified there, and set
$\Delta_\tau:=\min_{1\leq p\leq P}(\tau_p-\tau_{p-1})$.
The following properties hold.
\begin{enumerate}[label=(\roman*)]
\item There exist constants $C_0<\infty$ and $\kappa>0$,
independent of $\lambda$, such that, for all
$\lambda\in\R^{MP}$ and $(t,x)\in[0,T]\times\R^d$,
\begin{equation}\label{eq:feedback-uniform-bounds}
    |\alpha^\lambda(t,x)|+\|\sigma^\lambda(t,x)\|\leq C_0,
    \qquad
    \sigma^\lambda(t,x)\bigl(\sigma^\lambda(t,x)\bigr)^\top
    \succeq\kappa I_d.
\end{equation}
\item For every nonempty compact set $K\subset\R^{MP}$ and
every $\varepsilon\in(0,\Delta_\tau)$, there exists
$C_{K,\varepsilon}<\infty$ such that, for every
$p\in\{1,\ldots,P\}$ and $t\in[\tau_{p-1},\tau_p-\varepsilon]$,
\begin{equation}\label{eq:feedback-parameter-stability}
    |\alpha^{\lambda_1}(t,x)-\alpha^{\lambda_2}(t,x)|
    +\|\sigma^{\lambda_1}(t,x)-\sigma^{\lambda_2}(t,x)\|
    \leq C_{K,\varepsilon}|\lambda_1-\lambda_2|
\end{equation}
for all $\lambda_1,\lambda_2\in K$ and $x\in\R^d$, and
\begin{equation}\label{eq:feedback-spatial-lipschitz}
    |\alpha^\lambda(t,x)-\alpha^\lambda(t,y)|
    +\|\sigma^\lambda(t,x)-\sigma^\lambda(t,y)\|
    \leq C_{K,\varepsilon}|x-y|
\end{equation}
for all $\lambda\in K$ and $x,y\in\R^d$.
\end{enumerate}
The constants $C_0$ and $\kappa$ depend only on the dimension
and the uniform bounds and ellipticity constants in the hypotheses.
The constant $C_{K,\varepsilon}$ may additionally depend on
$K$, $\varepsilon$, the maturity times, the supremum norms of
the payoffs, and the structural regularity constants of $H$.
It is independent of $p$, $t$, and the spatial points in the
stated domains.
\end{lemma}

\begin{proof}
We first establish the uniform bounds and ellipticity
in~\textup{(i)} from the assumptions on the running cost and the Hamiltonian.
To prove~\textup{(ii)}, we control the terminal data of the HJB cascade
and their dependence on the multiplier. The interior PDE estimates then
yield Lipschitz bounds for the drift and covariance coefficients in both
the state variable and the multiplier, away from the maturity times.
Finally, uniform ellipticity allows us to transfer the covariance estimates
to the diffusion coefficients by taking matrix square roots.

\medskip
To prove~\textup{(i)}, we show that the feedback drift and
covariance belong to the effective domain of $F$, which gives their
uniform boundedness. The ellipticity assumption gives the required lower
bound for the covariance between maturity times. At the maturity times
themselves, the chosen values $\alpha^\lambda=0$ and $\sigma^\lambda=I_d$
also satisfy the boundedness and ellipticity estimates.
Recall from~\eqref{eq:primal-stability-feedback} that
$\beta^\lambda=\sigma^\lambda(\sigma^\lambda)^\top$ is the
covariance matrix associated with the diffusion coefficient
$\sigma^\lambda$.
Fix $\lambda\in\R^{MP}$ and a point
$(t,x)\in(\tau_{p-1},\tau_p)\times\R^d$, and set
$z:=(\Dx\phi_p^\lambda(t,x),\tfrac12\Dxx\phi_p^\lambda(t,x))$.
Condition~\ref{ass:dual-stability-regularity} ensures that
$H$ is twice continuously differentiable in its last two variables.
Moreover, Remark~\ref{rk:hamiltonian-duality-properties}
ensures that $H(t,x,\cdot,\cdot)$ is finite on
$\R^d\times\mathcal S^d$ under~\eqref{ass:convex-duality-d1}.
The interior regularity of the cascade in
\eqref{ass:hjb-slabwise-h1} ensures that $z$ is a point of this space.
By~\eqref{eq:primal-stability-feedback},
\begin{equation*}
    (\alpha^\lambda,\beta^\lambda)(t,x)
    =\nabla_{(a,b)}H(t,x,z).
\end{equation*}
Since $F(t,x,\cdot,\cdot)$ is proper, closed, and convex and
$H=F^*$,
the Fenchel equality criterion
\cite[Theorem~23.5]{rockafellar-1970a} gives
\begin{equation*}
    F\!\left(t,x,\alpha^\lambda(t,x),\beta^\lambda(t,x)\right)
    =\left\langle z,
       (\alpha^\lambda,\beta^\lambda)(t,x)\right\rangle
       -H(t,x,z).
\end{equation*}
Since $H(t,x,z)$ and the scalar product on the right-hand side
are finite, this equality implies that
$F(t,x,\alpha^\lambda(t,x),\beta^\lambda(t,x))<+\infty$.
Hence $(\alpha^\lambda,\beta^\lambda)(t,x)$ belongs to
$\operatorname{dom}F(t,x,\cdot,\cdot)$.
By~\eqref{ass:convex-duality-d1}, there is a constant $C_F<\infty$,
independent of $\lambda$, $p$, $t$, and $x$, such that
$|\alpha^\lambda(t,x)|\leq C_F$ and
$\|\beta^\lambda(t,x)\|\leq C_F$.
As $\beta^\lambda$ is positive semidefinite and
$\sigma^\lambda=(\beta^\lambda)^{1/2}$, the Frobenius norm satisfies
\begin{equation*}
    \|\sigma^\lambda(t,x)\|^2
    =\operatorname{Tr}\beta^\lambda(t,x)
    \leq\sqrt d\,\|\beta^\lambda(t,x)\|
    \leq\sqrt d\,C_F.
\end{equation*}
Moreover, the uniform ellipticity in
condition~\ref{ass:dual-stability-regularity} gives, on
$(\tau_{p-1},\tau_p)\times\R^d$, $\beta^\lambda(t,x)\succeq\kappa_p I_d$
for some $\kappa_p>0$ independent of $\lambda$, $t$, and $x$.
Since $\sigma^\lambda(\sigma^\lambda)^\top=\beta^\lambda$,
this is the required ellipticity bound there.

At each endpoint, the chosen values
$\alpha^\lambda=0$ and $\sigma^\lambda=I_d$ satisfy
$|\alpha^\lambda|+\|\sigma^\lambda\|=\sqrt d$ and
$\sigma^\lambda(\sigma^\lambda)^\top=I_d$.
Thus~\eqref{eq:feedback-uniform-bounds} holds on the whole
time interval with
\begin{equation*}
    C_0:=\max\{C_F+d^{1/4}\sqrt{C_F},\sqrt d\},
    \qquad
    \kappa:=\min\{1,\kappa_1,\ldots,\kappa_P\}>0.
\end{equation*}
This proves~\textup{(i)}.

\medskip
To prove~\textup{(ii)}, we first establish uniform bounds
for the terminal data of the cascade when the multiplier ranges over
a compact set, together with Lipschitz dependence on the multiplier
in the supremum norm. These bounds will allow us to apply the interior
PDE estimates with uniform constants.
Fix a nonempty compact set $K\subset\R^{MP}$. For each $p$, write
\begin{equation*}
    h_p^\lambda
    :=\phi_p^\lambda(\tau_p,\cdot)
    =\phi_{p+1}^\lambda(\tau_p,\cdot)+g_p^\lambda,
    \qquad
    g_p^\lambda=\sum_{m=1}^M\lambda_{i(p,m)}G_{i(p,m)},
\end{equation*}
with $\phi_{P+1}^\lambda(\tau_P,\cdot)=0$.
These terminal data belong to $\BUC(\R^d)$ by the cascade
construction. Set
\begin{equation*}
    C_G:=\left(\sum_{p=1}^P\sum_{m=1}^M
                       \|G_{i(p,m)}\|_\infty^2\right)^{1/2}.
\end{equation*}
After translating time, the uniform-norm stability estimate in
Proposition~3.1 of~\cite{HJB-property} gives
\begin{equation*}
    \|\phi_p^{\lambda_1}-\phi_p^{\lambda_2}\|_
        {\infty;[\tau_{p-1},\tau_p]\times\R^d}
    \leq\|h_p^{\lambda_1}-h_p^{\lambda_2}\|_\infty.
\end{equation*}
Using the terminal relation in~\eqref{eq:hjb-cascade-duality}
and iterating backward from $p=P$, we obtain, for all
$\lambda_1,\lambda_2\in K$,
\begin{equation}\label{eq:feedback-terminal-data-stability}
    \|h_p^{\lambda_1}-h_p^{\lambda_2}\|_\infty
    \leq\sum_{q=p}^P\|g_q^{\lambda_1}-g_q^{\lambda_2}\|_\infty
    \leq C_G|\lambda_1-\lambda_2|.
\end{equation}
The last inequality follows from Cauchy--Schwarz over the
payoff indices.

We also obtain a uniform bound on these terminal data.
Fix $\lambda_0\in K$. By~\eqref{eq:feedback-terminal-data-stability},
\begin{equation}\label{eq:feedback-terminal-data-bound}
    \sup_{\lambda\in K}\max_{1\leq p\leq P}\|h_p^\lambda\|_\infty
    \leq
    \max_{1\leq p\leq P}\|h_p^{\lambda_0}\|_\infty
    +C_G\sup_{\lambda\in K}|\lambda-\lambda_0|
    <\infty.
\end{equation}
The finiteness follows from $h_p^{\lambda_0}\in\BUC(\R^d)$
and the compactness of $K$.

\medskip
We now use the interior PDE estimates to control the
spatial derivatives of the HJB solutions and their dependence on the
multiplier, away from each maturity. Through the feedback
formula~\eqref{eq:primal-stability-feedback}, these estimates give the
required Lipschitz bounds for $\alpha^\lambda$ and $\beta^\lambda$.
Fix $\varepsilon\in(0,\Delta_\tau)$ and choose $R>0$ strictly
larger than the bound in~\eqref{eq:feedback-terminal-data-bound}.
After translating time on each time interval, Propositions~3.3 and~3.4
of~\cite{HJB-property} apply to the family
$(\phi_p^\lambda)_{\lambda\in K}$ with terminal data
$(h_p^\lambda)_{\lambda\in K}$.
Write
\begin{equation*}
    z_p^\lambda(t,x)
    :=\left(\Dx\phi_p^\lambda(t,x),
             \tfrac12\Dxx\phi_p^\lambda(t,x)\right).
\end{equation*}
The interior derivative bounds and the stability estimate in
Proposition~3.4, together with~\eqref{eq:feedback-terminal-data-stability},
give
\begin{align*}
    |z_p^\lambda(t,x)|&\leq C_{K,\varepsilon},\\
    |z_p^{\lambda_1}(t,x)-z_p^{\lambda_2}(t,x)|
    &\leq C_{K,\varepsilon}
        \|h_p^{\lambda_1}-h_p^{\lambda_2}\|_\infty
     \leq C_{K,\varepsilon}|\lambda_1-\lambda_2|.
\end{align*}
Moreover, the bound on $\Dx\phi_p^\lambda$ in the parabolic
$C^{2+\alpha}$ norm from Proposition~3.4 controls the second
and third spatial derivatives of $\phi_p^\lambda$, and hence
\begin{equation*}
    |z_p^\lambda(t,x)-z_p^\lambda(t,y)|
    \leq C_{K,\varepsilon}|x-y|.
\end{equation*}
These estimates hold for $\lambda,\lambda_1,\lambda_2\in K$,
$t\in(\tau_{p-1},\tau_p-\varepsilon]$, and $x,y\in\R^d$.
Since there are finitely many time intervals, their constants can be
chosen independent of $p$. Here and below, $C_{K,\varepsilon}$
may increase from one estimate to the next.

By~\eqref{eq:primal-stability-feedback},
$(\alpha^\lambda,\beta^\lambda)(t,x)
=\nabla_{(a,b)}H(t,x,z_p^\lambda(t,x))$.
The Lipschitz estimates for $\nabla_{(a,b)}H$ following from
Assumption~1 of~\cite{HJB-property}, and the composition property
in Remark~2.4 therein, combined with the preceding uniform
bounds on $z_p^\lambda$, yield
\begin{align*}
    |\alpha^{\lambda_1}(t,x)-\alpha^{\lambda_2}(t,x)|
    +\|\beta^{\lambda_1}(t,x)-\beta^{\lambda_2}(t,x)\|
    &\leq C_{K,\varepsilon}|\lambda_1-\lambda_2|,\\
    |\alpha^\lambda(t,x)-\alpha^\lambda(t,y)|
    +\|\beta^\lambda(t,x)-\beta^\lambda(t,y)\|
    &\leq C_{K,\varepsilon}|x-y|.
\end{align*}
At $t=\tau_{p-1}$, these two inequalities also hold because
the prescribed endpoint values are independent of both
$\lambda$ and $x$.

\medskip
It remains to obtain the same Lipschitz estimates for
$\sigma^\lambda=(\beta^\lambda)^{1/2}$. We use the uniform ellipticity
established in~\textup{(i)} to show that taking the symmetric positive
definite square root preserves these estimates.
For symmetric matrices $B_1,B_2\succeq\kappa I_d$, their
symmetric positive definite square roots satisfy
\begin{equation}\label{eq:feedback-square-root-lipschitz}
    \|B_1^{1/2}-B_2^{1/2}\|
    \leq\frac{1}{2\sqrt\kappa}\|B_1-B_2\|.
\end{equation}
Indeed, setting $S=B_1^{1/2}-B_2^{1/2}$ gives
$B_1-B_2=B_1^{1/2}S+SB_2^{1/2}$.
Taking the Frobenius inner product with $S$ and using
$B_i^{1/2}\succeq\sqrt\kappa I_d$ yields
\begin{align*}
    2\sqrt\kappa\,\|S\|^2
    &\leq \operatorname{Tr}(S^\top B_1^{1/2}S)
          +\operatorname{Tr}(S^\top S B_2^{1/2})\\
    &=S:(B_1-B_2)
    \leq\|S\|\,\|B_1-B_2\|.
\end{align*}
This proves~\eqref{eq:feedback-square-root-lipschitz},
including the case $S=0$.

By~\textup{(i)}, $\beta^\lambda(t,x)\succeq\kappa I_d$
for every $\lambda$, $t$, and $x$.
Applying~\eqref{eq:feedback-square-root-lipschitz} first to
$B_i=\beta^{\lambda_i}(t,x)$, and then to
$B_1=\beta^\lambda(t,x)$ and $B_2=\beta^\lambda(t,y)$,
transfers the preceding covariance estimates to
$\sigma^\lambda=(\beta^\lambda)^{1/2}$.
Together with the estimates for $\alpha^\lambda$, this gives
\eqref{eq:feedback-parameter-stability} and
\eqref{eq:feedback-spatial-lipschitz}, after multiplying
$C_{K,\varepsilon}$ by $\max\{1,(2\sqrt\kappa)^{-1}\}$.
This completes the proof of~\textup{(ii)}.
\end{proof}

\newpage
\addcontentsline{toc}{section}{References}
\bibliographystyle{alpha}
\bibliography{biblio}

\end{document}